\documentclass[tradiabstract,twocolumn]{aa} 
\usepackage{graphicx}
\usepackage{xcolor}
\usepackage{fontawesome}
\usepackage{txfonts}
\usepackage[colorlinks=true,linkcolor=blue,citecolor=blue, urlcolor=blue]{hyperref}
\newcommand{\kms}{km~s$^{-1}$}

\usepackage{url}

\begin{document}

   \title{Strong lens systems search in the Dark Energy Survey using Convolutional Neural Networks}


   \author{K. Rojas\inst{\ref{epfl}}\and
           E. Savary\inst{\ref{epfl}}\and
           B. Cl\'ement\inst{\ref{epfl}}\and
           M. Maus\inst{\ref{epfl}}\and
           F. Courbin\inst{\ref{epfl}}\and
           C. Lemon\inst{\ref{epfl}}\and
           J. H. H. Chan\inst{\ref{epfl}}\and
           G. Vernardos\inst{\ref{epfl}}\and
           R. Joseph\inst{\ref{princeton}}\and  
           R. Ca\~nameras\inst{\ref{maxplanck}}\and 
           A. Galan\inst{\ref{epfl}}}

   \institute{Institute of Physics, Laboratory of Astrophysics, Ecole Polytechnique 
F\'ed\'erale de Lausanne (EPFL), Observatoire de Sauverny, 1290 Versoix, 
Switzerland \label{epfl}
    \and
    Department of Astrophysical Sciences, Princeton University, Princeton, NJ 08544, USA \label{princeton}
    \and
Max-Planck-Institut für Astrophysik, Karl-Schwarzschild-Str. 1, 85748 Garching, Germany\label{maxplanck}}


 
\abstract{
We performed a search for strong lens galaxy-scale systems in the first data release of the Dark Energy Survey (DES), from a color-selected parent sample of 18~745~029 Luminous Red Galaxies (LRGs). Our search was based on a Convolutional Neural Network (CNN) to grade our LRG selection with values between 0 (non-lens) and 1 (lens). Our training set was data-driven, i.e. using lensed sources taken from HST COSMOS images and where the light distribution of the lens plane was taken directly from DES images of our LRGs. A total of 76~582 cutouts obtained a score above 0.9. These were visually inspected and resulted in two catalogs. The first one contains 405 lens candidates, where 90 present clear lensing features and counterparts, while the others 315 require more evidence, such as higher resolution images or spectra to be conclusive. A total of 186 candidates were totally new identified in this search. The second catalog  includes 539 ring galaxy candidates that will be useful to train CNNs against this type of false positives. For the 90 best lens candidates we carried out color-based deblending of the lens and source light without fitting any analytical profile to the data. The method turned out to be very efficient in the deblending, even for very compact objects and for objects with very complex morphology. Finally, from the 90 best lens candidates we selected 52 systems having one single deflector, to test an automated modeling pipeline which successfully modeled 79\% of the sample within an acceptable amount of computing time.}

 
\maketitle
%

\section{Introduction}\label{sec:intro}

Gravitational lensing is a phenomenon produced by the deflection of the light rays by a gravitational field. In strong lens systems it is possible to observe multiple images, arcs or rings of a distant source around a foreground galaxy, group or cluster. When this happens this effect serves as an important tool to study diverse and fundamental questions about the Universe. Some examples are the study of luminous and dark matter components of the deflector \citep{Kochanek2001,Oguri2002,Davis2003,Jimenez-Vicente2015a}, measure the Hubble constant H$_0$ using time delays \citep{Falco1997,Vuissoz2007,Bonvin2017,Wong2020,Millon2020}, constraint the dark energy equation of state \citep{Biesiada2010,Collett-Auger2014,Cao2012,Cao2015}, among many others. Nevertheless, most of these applications find a limit in their analysis because the number of confirmed systems up-to-date only reach hundreds. Therefore an effort on the detection and confirmation of more lenses is required. 

The discovery of these systems have evolved through time, from the serendipitous first lensed quasar discovered \citep{Walsh1979} to more dedicated searches using a variety of techniques. Source-selected sample of quasars to find multiple images of the same source \citep[e.g.][]{Oguri2006,Inada2012,Agnello2018}. Spectroscopic selection of Luminous Red Galaxies (LRG) containing emission lines from a background source \citep{Bolton2004,Bolton2006,Brownstein2012}. Algorithms based on lens features like Arc-finder, which searches for elongated structures \citep{Alard2006}, Ringfinder that searches for blue features blended with red light \citep{Gavazzi2014}, Principal Component Analysis (PCA) of potential lens galaxies to search lensed features in the residual images using machine learning \citep{Joseph2014,Paraficz2016}, and CHITAH which inspect the image configurations using lens modeling \citep{Chan2015}. In recent years the growing amount of available data has motivated the use of more automated techniques like Artificial Neural Networks (ANNs)  \citep{Rosenblatt1957} and in particular Convolutional Neural Networks \citep[CNNs;][]{LeCun1989}. This technique is based on supervised machine learning algorithms capable of solving complex problems such as pattern recognition or image classification when a proper training set is provided. 

Therefore, the big challenge of using CNNs for lens finding is to create a robust training set that contains diverse lens systems for positive examples and different types of galaxies, including some that can be mistaken as lenses like spirals, rings, and mergers, as negative examples. We currently lack both sufficient numbers of lens systems as well as catalogs to represent the diverse population of some possible false positive galaxies, like is the case of ring galaxies. The solution to create a robust sample of lenses is to simulate them as realistic as possible. This has been addressed following different strategies: fully simulated images using analytical profiles for both lens and source \citep{Jacobs2019B}, mix procedure with an analytical profile for the source but a real image of the lens \cite{Petrillo2019}, use of real data for both deflector and background galaxy \citep{Canameras2020}. The first method had the advantage of full space of free parameters to create a sample as varied as possible, but mimicking features from real images like artifacts, noise, and companions is difficult and is the advantage when using real images. 

At present, several searches for lens systems have been carried out. Each uses its own strategy, including the method for simulation, training in single or multiband and the architecture design. Some examples of lens finding, using machine learning, applied to different surveys are: the Canada-France-Hawaii Telescope Legacy Survey (CFHTLS) \citep{Jacobs2017}, the Kilo Degree Survey (KiDS) \citep{Petrillo2017,Petrillo2019B,Petrillo2019,He2020}, the Dark Energy Survey (DES) Year 3 \citep{Jacobs2019B,Jacobs2019A} , the Dark Energy Spectroscopic Instrument (DESI) Legacy Imaging Surveys \citep{Huang2020,Huang2021}, the Pan-STARRS 3$\pi$ survey \citep{Canameras2020}, the VST Optical Imaging of the CDFS and ES1 fields (VOICE survey) \citep{Gentile2021}. Overall, these studies have shown that CNN is a promising technique, listing thousands of new lens candidates, but all of them rely on human visual inspection afterwards to compile the final candidate list, and more effort is needed for confirmation, including spectroscopic data and/or higher resolution imaging. Thus, improving the training process with realistic lenses and diverse types of galaxies is of key importance when the next generation of surveys like Euclid Space Telescope \citep{Laureijs2011} and the Rubin Observatory Legacy Survey of Space and Time  \citep[LSST,][]{LSST2009,LSST2019}, will be operating. All the current lens finding efforts, including some performed in simulated data from these surveys \citep{Lanusse2018,Avestruz2019}, serve as preparation, as it is expected that over 100~000 new strong lensing systems will be discovered \citep{Collett2015}, and a visual inspection process is not affordable. In this new survey era, we anticipate that numerous applications of the strong lensing effect will be statistically benefited by this increase of confirmed systems. 

In this work our main aim is to find new strong lensing systems. As a secondary goal we want to characterize some of our false positives. In addition, we present a compilation of ring galaxy candidates that will help to improve future lens finding searches. We performed our search in the footprint of the Dark Energy Survey (DES) whose main scientific driver is the study of the accelerated expansion of the Universe, thought to be due to some unknown form of repulsive dark energy \citep{DES2005}. To look for these objects we select a sample of galaxies with colors and magnitudes similar to Luminous Red Galaxies (LRGs) from DES Data Release 1 (DES-DR1). The data description is in Section~\ref{sec:datasel}. To search for lenses in this sample we taught a CNN to distinguish between lensed systems and non-lens galaxies. As lensing is rare and we need thousands of examples for training we simulated strong lensing systems using real images where the background sources are high S/N cutouts from the Hyper Suprime Cam Subaru Strategic Program (HSC-SSP) and the Hubble Space Telescope (HST), while the foreground galaxies are taken from our DES LRGs selection, all the details about the simulation procedure are in Sect.~\ref{sec:simulations}. We trained our CNN using a balanced training set with half mock lenses and half random galaxies taken from the parent sample. See Sect.~\ref{sec:cnn} for details about training and validation. We grade the parent sample using the CNN model, and we select the ones with higher score to perform a dedicated visual inspection to identify the best lens candidates and subclassify false positives. In Sect.~\ref{sec:visualinspec} we enumerate all the steps followed. Finally in Sect.~\ref{sec:model} we show results of an automatic modeling tool on a sample of our best candidates.

\section{Data selection}\label{sec:datasel}

We selected our data from the Dark Energy Survey (DES) that uses the Dark Energy Camera \citep[DECam,][]{Honscheid2008,Flaugher2015} on the Blanco 4-m telescope at Cerro Tololo Inter-American Observatory (CTIO), Chile. DECam is a 570 megapixels camera with a field of view of 2.2 square degrees on-a-side and a pixel size of 0.27\arcsec. The observations are performed in the optical $grizY$ bands and are mostly used for cosmic shear measurements, i.e. getting the shape and photometric redshifts of millions of galaxies. The first DES Data release \citep[DES-DR1,][]{Abbott2018} contains images taken over the first three years of operation, covering an area of 5186 deg$^{2}$. The images from DR1 have been coadded, and each filter has been re-scaled to have a fixed zero point of 30~mag.

\begin{figure}[t!]
\centering
\includegraphics[width=9.6cm]{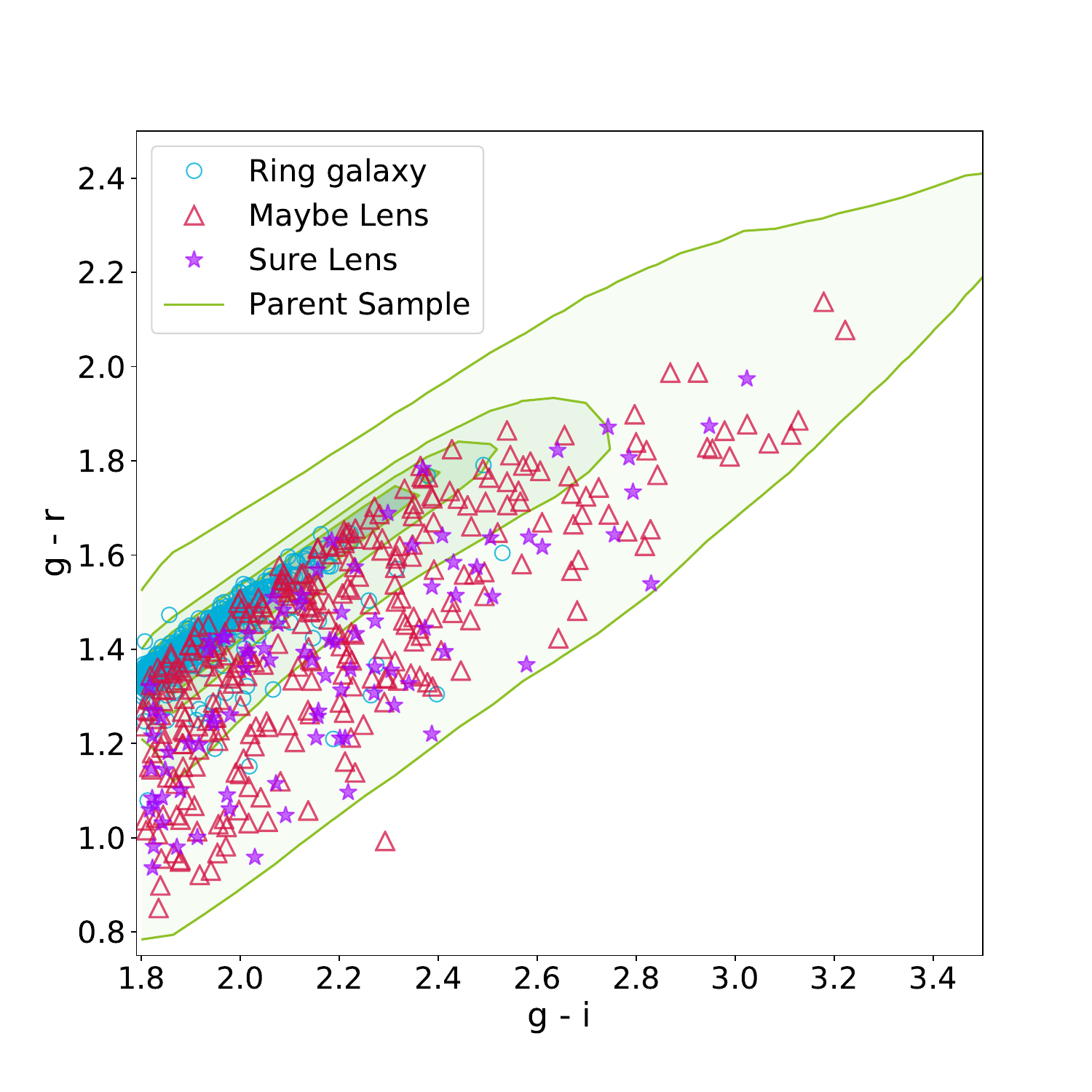}
\caption{Color-color diagram of our parent sample of $\sim$19 million galaxies. The extent of the plot displays a zoom in the ranges $g-i$ and $g-r$ in  Eq.~\ref{eq.1} to highlight where most of the selection is located. The green shaded area shows the density of galaxies from the parent sample in that color space. The green solid lines shows the 1, 2, 3, and 6$\sigma$ contours. The lens candidates and contaminants discussed in this work are shown in overlay : Sure Lens (purple star symbols), Maybe Lens (open red triangles), and ring galaxies (open cyan circles). See Sect. 5.3 for details.}
\label{Fig.colorplot}
\end{figure}

We used the NOAO Data Lab \citep{Fitzpatrick2016} service to build our sample from the {\tt des$\_$dr1.galaxies} catalog and selected a sample of Luminous Red Galaxies (LRGs) in order to maximize the lensing cross-section \citep[e.g.][]{Turner1984}. In doing so, we applied the following cuts in color and magnitude: 
\begin{eqnarray}
\begin{array}{l}
1.8 < g - i < 5, \\
0.6 < g - r < 3, \\
18 < r < 22.5, \\
g > 20, \\
i > 18.2,
\label{eq.1}
\end{array} 
\end{eqnarray}
\noindent where the magnitude system is the {\tt mag$\_$auto} column reported in the DES data release. Our color selection is summarized in Fig.~\ref{Fig.colorplot} and is similar to the one adopted by \cite{Jacobs2019A}. However, we slightly widened the $g-i$ range and adopted a brighter magnitude limit in the $r$-band. This selection allows us to account for the contamination of the lens red colors by the bluer color of any putative lensed source, but also increases the probability that other types of galaxies, e.g. mergers, spirals, or ring galaxies can be selected in the sample. The result is a sample of 18~745~029 galaxies located in 10~388 coadded tiles from DES DR1, which we refer to as the parent sample in this work. We downloaded all the tiles for the $g$, $r$, and $i$ bands for our search for galaxy-scale strong lenses. All generated cutouts have a size of $50 \times 50$ pixels, corresponding to $\sim 13 \arcsec\times 13 \arcsec$. When the Point Spread Function (PSF) is required for the simulation process and modeling, {\tt PSFEx} \citep{bertin2011} is run on the relevant coadd tile, extracting a model of the PSF from the FITS image, allowing us to retrieve a PSF at any position on the tile.

\section{Simulated galaxy-scale lenses}\label{sec:simulations}

Our search for galaxy-scale strong lenses is based on CNNs and therefore requires a training set that mimics in the best possible way the lenses we want to find and also includes a set of non-lensed objects. When simulating lenses, we adopt an approach where the training set is data-driven, in the sense that both the images of the lenses and the background sources are obtained from real data. The outline of our procedure is as follows. First, we select a sample of high-redshift background galaxies with available high-resolution imaging and accurately measured colors (Sect.~\ref{subsec:sourcesel}). Then, we select a sample of LRGs that will act as deflectors and match them to the background sources to create pairs of lens-source suitable for simulations (Sect.~\ref{subsec:lenssel}). Lastly, we act on the source through the lens equation whose parameters (deflection angles) are defined by the lens and produce a simulated image of a lens system that combines images from both samples (Sect.~\ref{subsec:lenssimu}).

\subsection{Background galaxies selection}\label{subsec:sourcesel}

We wanted realistic source galaxies, with real observed morphologies at HST resolution, with color information and high S/N. The background sources used in this work were originally compiled by \cite{Canameras2020}. 
In summary, the galaxies were selected from the Galaxy Zoo catalog \citep{Willett2017}, and are within the COSMOS2015 photometric catalog \citep{Laigle2016}. This survey is by far the one with the broadest galaxy properties and spanning a wide redshift range given the depth. All the objects categorized as galaxies in these catalogs were picked, with no previous selection in color or magnitude, as the depth is limited by the Galaxy Zoo selection, i.e. down to F814W $\sim$ 23.5. The sample was cleaned from stars and artifacts, but also extended galaxies and galaxies with nearby companions were removed, leaving a final sample of 52~696 objects. For all of them, spectroscopic redshifts were obtained from different surveys \citep{Lilly2007,Comparat2015,Silverman2015,LeFevre2015,Tasca2017,Hasinger2018}, and photometric redshift from \cite{Laigle2016} for the ones with no other available. 

To create high resolution $gri$-images of our sources, we created cutouts combining the morphological information from HST/ACS F814W high resolution images \citep{Leauthaud2007,Scoville2007,Koekemoer2007} and the color information from Hyper Suprime Cam (HSC) ultra-deep stack images \citep{Aihara2018}. The detailed procedure to combine the information from these two surveys is described in \cite{Canameras2020} who followed the steps described in \cite{Griffith2012}. In summary, HST/ACS F814W images were aligned and rescaled as if they were observed in HSC $i$-band. These HSC images were then resampled to the resolution of the HST/ACS F814W images and were multiplied by an illumination map obtained by dividing the HST/ACS F814W image by the HSC $i$-band image. At the end, each stamp has a size of $10 \arcsec \times 10 \arcsec$ and a pixel size of 0.03\arcsec, therefore they have the HST resolution and PSF but the relative flux between bands is determined from the HSC colors. The morphology of the source is therefore the same in each band, but not the flux. Since the HST PSF is much sharper than that of the ground-based DES images we do not deconvolve our stamps from the HST PSF, which would introduce noise and possible artefacts.

\subsection{Lens-source pair selection}\label{subsec:lenssel}

\begin{figure}[h!]
\centering
\includegraphics[width=8.9cm]{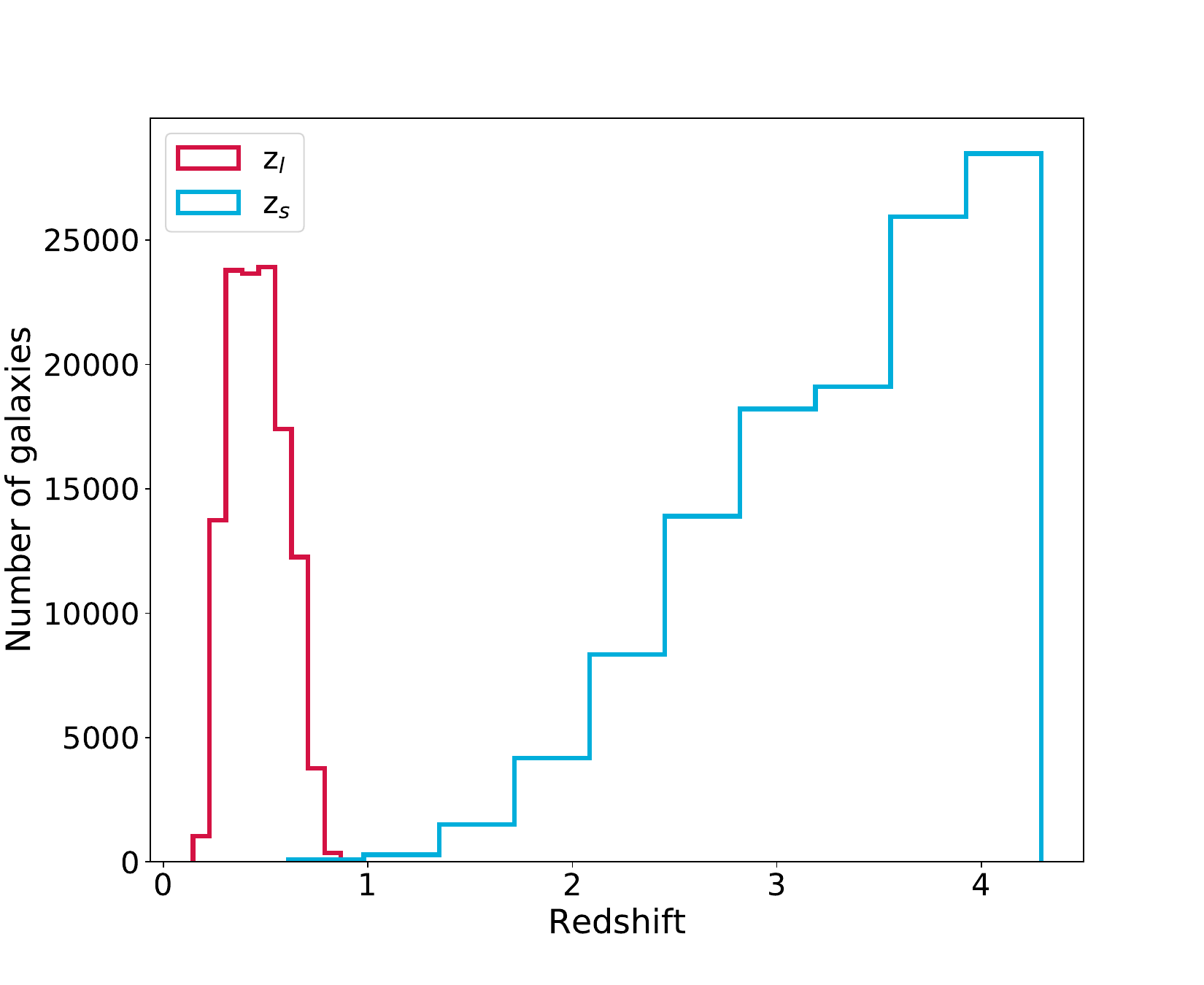}
\includegraphics[width=8.9cm]{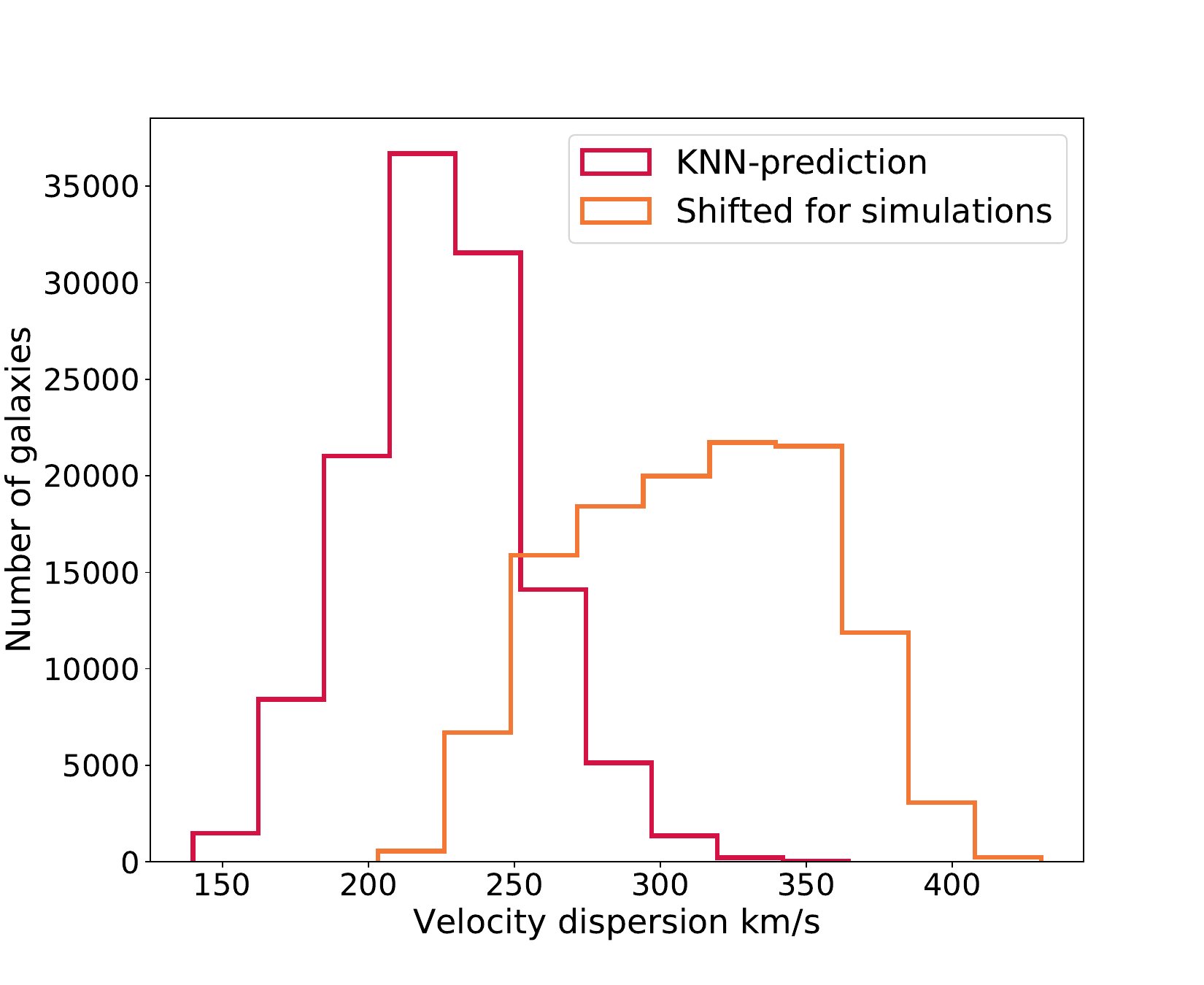}
\caption{{\it Top panel:} redshift distributions of the lenses (red histogram) and sources (blue histogram) in the simulated training set. {\it Bottom panel:} lens velocity dispersion distribution, predicted from the K-nearest neighbor algorithm (red histogram) and shifted to match the criteria of pairing lens and source (orange histogram), as described in Sec.~\ref{subsec:lenssel}. Note that the actual lens velocity dispersions used in the simulations are shifted to higher values, so that lensing features can be seen even at the DES resolution. \label{fig.zandvdisp}.}
\label{FigGam}%
\end{figure}

Most of the objects in our LRG selection are missing relevant parameters for lens galaxy selection and simulation, namely the redshift and velocity dispersion. To cope with this limitation, we perform a prediction of those parameters using a simple k-nearest neighbors (KNN) algorithm, assuming that other galaxies with similar $gri$ magnitudes will also have similar redshifts and velocity dispersions. The KNN algorithm stores all the possible cases from a training set to relate the $gri$ magnitudes with the redshift and velocity dispersion of the galaxies. It then predicts the parameters of the new data based on the k-objects with similar colors. We trained the algorithm with 1~400~000 SDSS galaxies that match the color-magnitude cuts of the parent sample and have redshift and velocity dispersion measurements available. We tested the model on an other set of 99~382 spectroscopically-confirmed SDSS galaxies, obtaining the predictions for the parameters from the ten nearest neighbors in the $gri$ color space of the training set. We found that the $rms$ scatter in the predictions was $\sigma_z=0.06$ for the redshift, and $\sigma_{vel}=69.02$ \kms for the velocity dispersion. Finally, we used this model to predict the most likely redshift and velocity dispersion for each of our galaxies in the parent sample using as base information the $gri$ magnitudes in our catalog. The distributions of the predicted redshift and velocity dispersion are shown in Fig.~$\ref{fig.zandvdisp}$.

As a preparation step before simulation and to select the sample of lens galaxies, we matched LRGs with source galaxies so our simulations have an uniform distribution in Einstein radii spanning $1.2\arcsec < \theta_E < 3.0\arcsec$. We choose a conservative lower limit on the $\theta_E$ because we noticed that beside the good average seeing in the corresponding bands, g = 1.12, r = 0.96, i = 0.88 \citep{Abbott2018}, simulations with $\theta_E < 1.2$ create lensing features that are too close or blended with the lens galaxy, that can be mistaken as e.g. galaxies with extended disk. To estimate the $\theta_E$  we relate the redshift of the source galaxy with the redshift of the lens and its velocity dispersion. To do our selection  we first take a random lens galaxy from the parent sample and we compute the distribution in Einstein radii, $\theta_E$, corresponding to the redshift distribution of all our source galaxies. We then form lens-source pairs that produce Einstein radii within our desired bounds. In case no lens-source pairs satisfy the $\theta_E$ conditions we artificially boost the velocity dispersion of the lens galaxy up to 1.5 times its original value. If still no pair satisfies our criteria we discard the LRG for simulation and use the next one in our list. Note that, as illustrated in Fig.~\ref{fig.zandvdisp}, this results in a high bias of the velocity dispersion distribution of the lenses. Although this procedure tends to produce lenses with dark matter halos larger than predicted, this ensures that the lensing features are clearly noticeable to the CNN.

Finally, we enforce that the final Einstein radii distribution for all simulations is uniform. In other words, our training set is not representative of the true distribution of Einstein radii on the sky, but gives equal probability to all possible values, allowing for more discriminating power in our trained CNN.

\subsection{Lensing simulation}\label{subsec:lenssimu}

We create mock strong lens systems based on real images of galaxies. The lens model that we adopt is a Singular Isothermal Ellipsoid (SIE), which is defined by the Einstein radius ($\theta_E$), the Position Angle (PA) and axis ratio converted into a complex ellipticity ($e_1$, $e_2$), and the central position ($x_1$, $x_2$). As is mentioned before, the Einstein radius follows a uniform distribution, while the other parameters are acquired individually according to the light distribution of each lens galaxy. Our simulation toolbox uses the python package {\tt Lenstronomy}\footnote{\url{https://github.com/sibirrer/lenstronomy}} \citep{Birrer2015,Birrer2018}. 
The first step is to find a simple but realistic representation for the mass of the lensing object. The Einstein radius is calculated using the lens and source redshift as well as the lens velocity dispersion of the lens derived in Sec.~\ref{subsec:lenssel}. The ellipticity and mass centroid were estimated from fitting an elliptical S\'ersic profile to the DES $r$-band image of the LRG. We optimized the fitting procedure using 50 iterations of Particle Swarm Optimization \citep[PSO;][]{Kennedy1995}, with 50 particles. This simple model provides us with parameters for a mass distribution that broadly follows the light distribution of the brightest object in the image. Even though it could produce a few lenses with exotic properties, e.g. very elliptical mass profiles or unusually large dark matter halos, it was found to be adequate to our goal of building realistic simulations in the vast majority of the cases.

The second step is to deflect the light rays from the source according to the mass model of the lens defined before. To ensure we can distinguish the final lensed source features in contrast with the lens galaxy we first boosted the original source brightness by one magnitude. To decide where in the source plane our background galaxy is located, we select a random position inside a square that encloses the caustic curves, that mark the location of the maximum magnification and delimit the region inside which a source will be multiply-imaged. Then, we perform a ray-tracing simulation to map the source image onto the image plane and we further convolve the resulting lensed source with the specific PSF provided for each DES stamp. To convert this image into the DES characteristic pixel resolution we down-sample the pixels from 0.03\arcsec (HST) to 0.27\arcsec (DES), and we re-scale the flux to match the DES zero points in each filter. As a last step, we add the convolved, resized, and flux-normalized image of the lensed source to the original image of the LRG lens. The latter has, by construction, the right DES PSF and noise properties. Thus, our simulations preserve the characteristics of the original image, such as background noise, seeing, the presence of artifacts, and neighbouring galaxies or stars in the field of view.

\begin{figure}[t!]
\centering
\includegraphics[width=8.9cm]{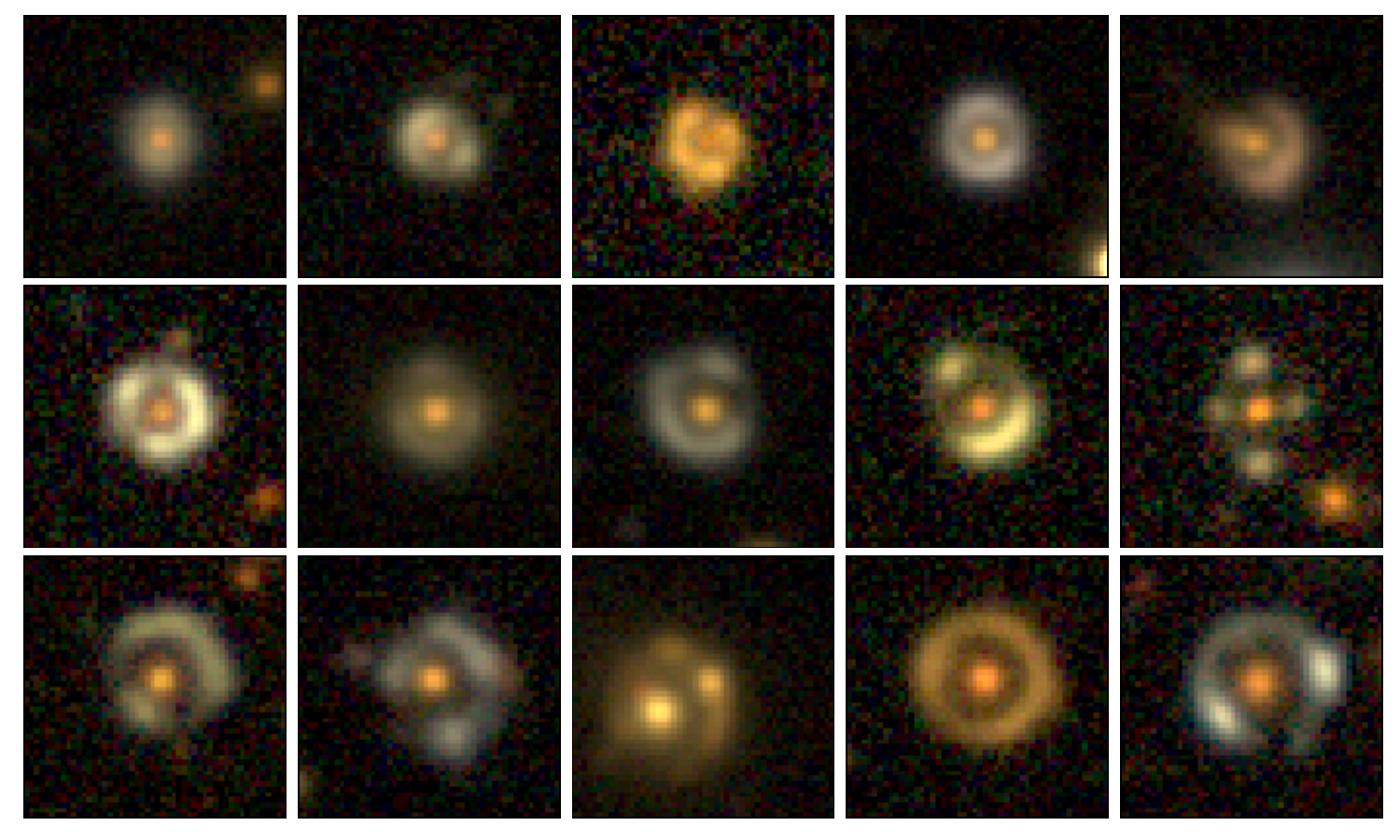}
\includegraphics[width=8.9cm]{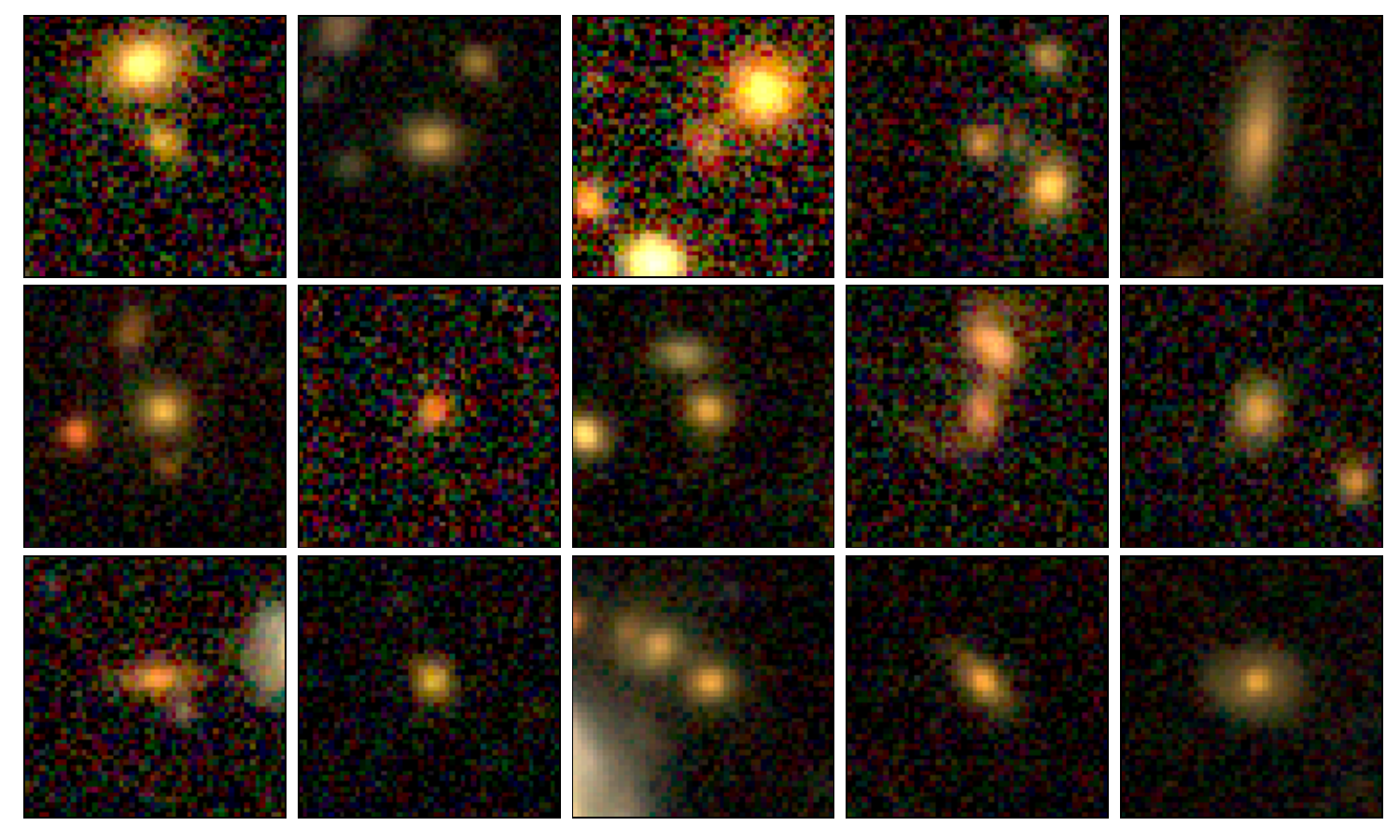}
\caption{{\it Top:} examples of simulated lenses based on real DES images. From top-left to bottom-right are  shown simulated stamps ordered by increasing Einstein radii. The top row corresponds to $\theta_E$=1.2\arcsec-1.8\arcsec, the middle row to $\theta_E$=1.8\arcsec-2.4\arcsec, and the bottom row to $\theta_E$=2.4\arcsec-3.0\arcsec. {\it Bottom:} examples of LRGs used either as non-lenses during the training of the CNNs (see Sec.~\ref{subsec:training}) or as objects on which we inject a lensed source to build simulated lens systems. All the cutouts are 50 pixels on-a-side, corresponding to 13\arcsec \label{fig.simulations}. }
\label{FigGam}%
\end{figure}

To build the multi-band $gri$ simulations we use the same mass model for all bands, with its parameters derived only from the $r$-band, and lens the source image in each band according to this model. We then add the lensed source in each band to the corresponding image of the lens taken from the DES images in $g$, $r$ and $i$ bands. Our final set of simulated galaxy-scale lenses consists of 100~000 systems with a uniformly distributed Einstein radius in the range 1.2\arcsec $< \theta_E <$ 3.0\arcsec. Examples of these stamps are shown in Fig.~\ref{fig.simulations} as well non-lens objects.

\section{Lens Finding using CNN}\label{sec:cnn}

Artificial Neural Networks \citep[ANNs;][]{Rosenblatt1957} consist of an interconnected group of nodes which are typically organized as so-called input, hidden, and output layers. In particular CNNs \citep{LeCun1989}, which are especially good in solving image classification problems \citep{He2015}, have hidden layers that are of key importance, as they highlight the patterns in the data using a series of convolutional, pooling, normalization and fully connected layers. The level of abstraction in the pattern features increases with the depth of the convolutional layers, helping to classify objects among different classes. Here, we train a CNN to recognize strong lens systems against isolated red galaxies. 

\subsection{CNN training}\label{subsec:training}

The training set, consisting of 50$\times$50 pixels cutouts in each of the $gri$-bands, was split into two equal subsets, one being the 100~000 simulated cutouts from Sect. \ref{sec:simulations} and the other containing LRGs that were not used in the simulation process. We labeled our data using 1 for lenses and 0 for non-lenses. We used 20$\%$ of this sample as a validation set. Before training the CNN we preprocessed our data by normalizing each image brightness in the range between 0 and 1 and augmenting it by flipping each image horizontally and vertically. Data augmentation increases the probability that the network correctly classifies different orientations of the same image, but it does not transform the CNN into a rotationally invariant one. To achieve this, a different architecture must be used that is not explored in this work. All the training process is performed using the Keras Deep Learning framework \citep{chollet2015keras}.

Our CNN used a model from the EfficientNet family \citep{tan2020efficientnet}, which has been designed to achieve better performance than other CNNs. To do so, the network used a compound coefficient to scale the depth, width, and resolution, which are key parameters to obtain better accuracy and efficiency. EfficientNet implementation in Keras counts with 8 different variants B0-B7, whose depth, width, and resolution parameters have been carefully selected and tested to produce good results. The complexity and requirements of the models grow as we move from B0 to B7. 
As running a more complex model also implies the use of more computational resources, we decided to use an EfficientNet-B0, whose architecture is described in \citep{tan2020efficientnet}, and is good enough for our classification task and the characteristics of our data resolution. After this CNN model we add a sequence of fully connected hidden layers, so at the end the network has a total of 4~182~205 trainable parameters. 

During training, the neural network learned how to grade images of galaxies and distinguish between lenses and non-lenses. In each iteration the network analyzed subsets of 32 images, when all iterations are completed through the entire training set it is counted as one epoch. During each epoch the accuracy and loss of the model is monitored using the validation set. We minimize a binary cross-entropy loss function using a stochastic gradient descent optimizer (Adam) with a learning rate of 0.0001, and stop the training if the loss value does not improve by more than 0.0001 during 10 epochs or if 100 epochs are reached.

\subsection{Evaluation of the CNN performance}\label{subsec:eval}

\begin{figure}[t!]
\centering
\includegraphics[width=9.6cm]{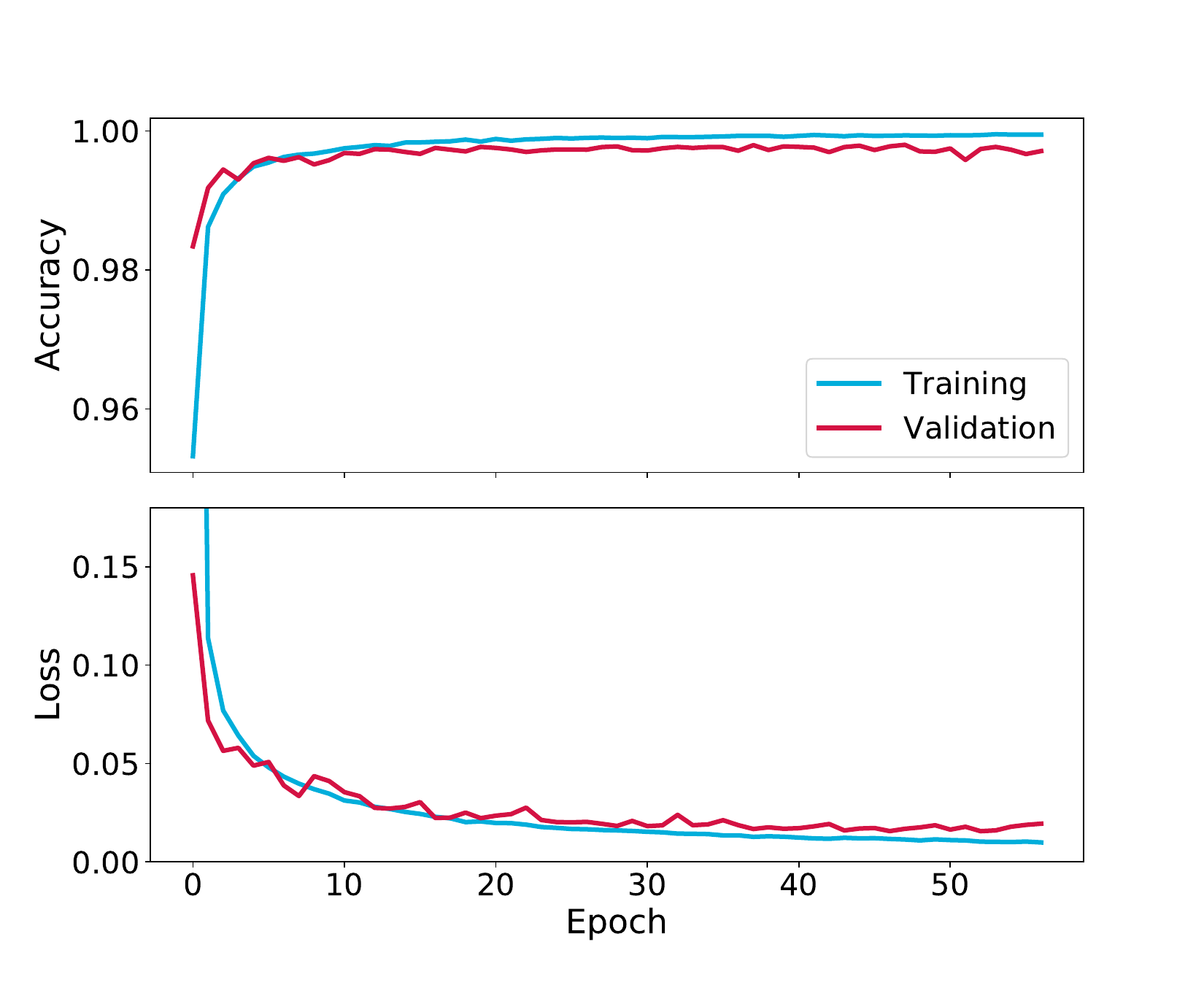}
\caption{Learning curves for the accuracy (top) and loss (bottom) for the training (blue) and validation (red) sets, as a function of epoch. In the case of loss we show a zoom in the y-axis, but the loss in the training set starts decreasing from 0.76.}
\label{Fig:accloss}
\end{figure}

The network provided a score, $S_{\text{CNN}}$, between 0 and 1, for each image that it is fed with. This means that those images classified as lenses obtain $S_{\text{CNN}} \sim 1 $ while non-lenses obtain $S_{\text{CNN}} \sim 0 $. The training process converged, within our criteria above, after 57 epochs and achieved a 99.9 (99.8) percent accuracy in the training (validation) sets and a loss of 0.01 (0.02). This perfect accuracy achieved in the training set can be understood as overfitting from a classical point of view, thus, to evaluate this possibility we compared the loss and accuracy learning curves for the training and validation sets (Fig. \ref{Fig:accloss}). Overall, in both cases we see that after 10 epochs the training set reaches a stability point with minimal changes, while the validation set follows the same trend with a small gap showing less accuracy and more loss than the training set, as is expected. The lack of overfitting signs, i.e. training loss continues decreasing and/or the validation loss starts increasing again after several epochs, leads us to the conclusion that our model is able to learn and generalize this classification problem.

\begin{figure}[t!]
\centering
\includegraphics[width=9.6cm]{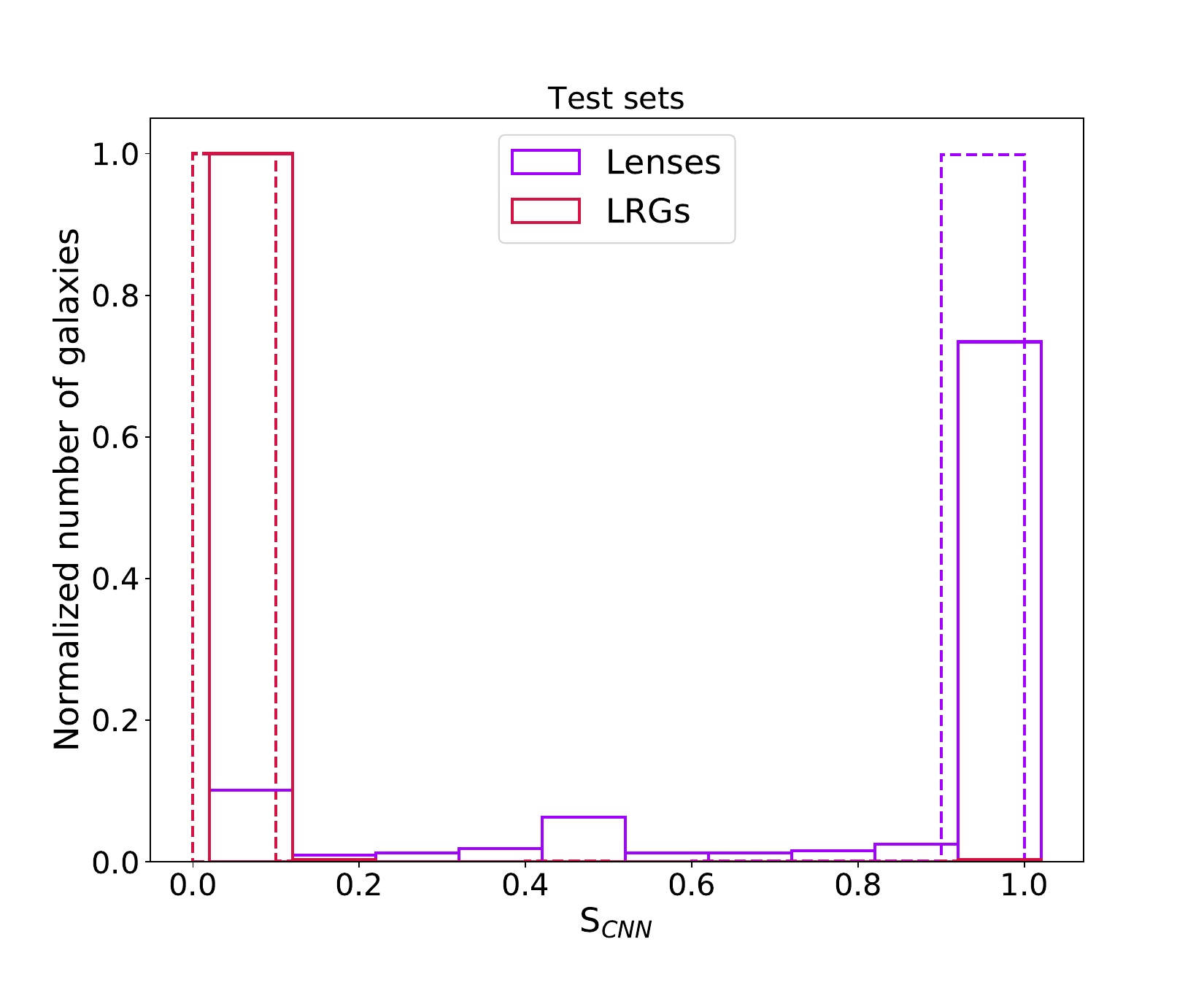}
\includegraphics[width=9.6cm]{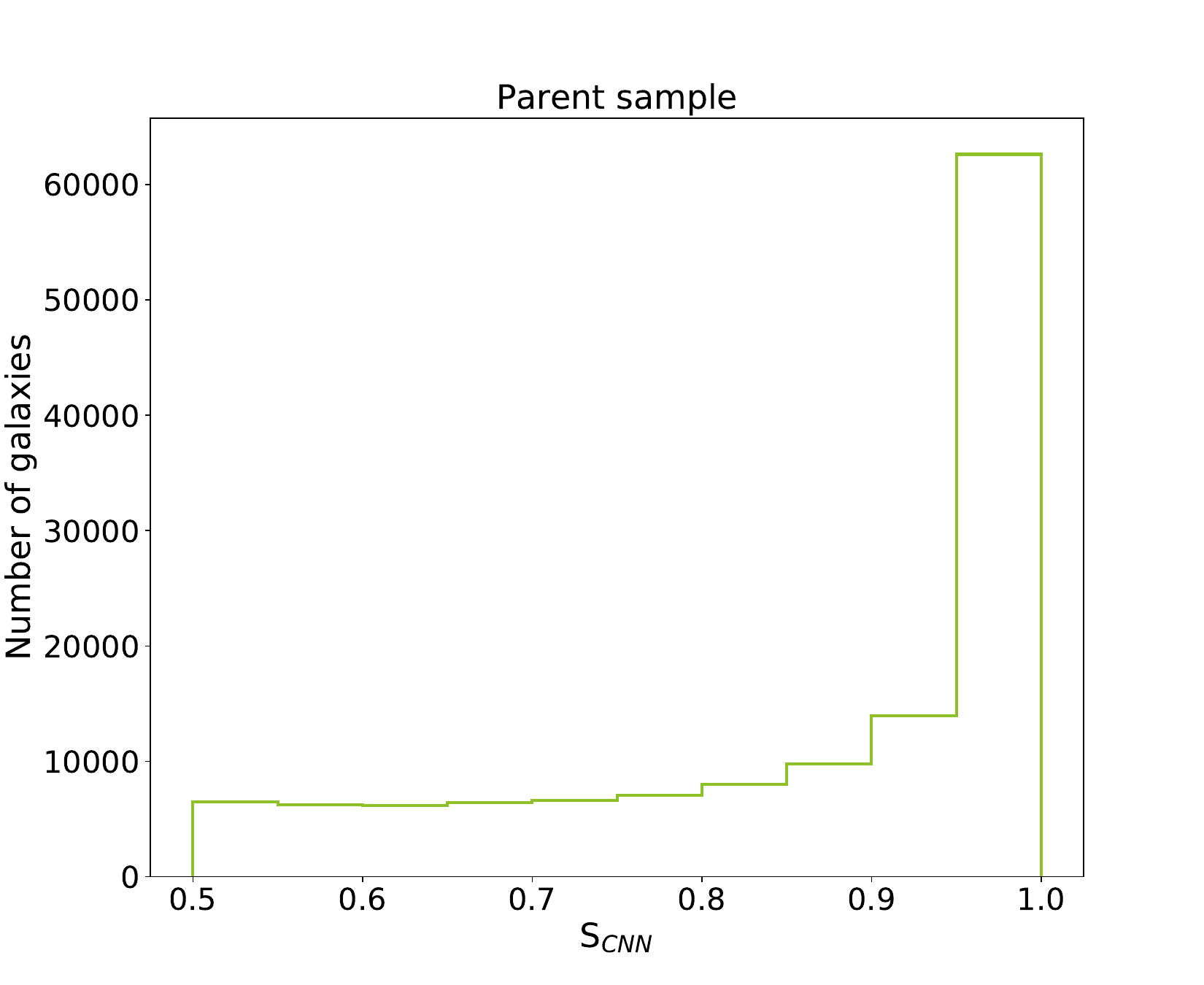}
\caption{CNN score ($S_{\text{CNN}}$) distribution for the different datasets.  The top-panel shows the $S_{\text{CNN}}$ distribution of both test sets: built with lens simulations (dashed line) and built with confirmed/candidate lenses (solid line), both datasets contain images labeled as lenses (purple lines) and real LRGs labeled as non-lenses (red lines). The two test-sets are normalized to their corresponding maximum value in the distribution. We shifted the x-axis of the second test set distribution by 0.02 for clarity. The bottom-panel shows the $S_{\text{CNN}}$ distribution for the objects in the Parent sample classified with scores above 0.5, which are the ones with  higher probability of being a lens system. We zoomed in the x-axis for visualization reasons as 99$\%$ of the sample is concentrated in values below 0.5, then they are more probable to not be a lens system hence not relevant for this study. }
\label{Fig:testset}
\end{figure}

In order to evaluate the performance of the CNN we built two test sets. The first one contains 40~000 cutouts and has the same characteristics as the training set, half simulated lenses and half LRGs. The second one has 636 cutouts where half are known lenses or lens candidates (visually selected to have noticeable lensing features) and half are LRGs not seen by the CNN during training, obtained from the same data as before. The known lenses are taken from the Master Lens Database\footnote{\url{http://admin.masterlens.org/index.php}} and the candidates from \cite{Jacobs2019A,Jacobs2019B}. The purpose of this second test set is to have a more realistic idea of the performance of the CNN in grading real strong lens systems instead of simulations. The distribution of $S_{\text{CNN}}$ for both test sets (Fig. $\ref{Fig:testset}$) shows that objects labelled as lenses are concentrated around $S_{\text{CNN}} > 0.9$ while non-lenses are around  $S_{\text{CNN}} < 0.1$, as was expected. 

\begin{figure}[t!]
\centering
\includegraphics[width=9.6cm]{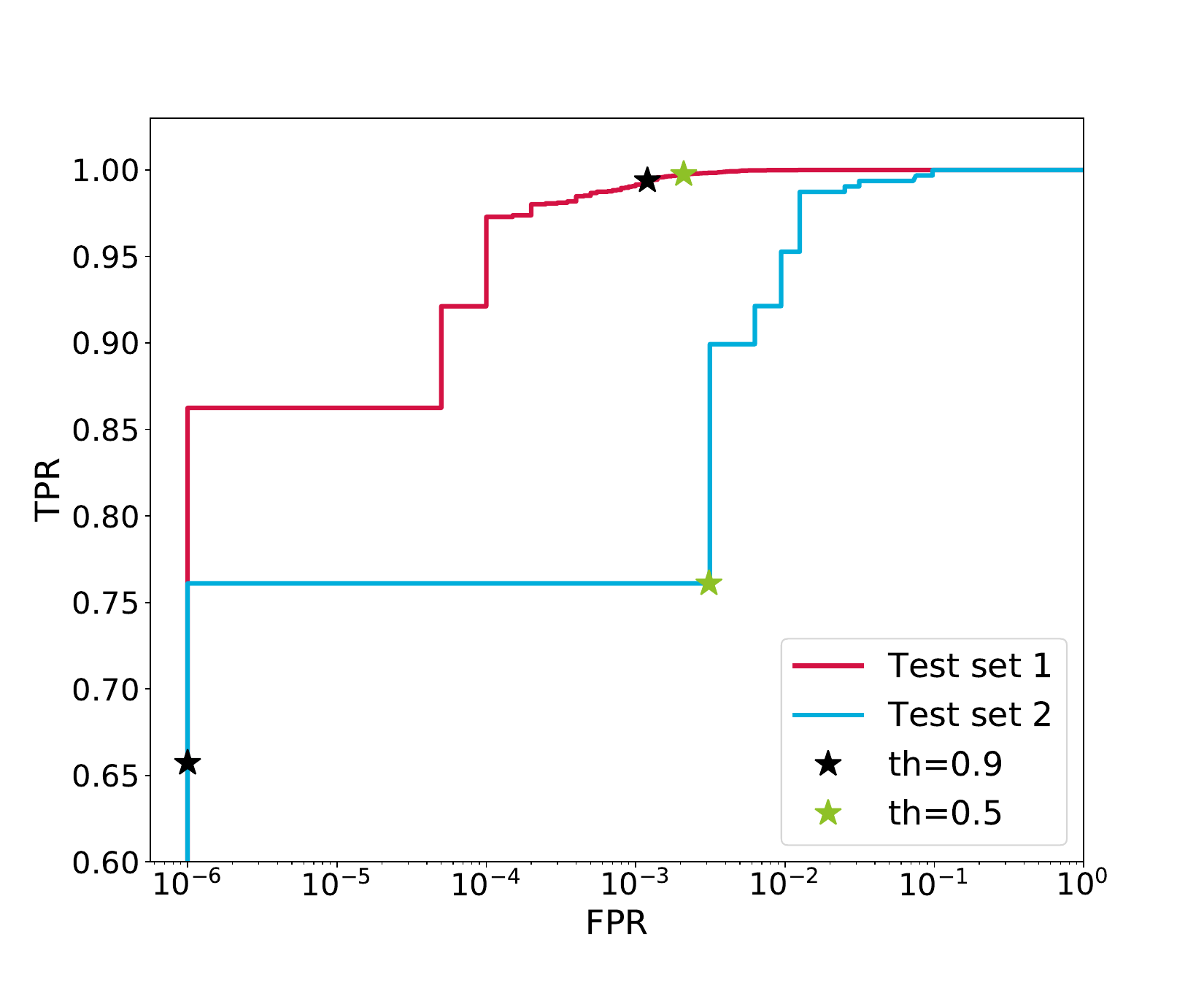}
\caption{ Receiver Operating Characteristic (ROC) curve for the test set containing simulations, (test set 1 in red) and confirmed/candidates lenses (test set 2 in blue), both datasets contain real LRGs as non-lens examples. The FPR is plotted in logarithmic scale for visualization reasons. The TPR and FPR for $S_{\text{CNN}}=0.5$ (green) and $S_{\text{CNN}} = 0.9$ (black) are also shown for each set.}
\label{Fig:roccurve}
\end{figure}

To evaluate the number of lenses correctly identified we use a Receiver Operating Characteristic (ROC) curve, (Fig. \ref{Fig:roccurve}) which shows the True Positive Rate (TPR) against the False Positive Rate (FPR) that are both functions of the decision threshold applied to the score. It illustrates the performance of a binary classifier in discriminating between the two classes as the decision threshold is varied. The first test set shows a very good performance reaching an accuracy of 99.7$\%$ and a loss of 0.02. From the ROC curve we see that choosing $S_{\text{CNN}} = 0.5 (0.9)$ we obtain a TPR = 99.8$\%$ (99.4$\%$) and a FPR = 0.21$\%$ (0.12$\%$). On the other hand, in the second test set the performance of the network decreases obtaining an accuracy of 89.6$\%$, and a loss of 0.44, while we derive TPR = 76.1$\%$ (65.7$\%$) and FPR= 0.31$\%$ (>0.01$\%$) for $S_{\text{CNN}} = 0.5 (0.9)$. Thus, whereas the accuracy in the second test set is still high and the network did not grade any LRG above 0.9, the loss and TPR are significantly worse than in a dataset with similar characteristics to the training set. We think that this decrease in the performance of the CNN is because it was trained to recognize lens simulations, which even though look very alike to real systems, they lack the diversity and uniqueness of some strong lens systems, e.g., multiple deflectors, distortions produced by substructures or external sources, among others. For example most of the false negatives in this second test set are compact lens systems or the lensing features are too faint to be properly recognized. Nevertheless, we found that our model is able to generalize and accomplish the goal of successfully classifying a high percentage of strong lens systems, although we are aware that in a realistic scenario we misclassify more objects compared with simulations, as the Fig. $\ref{Fig:testset}$ and Fig. $\ref{Fig:roccurve}$ show.

When we applied the CNN to our parent sample, we found that 98.6$\%$ of cutouts obtained $S_{\text{CNN}} \leq 0.1$. On the other hand, 133~322 obtained $S_{\text{CNN}} \geq 0.5$ (Fig.~\ref{Fig:testset}, bottom panel), being potential lens candidates. Out of these potential candidates, 76~582 cutouts obtained $S_{\text{CNN}} \geq 0.9$. The choice of $S_{\text{CNN}} \geq 0.9$ as the decision threshold has an impact on the resulting TPR and FPR as Fig. $\ref{Fig:roccurve}$ shows, but we consider that it is acceptable as the number of cutouts selected allows for human inspection to be conducted in a reasonable amount of time.

\section{Visual inspection}\label{sec:visualinspec}

The 76~582 cutouts scored above $S_{\text{CNN}}=0.9$ by the CNN were visually inspected by 7 authors of this work (K.R., E.S., B.C., F.C., C.L., J.C., and G.V.). 

\subsection{Visualization tools and guidelines}\label{subsec:vis_tool_guid}

We created two visualization tools\footnote{\url{https://github.com/esavary/Visualisation-tool}}: one to quickly select lens candidates from many objects displayed simultaneously in a mosaic configuration, and one to visual inspect each individual object in more detail and classify them in specific categories. 

\begin{figure}[h!]
\includegraphics[width=\hsize]{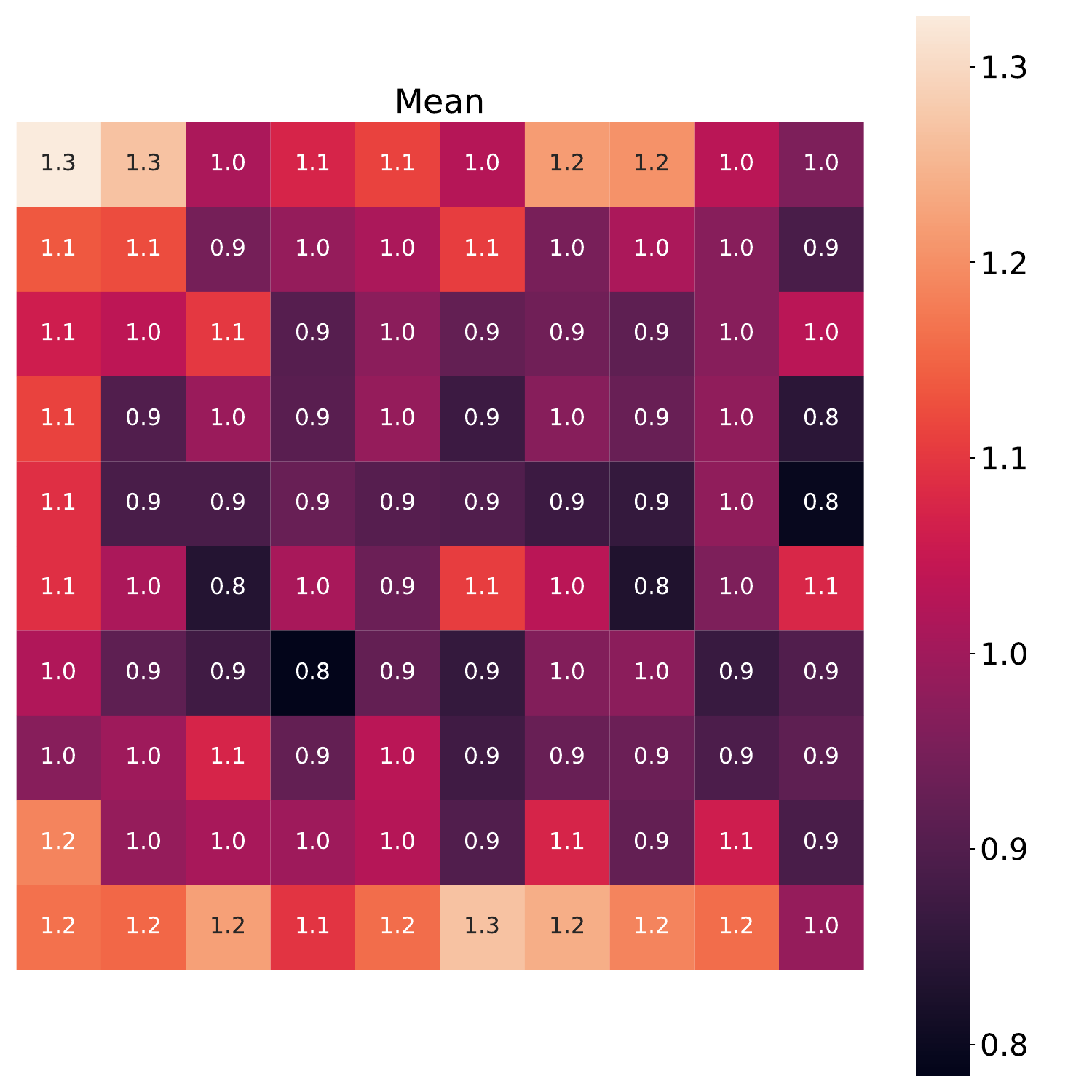}
\includegraphics[width=\hsize]{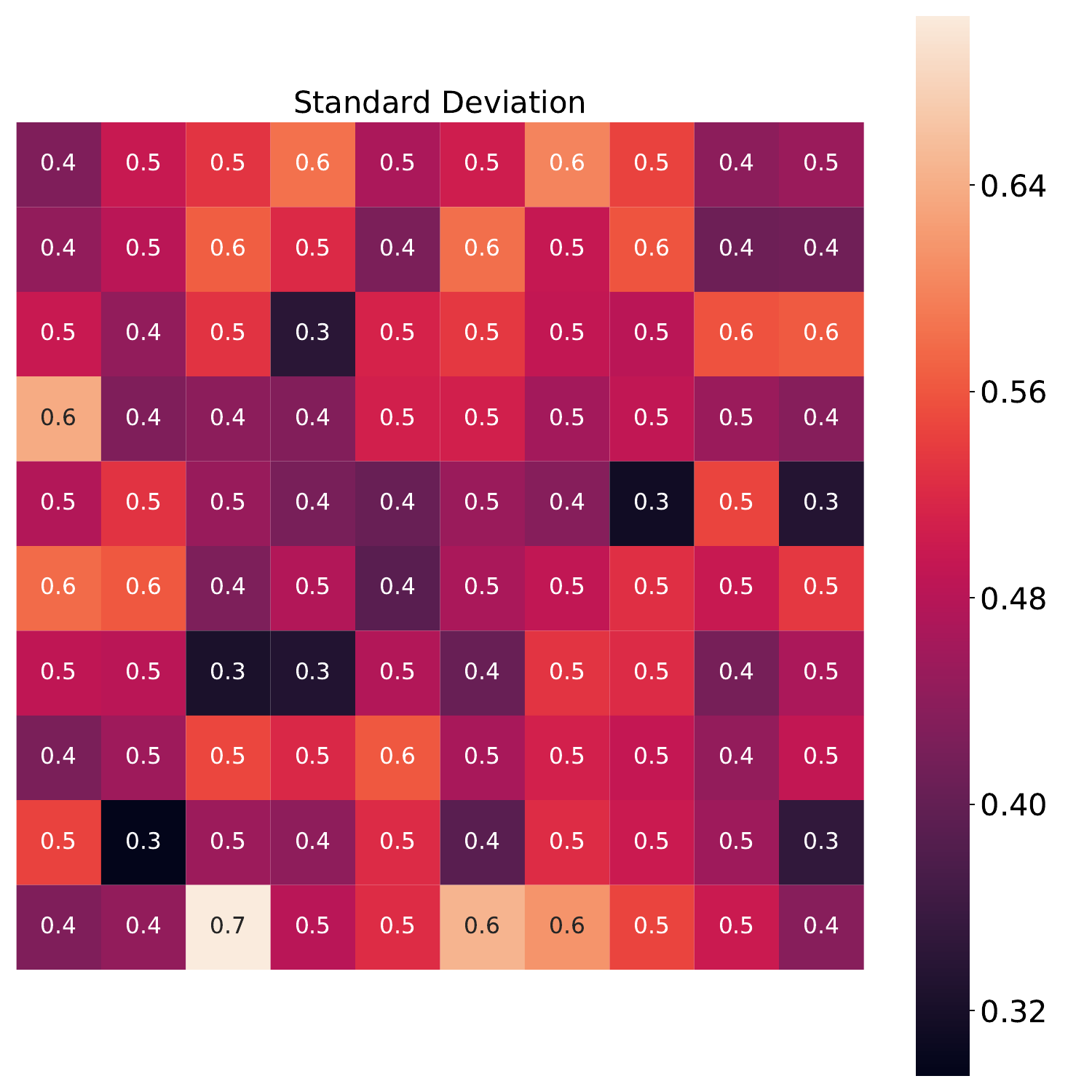}
\caption{Heat-maps for the normalized mean (top-panel) number of times that a determined cell in the Mosaic was clicked among all seven visual inspectors and its standard deviation (bottom-panel) during the Phase 1 of the visual inspection. The values in each cell were obtained by calculating the mean and the standard deviation of the total number of clicks per-cell among the 7 users, we normalized these values by a factor of 24.78 that represent the mean click in a cell overall the users for this specific classification.}
\label{fig:heatmap}
\end{figure}

The mosaic tool displays simultaneously 100 color cutouts on which the user can click on a cell to mark for selection. The user can choose a random seed for displaying the images in a random fashion on the grid, avoiding that all users see each object at the same location in the grid. This has the objective of preventing any possible bias against the position of the object on the mosaic due to the different level of concentration when looking at numerous mosaics in a row. This turned out to be very efficient, as illustrated by the "heat-maps" of user grid selections displayed in Fig.~\ref{fig:heatmap}, that are fairly flat, with a small bias towards selecting more objects from the top, bottom and left row for this particular example. With the mosaic tool, we classify in only two categories, i.e. objects that we select as displaying potential strong-lensing features, and the rest that we discard from any following step of the visual inspection. 

The second visualization tool allows us to inspect one by one all lens candidates selected with the mosaic tool. In doing so, we display the $gri$ color stamps allowing the user to change the display scale and color-map. With this tool, we classify each object among four categories: 1- Sure Lens, 2- Maybe Lens, 3- Flexion, and 4- Non Lens. In addition, we define as well 5 sub-categories when an object is not classified as a lens: 1- ring, 2- spiral, 3- elliptical, 4- disk and 5- merger.

In order to achieve a more consistent classification among users, we all agreed to follow the same guidelines for the four main categories. “Sure Lens” is selected when the cutout shows a clear strong lensing configuration without the help of a higher resolution image. This means that several clear multiple images can be identified or that there are signs of a counter-image. “Maybe Lens” is chosen if the object shows a promising lensing-like configuration but a clear identification of multiple images is not possible visually. This category also includes cases where several objects or a single arc-like object lie on one side of the central galaxy but no clear counter-image can be distinguished on the other side. In this case, high resolution imaging or spectroscopy will be required to decide whether it is a false positive or a genuine lens. When there is a single image object or a single arc far away from the central galaxy with signs of tangential distortion, the cutout goes to the category “Flexion”. Finally, everything else that could not enter in the other categories is classified as "Non Lens".

\subsection{Visual selection procedure}\label{subsec:vis_procedure}

We used both tools in 4 different phases to ensure that at the end we have a clean sample with potential lens candidates but also a sub-sample of contaminants, such as ring galaxies, which are a big source of confusion for CNNs and a matter of debate among visual inspectors. The steps carried out are described below:

\begin{enumerate}
\item Lens and ring galaxy selection. Using the mosaic tool we selected among the 76~582 cutouts all the objects that presented signs of lensing features or looked like ring galaxies in one category and we discarded the rest. An average of 2~478 cutouts were selected per visual inspector, the normalized mean distribution of clicks per cell are shown in Fig~\ref{fig:heatmap}. A total of 9~210 objects was selected by at least one user, while 89 of them were selected by all the users unanimously (see Figure \ref{Fig:vi_phase1and2}, top panel). 

\item Ring galaxies selection. We used the mosaic tool to select only ring galaxies from the 9~210 objects. A mean of 230 cutouts were selected per visual inspector, but only 71 were classified by all seven, while a total of 1~445 were selected by at least one (see Figure \ref{Fig:vi_phase1and2}, bottom panel).

\begin{figure}
\centering
\includegraphics[width=\hsize]{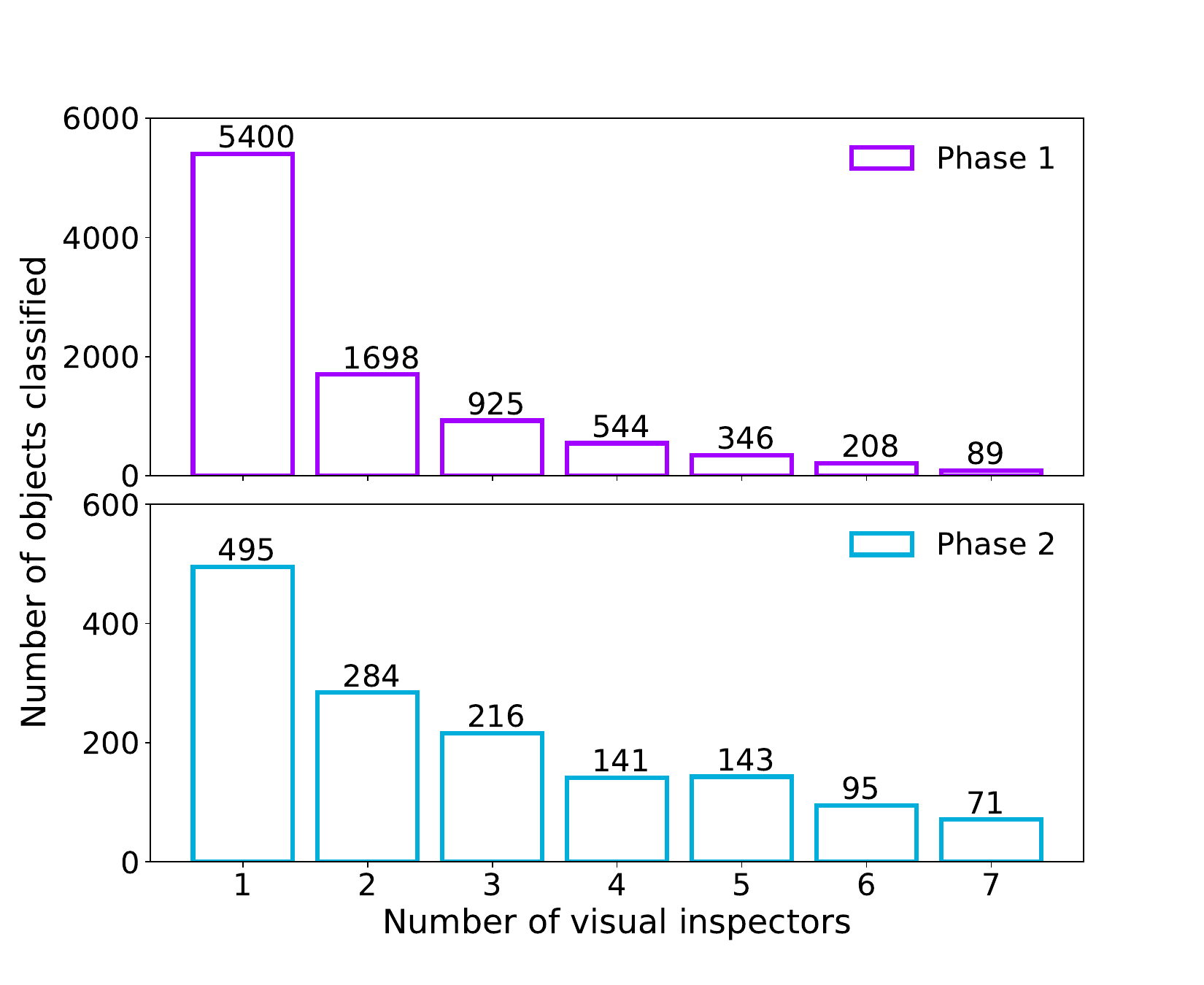}
\caption{Number of objects classified per certain number of visual inspectors in phase 1 (top panel) to select lenses and ring galaxies, and phase 2 (bottom panel) to select only the ring galaxies from phase 1. Objects selected by 7 visual inspectors represent a 100$\%$ in agreement among users, while the sum along the different bins give us the union, thus the amount of objects selected by at least one user. The exact number in each bin is shown on the top of each bar.}
\label{Fig:vi_phase1and2}
\end{figure}

\item Lens systems classification. We visually inspected all the 9~210 objects selected in phase 1 using the one by one visualization tool and looking for lens systems. Here, we showed again the classified ring galaxies from the previous step as a consistency check (users should re-classify them as rings, or at least not classify them as lenses). We classified each object among: Sure lens, Maybe lens, Flexion and Non-lens. Optionally, in the case a Non-lens was clearly identified by the user as spiral, merger or ring galaxy the object was sub-classified in the corresponding category. From this visual inspection we obtained a total of 275 Sure lens, 2~666 maybe lens, 2~602 flexion and 9~125 non lenses, classified at least by one visual inspector. In an unanimous agreement among the 7 visual inspectors we counted only 6 sure lenses, 1 maybe lens, 1 flexion and 4~716 non lenses. On the other hand K.R., E.S., B.C., F.C., and J.C sub-classified 359 ring galaxies, 22 mergers and 49 spirals with an agreement of 50\% among the visual inspectors. In Table~\ref{table:visualinspection} we summarize the individual classification by category and subcategory of each user and in Fig.~\ref{Fig:mosaic_MLFRSM} there are examples of objects classified in all categories and sub-categories different of "sure lens".

\begin{table*}
\caption{Classification and sub-classification details per visual inspector during Phase 3}             
\label{table:visualinspection}      
\centering                          
\begin{tabular}{c c c c c c c c}        
\hline\hline                 
Classification & User 1 & User 2 & User 3 & User 4 & User 5 & User 6 & User 7 \\    
\hline                        
Sure Lens   & 116   & 41     & 79    & 146   & 120   & 19    & 90    \\      
Maybe Lens  & 612   & 1355   & 492   & 691   & 849   & 203   & 141   \\
Flexion     & 654   & 1421   & 540   & 300   & 812   & 473   & 26    \\
Non Lens    & 7828  & 6393   & 8099  & 8073  & 7429  & 8515  & 8953  \\
\hline
Sub-classification      &  &    &    &     &      &     &      \\ 
\hline
Spiral      & 33    & 100    & 651   & 59    & -     & 35    & -     \\ 
Ring        & 713   & 111    & 393   & 364  & -     & 563   & -     \\ 
Merger      & 112   & 9      & 246   & 70    & -     & 15    & -     \\ 
\hline                                   
\end{tabular}
\end{table*}

\begin{figure*}
\centering
\includegraphics[width=\hsize]{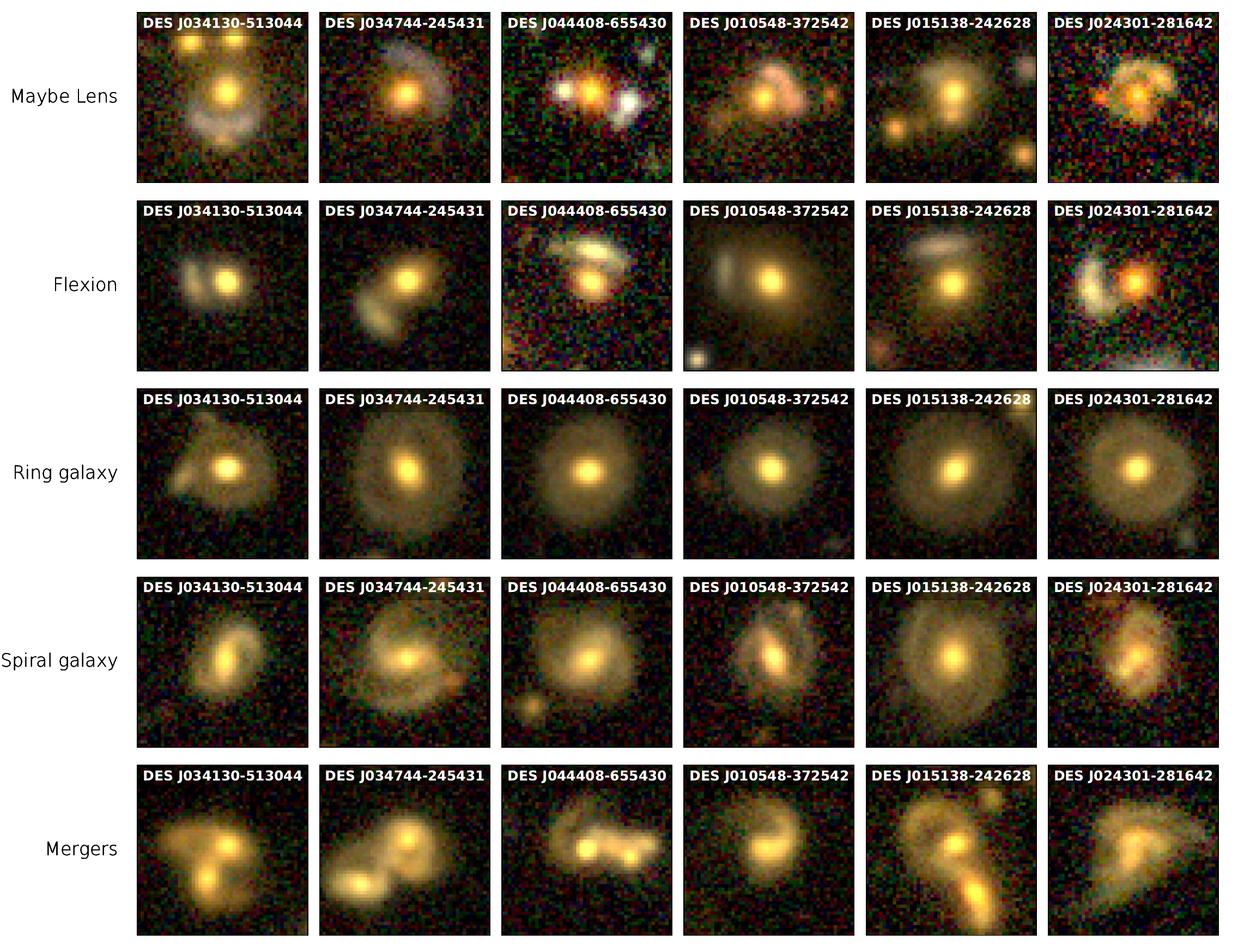}
\caption{Example of objects classified as "Maybe Lens" or "Flexion", and sub-classified as "Ring galaxy", "Spiral", or "Merger", "Sure Lens" category is shown in Fig.~\ref{Fig:mosaic_sl1}. In the top of each image is displayed the name, the CNN score and the visual inspection score (VIS) for all the system displayed is 1.0. The VIS is calculated for each case as follow: VIS$_L$ and VIS$_R$ are used for "Maybe Lens" and "Rings" respectively, see Subsection \ref{subsec:final_cat} for details, in the cases of cutouts classified as "Flexion", "Spiral" and "Merger" the VIS correspond to the percentage of visual inspectors that classified the object in the respective category.}
\label{Fig:mosaic_MLFRSM}
\end{figure*}

\item Final group visual inspection. This step was performed by K.R., E.S., B.C., F.C., J.C, and G.V. all together. Using the one by one visualization tool we revised 2~690 objects selected as Sure or/and Maybe lens by at least one visual inspector. This aimed to obtain a final selection of pure potential candidates that can be suitable for follow-up high-resolution imaging and spectroscopic confirmation avoiding spending telescope time in false positives. We classified them in the two main categories, selecting 81 sure lenses and 296 maybe lenses. This represents 0.5$\%$ of the sample with $ S_{\text{CNN}} \geq 0.9 $ and 0.002$\%$ of the initial LRG selection sample.

\end{enumerate}

An extra visual inspection was performed for the cutouts classified by the CNN with scores between $0.8< S_{\text{CNN}} < 0.9 $ (hereafter referred to as bin80) by K.R., E.S and B.C. with the purpose of quantifying how many objects we could have missed by selecting only those with $ S_{\text{CNN}} > 0.9 $. Similarly to the previous analysis we first used the mosaic visualization tool to inspect the 17~779 cutouts, selecting 190 potential lens candidates, which were then inspected one by one. A set of 24 objects were classified as lenses at least by one visual inspector and 115 as maybe lenses. Finally, K.R., E.S., B.C, F.C, J.C, and G.V. conducted a group visual inspection to the 190 firstly selected candidates to compile a final sample with 9 sure lenses and 19 maybe lenses that were then added to our candidate list. In total, only a 0.2$\%$ of the data visually inspected in the bin80 was considered as a lens candidate, while for all the objects with $S_{\text{CNN}} \geq 0.9 $ we selected 0.5$\%$ of the cutouts in the categories sure or maybe lens. Furthermore, taking into account that the amount of cutouts classified in the bin80 is about of 4 times smaller than the ones with $S_{\text{CNN}} \geq 0.9$, we conclude from this exercise that the number of expected candidates with $0.8 < S_{\text{CNN}} < 0.9 $ was very low, showing that we reached a point of diminishing returns which would make the additional human visual inspection of images classified with $S_{\text{CNN}} < 0.8$ ineffective.

\subsection{Final Catalogs} \label{subsec:final_cat}

As a final product we present two main catalogs: one containing lens and one containing ring galaxy candidates. We assigned a Visual Inspection Score (VIS) to each candidate, computed using the percentage of visual inspectors that classified it in a certain category. In the case of lens candidates, we used the percentage of users that classified a system as either a "Sure lens" and as "Maybe lens". We summed these percentages to obtain a "strong lensing percentage". Then, we considered this percentage as the final visual inspection score for lens systems (VIS$_{L}$). In the case of ring galaxies we had candidates from step 2 (using the mosaic tool) and/or step 3 (the one by one method). We average the percentage of users who classified each object as a ring in each step, and present this as the final visual inspection score for Rings (VIS$_{R}$). If the candidate was selected only by one of the tools the final score obtained is the one corresponding to that classification (i.e. not average).

The ultimate catalog of lens systems can be split in two categories, "Sure lens" (SL) with 90 systems (Fig.~\ref{Fig:mosaic_sl1}~to~\ref{Fig:mosaic_sl6}) with prominent lensing features and counterpart images, and "maybe lens" (ML) with 315 systems that show promising lensing features but for which more evidence such as higher resolution imaging and spectra is needed (see Fig. \ref{Fig:mosaic_MLFRSM} for examples). 

\begin{figure}
\centering
\includegraphics[width=\hsize]{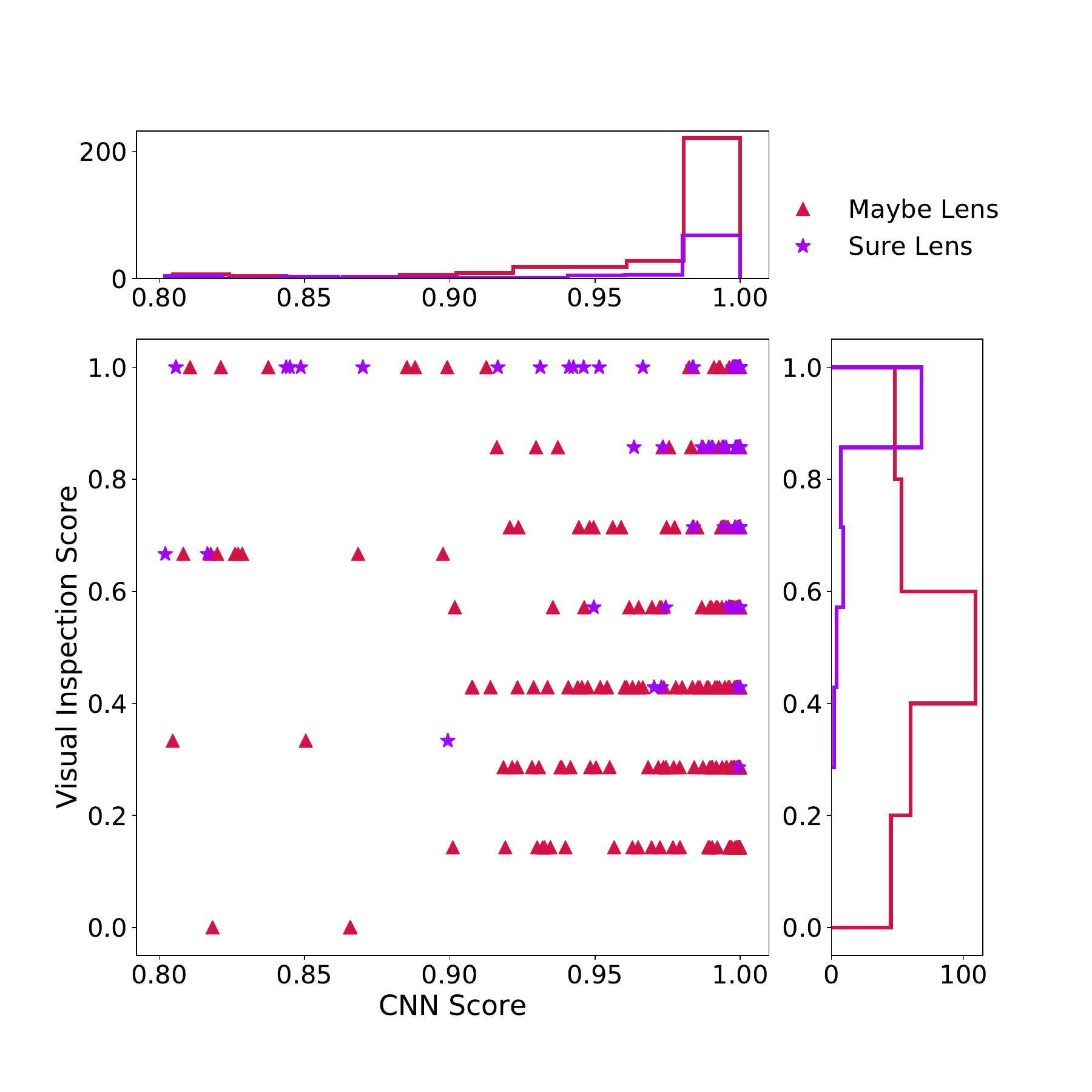} 
\caption{CNN score (S$_{CNN}$) against the visual inspection score (VIS$_L$), with their respective distributions for the final catalog of lens candidates containing 405 systems. The distribution of the category Sure lens is in purple, and Maybe lens is in red, while in the scatter plot Sure lenses are represented with a star and Maybe lenses with a triangle.}
\label{Fig:cnn_vs_vi}
\end{figure}

The CNN and visual inspection scores of both the SL and ML candidates is shown in Figure \ref{Fig:cnn_vs_vi}. Here, we clearly see that the most Sure Lenses are clustered towards the upper right corner, indicating that in general they obtained a high score from both methods, while very few of them had either CNN scores below 0.95 or visual inspection scores below 0.5. On the other hand, a large majority of maybe lenses did not receive a high visual inspection score, including two that originally were rejected by visual inspectors, but upgraded after the group visual inspection. Although several of the Maybe lens objects still got very high scores from the CNN, indicating that the visual inspection step is needed to refine the final catalog.

In order to identify lens candidates that were not previously published, we cross-matched our final catalog with available astronomical databases such as Vizier \citep{vizier}, Simbad \citep{simbad}, the Master Lens database, and other lens finding papers including \cite{Wong2018,Jacobs2019A,Jacobs2019B,Petrillo2019,Canameras2020,Jaelani2020,Huang2020,Huang2021}. As a result, we found that our catalog contains 219 previously identified candidates (74 SL, 145 ML), including at least 5 spectroscopically-confirmed systems, and 186 new candidates (16 SL, 170 ML). The detailed information for these systems can be found in Table \ref{tab:maybelens},  available at the CDS.

\begin{table*}
\caption{Excerpt of the Maybe lens catalog available online.}             
\label{tab:maybelens}      
\centering          
\begin{tabular}{l l l l c c }     
\hline\hline       
Candidate &          RA &        Dec & S$_{CNN}$ & VIS$_{L}$\tablefootmark{a} &                References \\
\hline                    
DES J034130$-$513044 &   55.378331 & -51.512411 &      1.00 &                    1.00 &                 [7]  [10] \\
DES J034744$-$245431 &   56.935562 & -24.908741 &      1.00 &                    1.00 &                 [9]  [10] \\
DES J044408$-$655430 &   71.034707 & -65.908598 &      1.00 &                    1.00 &                 This work \\
DES J010548$-$372542 &   16.450174 & -37.428457 &      1.00 &                    1.00 &                      [10] \\
DES J015138$-$242628 &   27.909990 & -24.441314 &      1.00 &                    1.00 &                      [18] \\
DES J024301$-$281642 &   40.754315 & -28.278515 &      1.00 &                    1.00 &                 This work \\
DES J025052$-$552411 &   42.717809 & -55.403251 &      1.00 &                    1.00 &                      [10] \\
DES J014358$-$470037 &   25.995764 & -47.010469 &      1.00 &                    1.00 &                 This work \\
DES J001718$+$015818 &    4.325557 &   1.971828 &      1.00 &                    1.00 &                      [10] \\
DES J040349$-$352601 &   60.955780 & -35.433763 &      0.99 &                    1.00 &                 This work \\
DES J225146$-$441220 &  342.943254 & -44.205688 &      0.99 &                    1.00 &                       [7] \\
DES J002056$-$594016 &    5.236669 & -59.671225 &      0.99 &                    1.00 &                 This work \\
DES J011758$-$052717 &   19.494766 &  -5.454924 &      0.98 &                    1.00 &                [10]  [13] \\
DES J015904$-$345009 &   29.767747 & -34.835994 &      0.98 &                    1.00 &                 This work \\
DES J015009$-$030438 &   27.537943 &  -3.077297 &      0.98 &                    1.00 &                 [9]  [10] \\
\hline                  
\end{tabular}
\tablefoot{
\tablefoottext{a}{Visual Inspection Score for Strong lens systems}}
\tablebib{[1]~\cite{Cabanac2007}, [2]~\cite{Limousin2009}, [3]~\cite{More2012}, [4]~\cite{maturi2014}, [5]~\cite{More2016}, [6]~\cite{Paraficz2016}, [7]~\cite{Diehl2017}, [8]~\cite{Wong2018}, [9]~\cite{Jacobs2019A}, [10]~\cite{Jacobs2019B}, [11]~\cite{Petrillo2019}, [12]~\cite{Canameras2020}, [13]~\cite{Huang2020}, [14]~\cite{Jaelani2020}, [15]~\cite{lemon2020}, [16]~\cite{Li2020}, [17]~\cite{Nord2020}, [18]~\cite{Huang2021}}
\end{table*}

Our second catalog is composed of ring galaxy candidates classified in two different steps of the visual inspection process. We identified 1~445 ring galaxy candidates during the second visual inspection step using the mosaic tool, while 985 galaxies were classified in this category by at least one visual inspector using the one by one tool in the third step. A cross-match between these two selections gave an intersection of 854 galaxies for a total of 1~576 ring galaxies selected by at least one user using either of the two methods. The final catalog was built by selecting the objects picked by at least 50$\%$ of the visual inspectors using the mosaic or the one by one tool, resulting in 539 ring galaxy candidates. In Fig.~\ref{Fig:mosaic_MLFRSM} we present the 6 top-graded candidates and in Table~\ref{tab:ringcat}, available at the CDS, we detail the information for the full sample.

\begin{table*}
\caption{Excerpt of the ring galaxy candidates catalog available online.}    
\label{tab:ringcat}      
\centering          
\begin{tabular}{l l l l c }     
\hline\hline       
Candidate & RA & Dec & S$_{CNN}$ & VIS$_{R}$\tablefootmark{a} \\
\hline                    
DES J013040$-$160110 &   22.666806 & -16.019599 & 1.00 &  1.00 \\
DES J012733$-$151618 &   21.888203 & -15.271692 & 1.00 &  1.00 \\
DES J010723$-$151315 &   16.847733 & -15.221047 & 1.00 &  1.00 \\
DES J004346$-$304929 &   10.942121 & -30.824795 & 1.00 &  1.00 \\
DES J033913$-$260914 &   54.805084 & -26.154158 & 1.00 &  1.00 \\
DES J045112$-$262143 &   72.804088 & -26.362060 & 1.00 &  1.00 \\
DES J012542$-$231630 &   21.427496 & -23.275137 & 1.00 &  1.00 \\
DES J003809$-$224742 &    9.537798 & -22.795153 & 1.00 &  1.00 \\
DES J012843$-$350926 &   22.183211 & -35.157252 & 0.99 &  1.00 \\
DES J041502$-$404547 &   63.762073 & -40.763330 & 0.99 &  1.00 \\
DES J010902$-$450634 &   17.258607 & -45.109657 & 0.98 &  1.00 \\
DES J012746$-$444820 &   21.942424 & -44.805651 & 0.98 &  1.00 \\
DES J004837$-$330630 &   12.156848 & -33.108576 & 0.98 &  1.00 \\
DES J024746$-$243851 &   41.941910 & -24.647757 & 0.97 &  1.00 \\
DES J021101$-$315721 &   32.757447 & -31.956016 & 0.97 &  1.00 \\
\hline  
\end{tabular}
\tablefoot{
\tablefoottext{a}{Visual Inspection Score for Ring galaxy candidates.}}
\end{table*}

Finally a catalog using the classification "Flexion" will be created after a more detailed analysis of the objects classified in this category, but this is out of the goal of the current work. We expect that this catalog could serve as a probe for works like \cite{Birrer2021-arcs}.

\subsection{Lens-source decomposition}\label{subsec:decomp}

In most strong lensing systems, especially with ground-based observations, the light from the source and lens galaxies are seen blended. In order to better visualize our 90 lens candidates in the Sure Lens catalog, we designed a prototypical automated procedure for deblending the light from lens and source galaxies based primarily on their colors. 

Due to the complexity of the light profile of the lensed sources we choose to represent them in a non-parametric way using undecimated isotropic wavelets \citep[starlets,][]{Starck2007}, as implemented in the Multi-band morpho-Spectral Component Analysis Deblending Tool \href{https://github.com/herjy/MuSCADeT}{\faGithub \tt MuSCADeT} \citep{Joseph2016}. Estimating both color and morphology of sources requires a large number of parameters, larger than the number of pixels in the starlet-decomposed image, making it a degenerate problem. To overcome this we use a combination of the \href{https://github.com/pmelchior/scarlet}{\faGithub \tt scarlet} \citep{Melchior2018} and {\tt MuSCADeT} algorithms. In both methods multi-band images are modeled as sums of factorised components, where each object $i$ in an image has a 2-Dimensional surface brightness $S_i$ with as many pixels as there are in the image bands, and a spectrum $A_i$ with as many entries as there are bands \citep[see details in][]{Melchior2018, Joseph2016}, such that:

\begin{equation}
    Y = \sum_{i<o}A_i S_i + N,
\end{equation}

\noindent where $Y$ is a multi-band cube of images, $o$ is the number of objects in the scene and $N$ is the noise map.

The strategy implemented in {\tt MuSCADeT} only allows for crude estimates of source colors, based on principal component analysis of pixel fluxes. Instead, the {\tt scarlet} software is able to estimate the colors of each source in the field provided that the morphology is constrained to be a monotonic profile. Monotonicity of galaxy profiles from the center out does not suit the description for complex lensed sources, hence the need for {\tt MuSCADeT} to model strongly lensed galaxies in a non-parametric way coupled with sparse regularization.

{\tt Scarlet} requires detection of the brightest pixel of each source to model, which is a challenging and ill-defined problem in the case of strongly lensed galaxies, where lensed features are often multi-modal and strongly blended with the deflector's light. In order to circumvent this issue and make sure we capture (lensed) sources with a bluer spectrum than the central LRGs, we allow that {\tt scarlet} models one source with Starlets, initialised with a "blue" spectrum. This allows {\tt scarlet} to capture blue features with complex morphologies that might not have been detected due to blending, while limiting degeneracies with other sources. The blue, normalised spectra used for initialisation are empirically set to $[0.4, 0.4, 0.2]$, where each of the three values reflect the relative contributions to $g$, $r$ and $i$ bands respectively. 
Among the spectra obtained through {\tt scarlet}, we then select the bluest and reddest spectra by finding the spectra that maximizes the scalar product between the normalized spectra $[0.667,0.333,0]$ (for blue) and $[0,0.333,0.667]$ (for red). This ensures that two components with different colors are extracted, with the expectation that the red component features the morphology of the LRG and its neighbours, while the blue component extracts the morphology of the lensed star-forming background galaxies. 

The results for our best lens candidates are shown in Figs.~\ref{Fig:mosaic_sl1}-\ref{Fig:mosaic_sl6} and display for each system the red residual, i.e. the data from which the blue model has been subtracted and blue residuals, i.e. the data from which the red model has been subtracted. In the following we refer to these as $R_r$ and $R_b$ respectively, defined as: 
\begin{eqnarray}
R_b  =  Y-A_r S_r,\\
R_r  =  Y-A_b S_b,
\end{eqnarray}
Where $A_r S_r$ and $A_b S_b$ are the models for the red and blue components expressed as the product between their spectra and morphologies. The summary of the procedure for deblending strong gravitational lens candidates is as follows:

\begin{itemize}
    \item Detection of sources in the image using the source extraction package, \href{https://github.com/kbarbary/sep}{\faGithub \tt sep} \citep{Bertin1996} on a starlet-filtered version of the image where only the first two levels of the starlet decomposition are used.
    \item Initialization of {\tt scarlet} sources: one extended source per detected object plus one starlet component with blue spectra.
    \item Run {\tt Scarlet} and extract spectra for each source in the field of view.
    \item Identify the bluest and reddest sources through scalar product with predefined red and blue spectra.
    \item Run {\tt MuSCADeT} with the red and blue spectra.
    \item Extract red and blue components by computing the difference between the multi-band images and the model for each {\tt MuSCADeT} component.
\end{itemize}

The results in Figs.~\ref{Fig:mosaic_sl1}-\ref{Fig:mosaic_sl6} show that the lens and source light can be deblended efficiently without fitting any analytical profile. The effectiveness of the method to deblend the profiles comes mostly from the spectral decomposition of the objects and on their representation on an array of pixels to which we apply sparse regularization with wavelets (starlets). This procedure is well suited to an automated use in a pipeline but assumes that lensed sources are significantly bluer than the lens light. This is the case for most of our lenses as by construction our lens finding method is based on a preselection of objects that favours such configuration. Still, we do have objects where the lens-source color contrast deviates significantly from our assumption. In this case the deblending works less efficiently and we see leakage of flux between the lens and source. 
Another case of leakage, leading to sub-optimal deblending can be observed in systems where the image contains sources with different colors than that of the lenses or sources. In this case, since the whole image is modeled as 2 fields of light, the spectra of the color components tends to offset towards an average spectra that better matches all the colors in the patch. This can be observed in systems DES J013522-423223, DES J024911+004848 or DES J010826-262019, where the blue components contain light from the lens galaxies and where object with colors different than that of the main deflector. 
These shortcomings are motivation enough for further refinement of our deblending in particular focused on using {\tt scarlet} better model individual, non-lensed sources, which is outside of the scope of this paper. 
Finally it is important to emphasize that our light deblending confirms our visual grading and does not discard any of our best candidates.

\section{Model}\label{sec:model}

We developed an automated modeling pipeline in order to further explore the highest rated lens candidates obtained from the visual inspection. Our candidate sample is very heterogeneous, containing galaxy, group and cluster scales systems. Thus, in order to perform this automatic modeling we split the sample and selected only the images in which there appeared to be just a single lens galaxy as a deflector. The $52$ images selected for modeling are labeled with an "M" in the mosaics of Figs. \ref{Fig:mosaic_sl1} - \ref{Fig:mosaic_sl6}. 

This pipeline allows us to efficiently model large samples of lens candidates acquired in current and future lens finding efforts, and to explore the model parameter distributions in search of meaningful trends. 

\subsection{Automated Modeling Pipeline}

We modeled the images using single elliptical S\'ersic profiles for the light distributions of both the deflector and source. For the mass distribution of the deflector, we use a Singular Isothermal Ellipsoid profile (SIE) along with an additional external shear component ($\gamma_{\text{ext}}$). The simplicity of these profiles allows us to model many lens candidates efficiently, while still fitting most images well enough for us to observe large-scale trends in the properties of the sample. The pipeline supports multi-band fitting, so we fit the DES lens candidates using images in the $g$, $r$, and $i$ bands. We used a separate elliptical S\'ersic profile for each of the three photometric bands when fitting the deflector and source light components, and fixed only the center positions between bands. The deflector mass profile is shared across all bands. 

The modeling pipeline was entirely written in Python and makes use of the {\tt Lenstronomy} lens modeling package \citep{Birrer2015,Birrer2018}. For parameter optimization, we used the Particle Swarm Optimization (PSO) \citep{Kennedy1995} as well as Markov Chain Monte Carlo (MCMC) methods. For each image, the pipeline first performed a chain of pre-sampling PSOs before running the sampling with the MCMC. The MCMC is performed using an affine-invariant Markov chain Monte Carlo ensemble sampler \citep{Goodman2010,emcee}, which is implemented using the {\tt emcee}\footnote{\url{https://github.com/dfm/emcee}} Python package. 

\begin{figure}
\includegraphics[width=\linewidth]{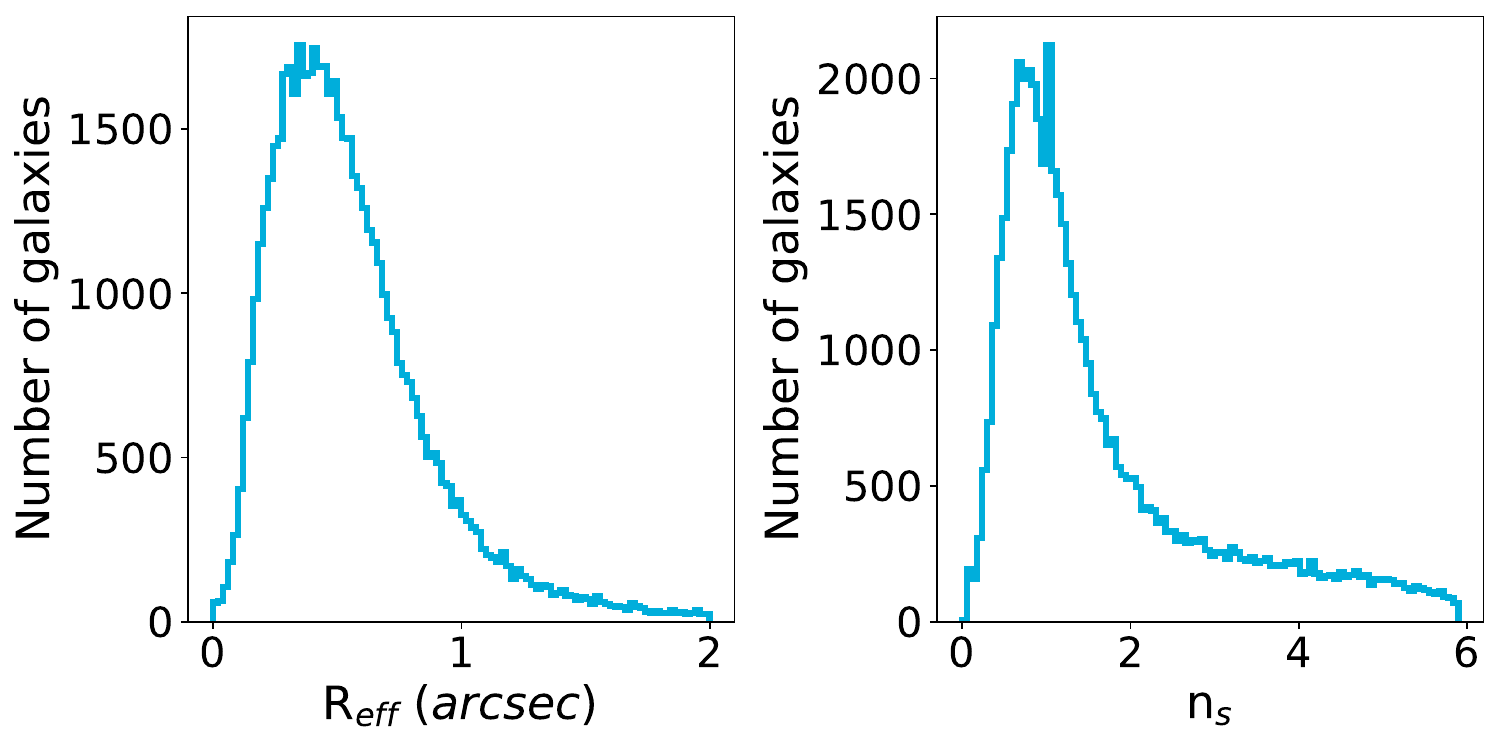}
\caption{Prior distributions for the effective radius ($R_{\text{eff}}$) and the S\'ersic index ($n_{\text{s}}$) used for constraining source light parameter values.}
\label{Fig: source priors}
\end{figure}

In order to obtain realistic results in the parameters, we introduced priors that punitively discourage extremes in the model parameter values. While we can not assume anything about the position angles of the S\'ersic or SIE profiles, we use Gaussian priors on the ratio between the semi-minor and semi-major axes
, $q$. The Gaussian prior was centered on a value of $\bar{q} = 0.8$, in accordance with the distributions of $138\;269$ galaxies from the Galaxy And Mass Assembly (GAMA) database that were modeled in \citet{Kelvin2012}. We also used a similar Gaussian prior method to constrain the deflector mass eccentricity and position angle to values close to those of the deflector light, as well as applying this method to allow only small variations between photometric bands in the light components of the model. Lastly, we also applied a prior distribution for the effective (half-light) radius, $R_{\text{eff}}$, and S\'ersic index $n_{\text{s}}$ parameters of the source light. The source priors we used came from the S\'ersic parameter distributions of $56\; 062$ galaxies from the COSMOS survey. This data was used as a training set in the development of the {\tt GalSim}\footnote{\url{https://github.com/GalSim-developers/GalSim}} software. \citep{Rowe2015}. 
We show these distributions in Fig.~\ref{Fig: source priors}.

When modeling the lens candidates, it is common for image cutouts to contain neighboring objects in the field of view that are unrelated to the lens system. Light contamination from these "satellites" can be mistaken as originating from the lensed source if not masked properly. This problem is handled differently by various authors. \citet{Shajib2020} excluded systems with contaminating satellites in their sample, and modeled only isolated lenses from the SLACS survey \citep{Auger2009}. \citet{Night2018} did not pre-select isolated lenses, but instead masked all pixels outside of a circular region with a fixed radius of $3.9\arcsec$.

\begin{figure}
\centering
\includegraphics[width=\linewidth]{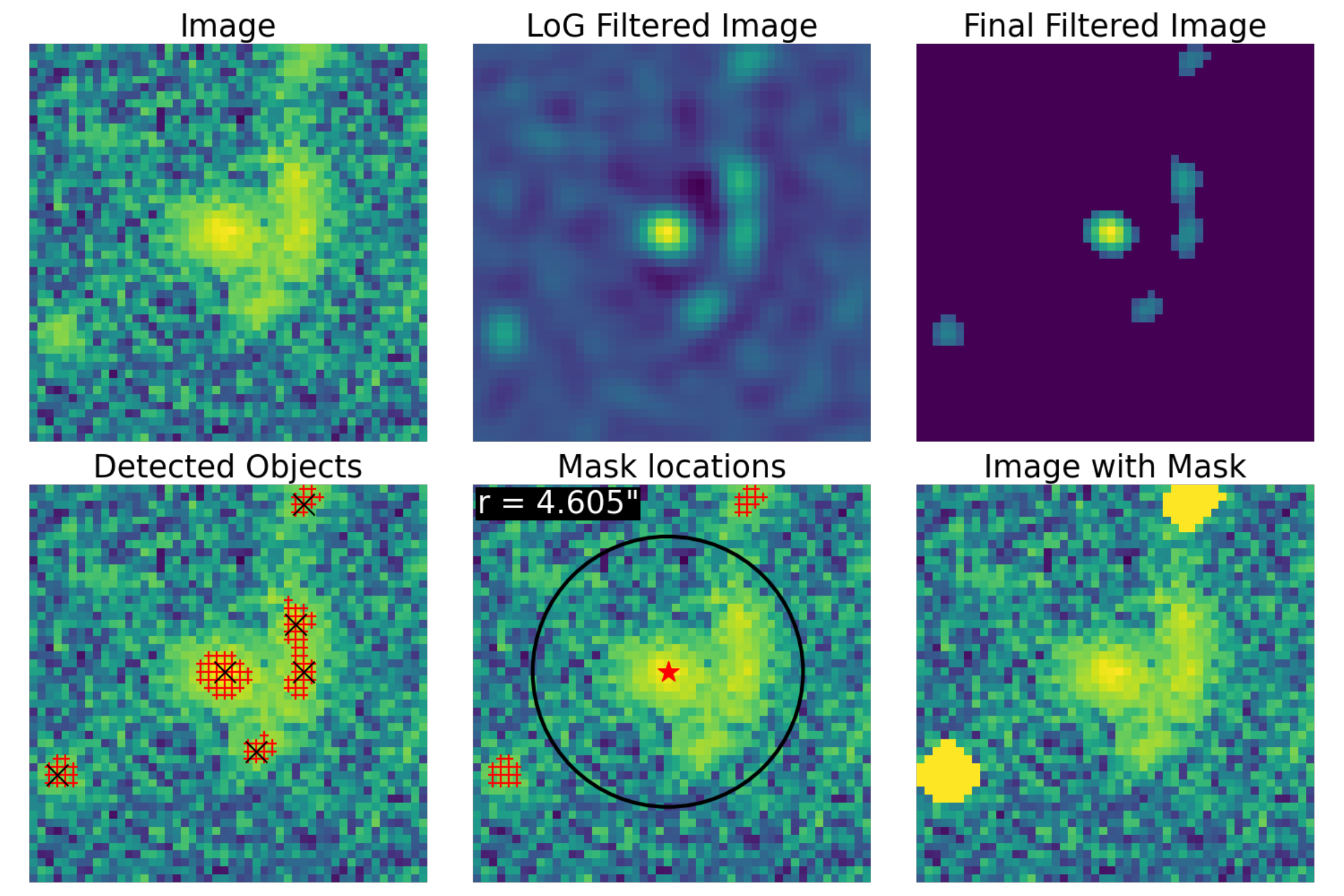}
\caption{Illustration of the automated masking procedure using an example DES image. In the upper row we show (from left to right) the original image, the image after applying an LoG filter, and finally the result of setting pixels with flux below the threshold to zero. In the bottom row we show in the leftmost frame the original image annotated with the remaining pixels from the filtering step (red '+' marks), along with the detected peaks (black 'x' marks). In the middle we show the estimated size of the lens system with the black circle, as well as the detected bright pixels that are considered contaminant light. These pixels are then used for the mask, and we show the areas covered in the rightmost panel by setting the corresponding pixels to a large constant value.}
\label{Fig: Masking}
\end{figure}

In our case, we designed the pipeline to be flexible in handling a large variety of lens system configurations and sizes. The steps to our masking procedure is illustrated in Fig.~\ref{Fig: Masking}, and begins with applying filters in order to identify the brightest regions in the image as well as their centroid locations. We first applied a Laplacian of Gaussian (LoG) filter to detect areas with rapid changes in flux. Next we took all remaining pixels with flux less than a threshold of six times the rms background, and set them to zero. This results in a final filtered image with only the areas of the image containing the most light. We find the centroid locations of these areas by finding the local maxima, or peaks, in the final filtered image. For our masking algorithm we made use of both, these peak locations as well as the pixels containing their light, and these are labeled with black and red markings in the bottom left panel of Fig.~\ref{Fig: Masking}, respectively. The peak locations are used first for determining the center of the lens system, i.e. the position of the deflector galaxy, and assume that this is the peak detected object nearest to the image center. Because the deflector is assumed to be an LRG, we use the reddest available band ($i$-band) for this step. Next, we take the detected peaks in the bluest band (i.e. the $g$-band) to estimate the lens system size. This is because the source galaxies in lens systems are usually younger, more active galaxies, meaning that the lensed source light will be more prominent in the bluest image band. We assume that the second closest detected peak to the center is the first of the lensed images of the source. We also assume that the furthest lensed image from the deflector is not more than 1.5\arcsec further out than the nearest one. Therefore, our estimated lens system radius is the distance from the deflector to the closest lensed source object plus 1.5\arcsec. We show this estimation as a black circle enclosing the lens system in the bottom middle panel of Fig.~\ref{Fig: Masking}.

Using the estimated size of the lens system from the $g$-band image and the location of the deflector obtained from the $i$-band, we created a circular mask for each band that is centered on the deflector location and only covers detected bright pixels outside of the circular region with our determined size. The mask itself is a boolean array with the same shape as the original data, and has the value of zero at any pixel that is to be ignored in {\tt Lenstronomy} computations and ones everywhere else. In the bottom right panel we illustrate the coverage of the mask by setting all of the "ignored" pixels of the original image to a large constant value. 

On average our pipeline took $4.3$ hours to model a $gri$ DES system. This includes reading data, masking process, and performing the modeling sequence to find the best parameters that describe the lens candidate.

\subsection{Modeling Results}

\begin{figure}
\centering
\includegraphics[width=\linewidth]{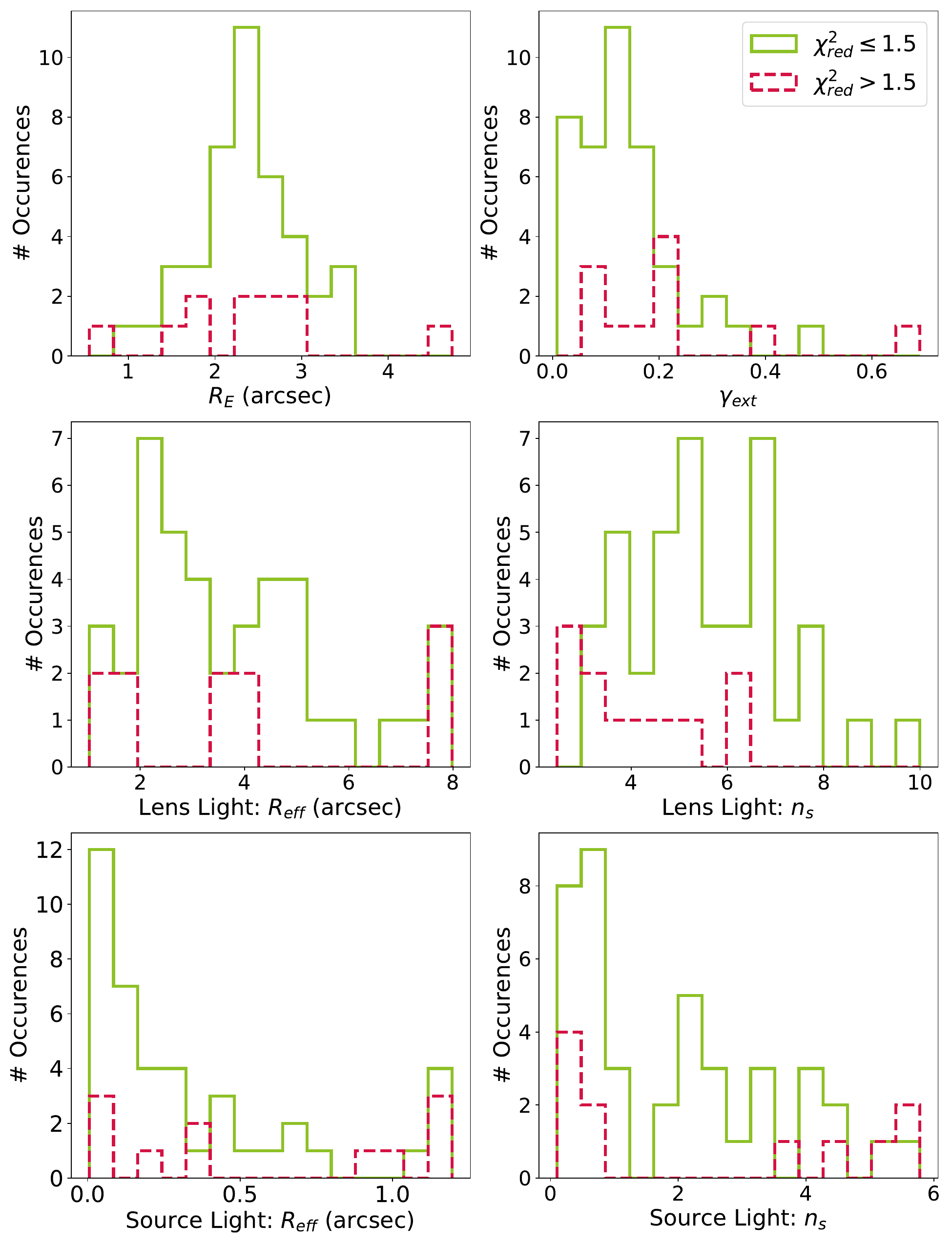}
\caption{Model best-fit parameter distributions after run the automatic pipeline for the lens mass, lens light, and source light profiles. Results for which $reduced \; \chi^2 \leq 1.5$ are shown in green and those with $reduced \; \chi^2 > 1.5$ are displayed in red. Top left: Einstein radii in arcseconds. Top right: External Shear strength, $\gamma_{\text{ext}}$. Middle left: S\'ersic half-light radii of the lens light. Middle right: S\'ersic indices of the lens light. Bottom left: S\'ersic half-light radii of the source light. Bottom right: S\'ersic indices of the source light. }
\label{Fig: param histograms}
\end{figure}

Using our automated pipeline, we modeled 52 of the lens candidates in the "Sure Lenses" catalog that appeared to have only a single galaxy as a deflector. We show in Figs.~\ref{Fig: modeling mosaic 1}$-$\ref{Fig: modeling mosaic 12} a sequence of images to visualize the modeling results in the $r$-band, including the corresponding image, a reconstructed image, normalized residuals, convergence map, and the reconstructed source light. In Tab.~\ref{tab: modeling results table} we present the best model parameters obtained for each system, and we show the obtained distributions for the Einstein Radii, the external shear, the effective radius and S\'ersic index for the lens and source light, in the histograms in Fig.~\ref{Fig: param histograms}. 

As result we obtained acceptable fits for 41 systems, which represent 79\% of the sample, and we observed 11 failures in the fitting, which we define as fits with mean reduced $\chi^2$ per pixel above the threshold of $\chi^2 = 1.5$. 
In the lens mass components, we observed Einstein Radii, $R_E$, distributed between $\sim1\arcsec$ and $\sim3.5\arcsec$. For the external shear strengths we observed $\gamma_{\text{ext}} \lesssim 0.47$ for all lenses except for one in which the fit failed. For the Effective Radii, $R_{\text{eff}}$ and S\'ersic Indices, $n_{\text{s}}$, of the lens light profiles in the $r$-band, we observed peaks at $R_{\text{eff}} \sim 2\arcsec$ and $n_{\text{s}} \sim 5$, respectively. Because the CNN searches selected lens systems from a catalogue of LRGs, we expect to obtain deflector light parameters that are typical for LRGs, and that is indeed what we see here. For the parameter distributions of the source light, we observe the Effective Radii and S\'ersic Indices peaking at $R_{\text{eff}} \sim 0.2$ and $n_{\text{s}} \sim 1$, respectively. This is also expected behavior for smaller, low-mass galaxies that are usually dominating the lensed galaxy population.

When modeling these lenses, the primary source of failures lies in the masking procedure. For example the estimated size for the lens system is either slightly too small or too large, resulting in parts of the lensed source light being masked, or neighboring contaminants not being masked and instead treated as lens features. This happened for four systems with failure fitted models (DES~J060653$-$585843, DES~J015216$-$583842, DES~J032216$-$523440, and DES~J051047$-$263222) and for two considered as with acceptable fits (DES~J034713$-$453506, DES~J040822$-$532714). Since it is common for images to contain companion objects very close to the lens systems, there is a small margin for error in determining the lens system size. For the system, DES~J012042$-$514353, the contaminant is actually residing among the lensed images of the source, a situation which can not simply be handled with a more precise measurement of the lens size. A method would be needed for better untangling the contaminant light from the lens features. Finally, there are two lens systems (DES~J010553$-$053419, DES~J041809$-$545735) in which the contaminant light distributions were spread out enough that the mask failed to adequately cover all this light. In general we need to improve our masking method to avoid these problems during an automatic fitting of the lens. In the meantime, for all of these systems for which the masking algorithm did not perform well, we recreated masks by hand and performed the modeling a second time. These results are shown in the rows directly below the original results for the specific system, and both sets of results are enclosed in a red dashed box in Figs.~\ref{Fig: modeling mosaic 1}$-$\ref{Fig: modeling mosaic 12}. Each time, we see a significant improvement in residuals after using the better mask. On the other hand, the rest of the systems considered as failure (DES~J010659$-$443201, DES~J021159$-$595624, DES~J024803$-$061606, DES~J202855$-$523118) do not show an obvious reason for it, but likely can be due to compactness of the system, faint lens features and complexity in the shape of the source. For these cases we need further investigation to find a general solution to improve their models in the automatic pipeline.

\section{Conclusions}\label{sec:conclu}

We used DES-DR1 to look for galaxy-scale strong lensing systems using a convolutional neural network that carries out a binary classification of optical images in the  $g$, $r$ and $i$ bands. In doing so, we targeted massive galaxies i.e. LRGs, which were selected using a wide color-magnitude cut accounting for a realistic color contamination of the putative LRG lenses by the background star-forming blue galaxies.

The design of our training set was data-driven in the sense that real DES images of LRGs were used to mock the light distribution of the lens plane. Real images of galaxies from the COSMOS HST were used to mock lensed sources. This helps ensure diversity in colors and morphologies for the sources and lenses, but also preserves the sky background characteristics, galaxies and/or stars acting as companions, and any artifacts in the images.

We used these data-driven simulations as positive examples to train a Convolutional Neural Network, while we used a portion of the LRG sample as negative examples. The CNN was trained and validated using a total of 200~000 images, half of them being mocked lens systems labeled as 1, and the other half being LRGs labeled as 0. Our model evaluated in a test set built from images with the same characteristics of the training set gave us an accuracy of 99.7\%. On the other hand a small test set built with 300 lens candidates and the same proportion of LRGs, gave us a more realistic evaluation reaching an accuracy of 89.6\%.

Applied to the 18~745~029 LRGs drawn from our color-magnitude selection, we obtained 76~582 images with CNN scores above or equal to 0.9, that we all visually inspected. To do so, we created guidelines to separate them in the different categories: “Sure Lens”, “Maybe Lens”, “Flexion”, “Non Lens”, and subcategories: “Ring galaxy”, “Spiral galaxy”, “Merger” for objects falling in the “Non Lens” category. To do the classification we used a mosaic visualization tool displaying 100 images at once, as well as a one-by-one visualization tool that displayed the color composite image and each band for one object at the time. We classified 0.5\% of the 76~582 images as lens candidates, 81 falling in the “Sure Lens” category and 296 in the “Maybe Lens” category. Additionally we inspected the 17~779 cutouts with a CNN score in the range $0.8 < S_{\text{CNN}} < 0.9$, with only 0.2\% of the images classified as lens candidates, i.e. 9 “Sure Lens” objects and 19 “Maybe Lens”. The visual inspection of these low-score lenses allowed us to conclude that the reward for inspecting images with scores below 0.9 was very poor compared with the amount of work. We therefore did not consider lower scores at all. 

From our visual inspection we created two main catalogs: a lens candidates catalog and a ring galaxy candidates catalog, the latter being our main source of contaminants. The first catalog contains a total of 405 lens systems candidates, i.e. 90 “Sure Lens” and 315 “Maybe Lens”. Out of these, 186 were totally new systems and 219 were identified (but not necessarily confirmed) in previous searches. We deblended the lens and source light for our 90 “Sure Lens” using the {\tt  MuSCADeT} software, which does not involve any profile fitting and uses the color contrast between the lens and source together with sparse regularisation. This was successful in most of the cases, where there were clear differences in the colors of the lens and source. The second catalog contains 539 ring galaxy candidates. We expect to use this ring catalog in the future to improve the training of machine learning algorithms in the recognition between lenses and ring galaxies. Still, 539 objects is not much to train CNNs and further work, e.g. with Generative Adversarial Networks is likely to be needed. 

Finally we selected from the “Sure Lens” category the 52 systems that apparently had one well-defined galaxy as a deflector to test an automated modeling pipeline. The relatively simple SIE + $\gamma_{\text{ext}}$ and Elliptical S\'ersic profiles used in the modeling appeared to be sufficient in describing these lens systems, and additional complexity is not necessary for the purposes of this automated modeling pipeline, at least with the image quality of DES-DR1. We successfully modeled 41 of these systems, while the other 11 failed mainly due to problems in the masking algorithm, especially in the estimation of the lens system size. To address the failures we plan improvements in a future version of the pipeline including the use of the decomposed images from {\tt  MuSCADeT} to initialize the code and find the right position and size of the system. 

The outcomes of this lens finding work in DES-DR1 includes a catalog with 405 lens candidates very meticulously selected that can serve as a good start for spectroscopic confirmation. In our selection, we did our best to privilege quality of the candidates over their quantity. The methods and tools studied, developed and presented here have room for improvements but this work serves as a preview of what the future generation of surveys, i.e. LSST, Roman telescope and the Euclid mission, will be achieved in the very near future. 


\begin{figure*}
\centering
\includegraphics[width=\hsize]{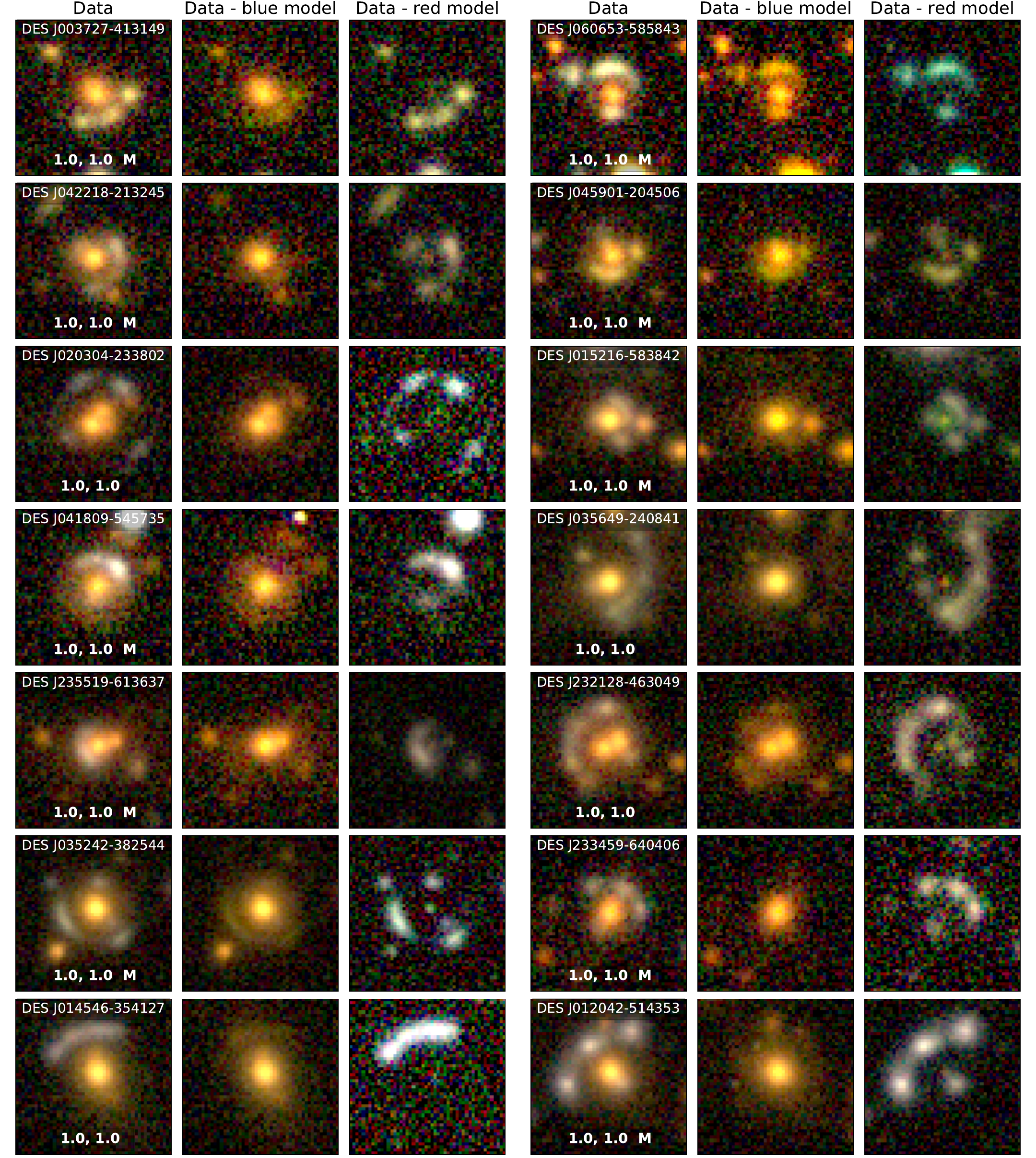}
\caption{Images for the 90 lens candidates in the category Sure lens and their corresponding decomposition performed with {\tt  MuSCADeT}. In the first and fourth columns we have the gri-composite image of the system, the name is on the top, while the CNN score and the visual inspection score (VIS$_{L}$) are displayed at the bottom of each image. Additionally we marked with a "M" those that we modeled in sect.~\ref{sec:model}. Columns 2 and 5 show the subtraction of the blue model to the respective data. Columns 3 and 6 show the subtraction of the red model to the respective data, isolating the lensing features. }
\label{Fig:mosaic_sl1}
\end{figure*}

\begin{figure*}
\centering
\includegraphics[width=\hsize]{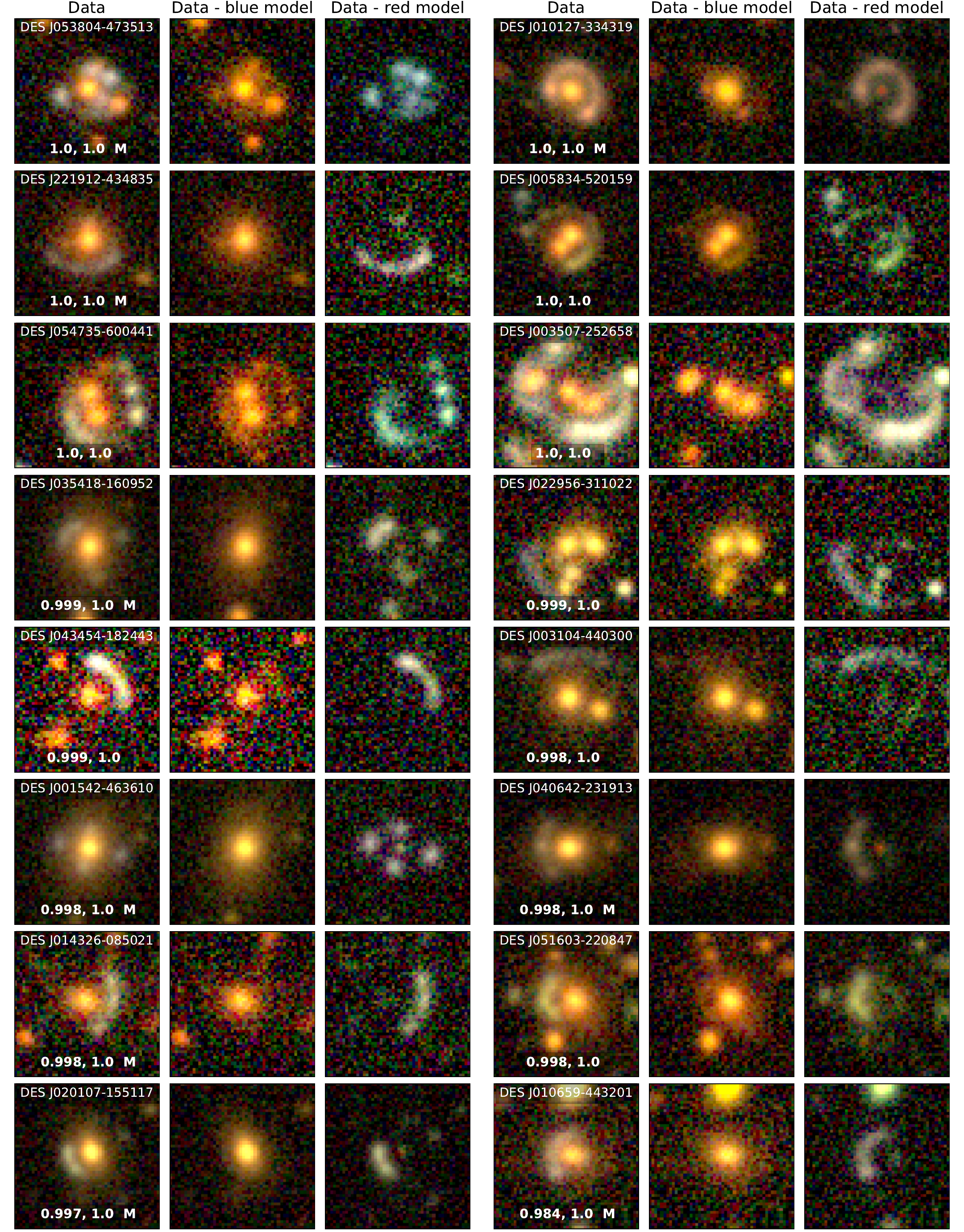}
\caption{Continued from Fig.~\ref{Fig:mosaic_sl1}}
\label{Fig:mosaic_sl2}
\end{figure*}

\begin{figure*}
\centering
\includegraphics[width=\hsize]{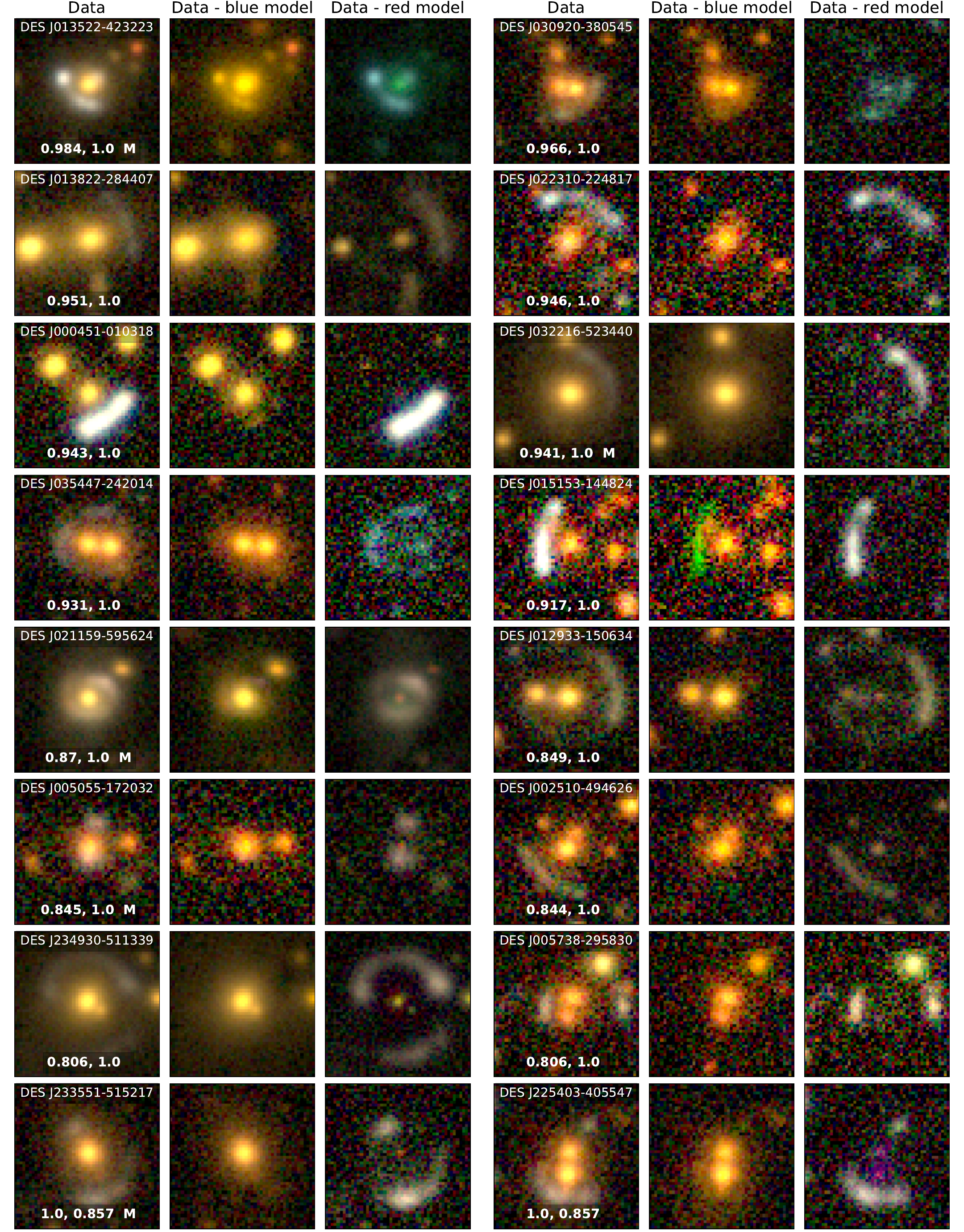}
\caption{Continued from Fig.~\ref{Fig:mosaic_sl2}}
\label{Fig:mosaic_sl3}
\end{figure*}

\begin{figure*}
\centering
\includegraphics[width=\hsize]{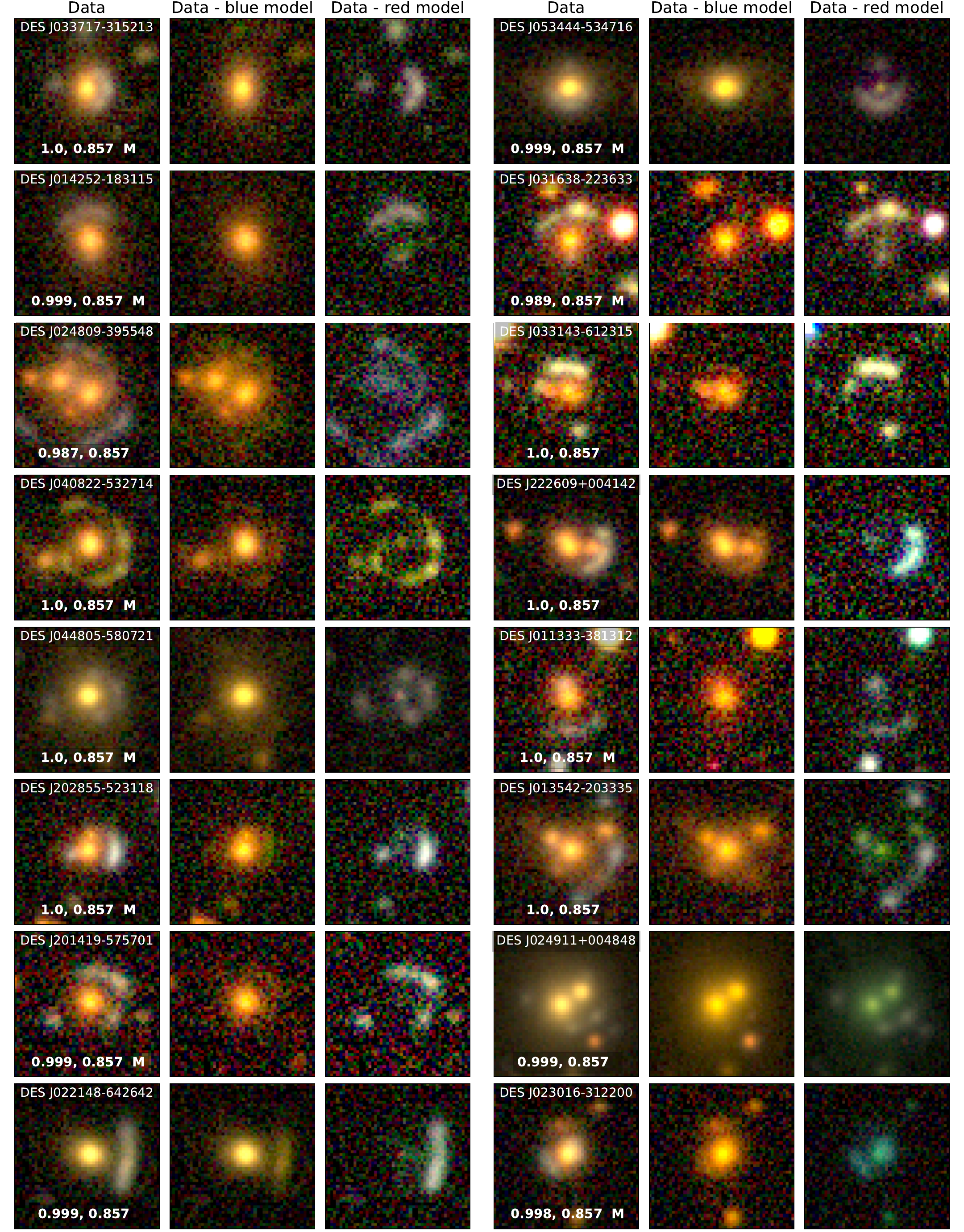}
\caption{Continued from Fig.~\ref{Fig:mosaic_sl3}}
\label{Fig:mosaic_sl4}
\end{figure*}

\begin{figure*}
\centering
\includegraphics[width=\hsize]{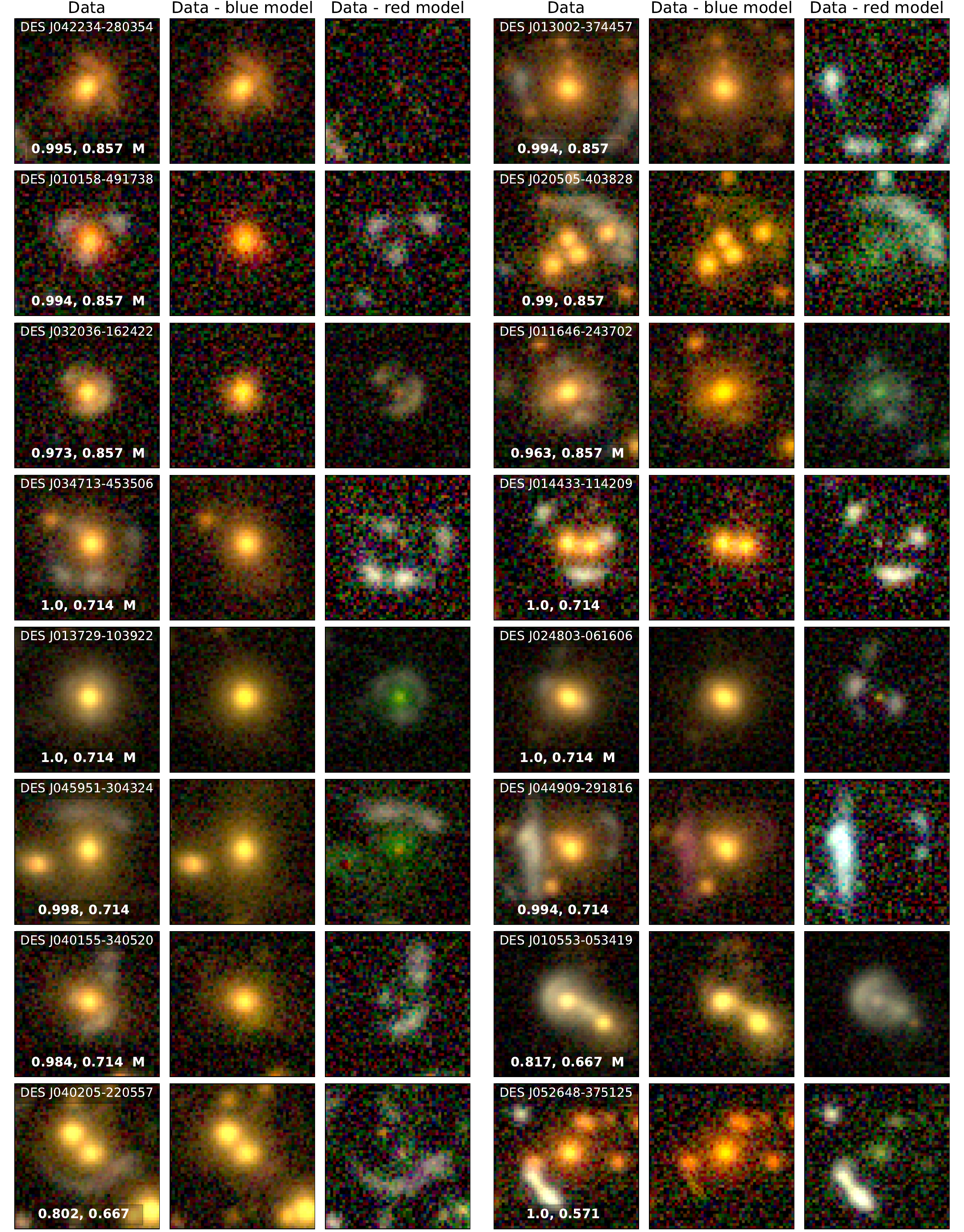}
\caption{Continued from Fig.~\ref{Fig:mosaic_sl4}}
\label{Fig:mosaic_sl5}
\end{figure*}

\begin{figure*}
\centering
\includegraphics[width=\hsize]{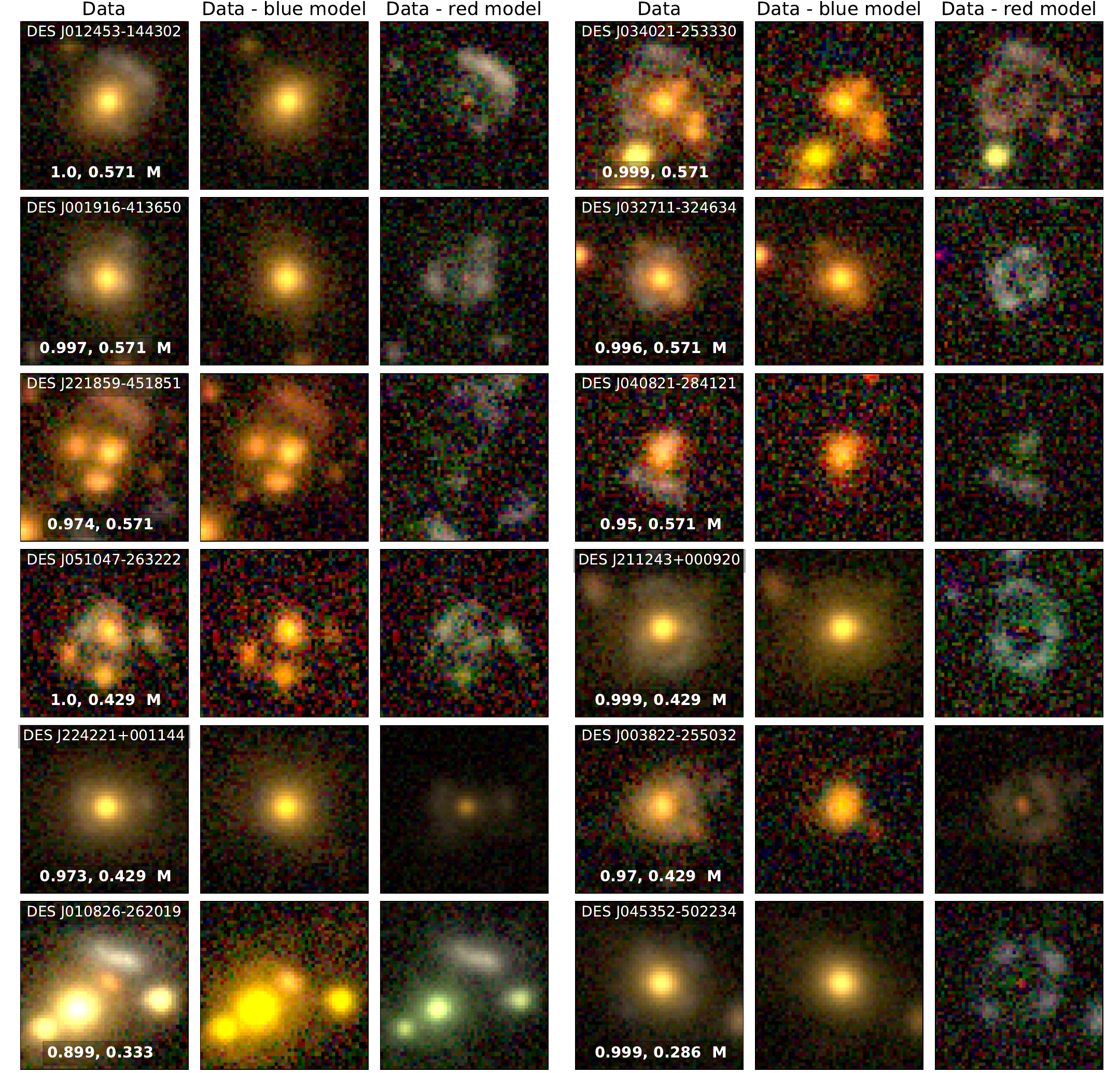}
\caption{Continued from Fig.~\ref{Fig:mosaic_sl5}}
\label{Fig:mosaic_sl6}
\end{figure*}

\longtab{
\begin{longtable}{lccccc}
\caption{Sure lens catalog. 
}\\
\label{tab:surelenscat} \\
\hline\hline
Candidate   &   RA  &   Dec &   S$_{CNN}$  &   VIS$_{L}$\tablefootmark{a}  &   References \\
\hline
\endfirsthead
\caption{continued.}\\
\hline\hline
Candidate   &   RA  &   Dec &   S$_{CNN}$  &   VIS$_{L}$\tablefootmark{a}  &   References \\
\hline
\endhead
\hline
\endfoot
 DES J003727-413149 &    9.362803 & -41.530542 &      1.00 &                    1.00 &                       [2]  [5] \\
 DES J060653-585843 &   91.721422 & -58.978786 &      1.00 &                    1.00 &                      This work \\
 DES J042218-213245 &   65.575901 & -21.546084 &      1.00 &                    1.00 &                       [4]  [5] \\
 DES J045901-204506 &   74.756099 & -20.751891 &      1.00 &                    1.00 &                            [5] \\
 DES J020304-233802 &   30.766707 & -23.634045 &      1.00 &                    1.00 &                  [4]  [5]  [7] \\
 DES J015216-583842 &   28.068067 & -58.645086 &      1.00 &                    1.00 &                            [5] \\
 DES J041809-545735 &   64.541168 & -54.959729 &      1.00 &                    1.00 &                 [2]  [5]  [11] \\
 DES J035649-240841 &   59.204383 & -24.144756 &      1.00 &                    1.00 &                            [5] \\
 DES J235519-613637 &  358.829823 & -61.610291 &      1.00 &                    1.00 &                            [5] \\
 DES J232128-463049 &  350.368208 & -46.513706 &      1.00 &                    1.00 &                 [2]  [5]  [11] \\
 DES J035242-382544 &   58.176701 & -38.429152 &      1.00 &                    1.00 &                            [5] \\
 DES J233459-640406 &  353.746649 & -64.068597 &      1.00 &                    1.00 &                            [5] \\
 DES J014546-354127 &   26.444934 & -35.690931 &      1.00 &                    1.00 &                            [5] \\
 DES J012042-514353 &   20.175973 & -51.731411 &      1.00 &                    1.00 &                 [2]  [5]  [11] \\
 DES J053804-473513 &   84.519228 & -47.587152 &      1.00 &                    1.00 &                 [2]  [5]  [11] \\
 DES J010127-334319 &   15.366041 & -33.722010 &      1.00 &                    1.00 &                  [1]  [5]  [6] \\
 DES J221912-434835 &  334.801660 & -43.809752 &      1.00 &                    1.00 &                            [5] \\
 DES J005834-520159 &   14.644654 & -52.033230 &      1.00 &                    1.00 &                       [4]  [5] \\
 DES J054735-600441 &   86.898069 & -60.078194 &      1.00 &                    1.00 &                            [2] \\
 DES J003507-252658 &    8.780570 & -25.449594 &      1.00 &                    1.00 &                            [5] \\
 DES J035418-160952 &   58.576136 & -16.164500 &      1.00 &                    1.00 &                       [4]  [5] \\
 DES J022956-311022 &   37.484396 & -31.172971 &      1.00 &                    1.00 &                     [10]  [12] \\
 DES J043454-182443 &   68.728380 & -18.412014 &      1.00 &                    1.00 &                      This work \\
 DES J003104-440300 &    7.770341 & -44.050039 &      1.00 &                    1.00 &                            [2] \\
 DES J001542-463610 &    3.928313 & -46.603047 &      1.00 &                    1.00 &                            [5] \\
 DES J040642-231913 &   61.676960 & -23.320485 &      1.00 &                    1.00 &                           [12] \\
 DES J014326-085021 &   25.862225 &  -8.839247 &      1.00 &                    1.00 &                       [4]  [5] \\
 DES J051603-220847 &   79.013218 & -22.146421 &      1.00 &                    1.00 &                            [5] \\
 DES J020107-155117 &   30.283189 & -15.854734 &      1.00 &                    1.00 &                       [5]  [8] \\
 DES J010659-443201 &   16.746389 & -44.533731 &      0.98 &                    1.00 &                            [5] \\
 DES J013522-423223 &   23.845122 & -42.539867 &      0.98 &                    1.00 &                       [2]  [5] \\
 DES J030920-380545 &   47.335761 & -38.096044 &      0.97 &                    1.00 &                            [5] \\
 DES J013822-284407 &   24.595671 & -28.735547 &      0.95 &                    1.00 &                      [5]  [12] \\
 DES J022310-224817 &   35.794564 & -22.804826 &      0.95 &                    1.00 &                           [12] \\
 DES J000451-010318 &    1.215541 &  -1.055067 &      0.94 &                    1.00 &                       [2]  [9] \\
 DES J032216-523440 &   50.568366 & -52.577897 &      0.94 &                    1.00 &                 [2]  [5]  [11] \\
 DES J035447-242014 &   58.697981 & -24.337490 &      0.93 &                    1.00 &                       [5]  [7] \\
 DES J015153-144824 &   27.972236 & -14.806869 &      0.92 &                    1.00 &                            [8] \\
 DES J021159-595624 &   32.997544 & -59.940266 &      0.87 &                    1.00 &                      This work \\
 DES J012933-150634 &   22.388682 & -15.109614 &      0.85 &                    1.00 &                      This work \\
 DES J005055-172032 &   12.730170 & -17.342350 &      0.84 &                    1.00 &                      This work \\
 DES J002510-494626 &    6.294459 & -49.774008 &      0.84 &                    1.00 &                      This work \\
 DES J234930-511339 &  357.375235 & -51.227509 &      0.81 &                    1.00 &                      This work \\
 DES J005738-295830 &   14.411528 & -29.975133 &      0.81 &                    1.00 &                      This work \\
 DES J233551-515217 &  353.966362 & -51.871614 &      1.00 &                    0.86 &                       [2]  [5] \\
 DES J225403-405547 &  343.512608 & -40.929780 &      1.00 &                    0.86 &                       [2]  [5] \\
 DES J033717-315213 &   54.321830 & -31.870431 &      1.00 &                    0.86 &                            [5] \\
 DES J053444-534716 &   83.686779 & -53.787850 &      1.00 &                    0.86 &                            [5] \\
 DES J014252-183115 &   25.720295 & -18.521051 &      1.00 &                    0.86 &                       [4]  [5] \\
 DES J031638-223633 &   49.161789 & -22.609246 &      0.99 &                    0.86 &                            [5] \\
 DES J024809-395548 &   42.039715 & -39.930079 &      0.99 &                    0.86 &                            [5] \\
 DES J033143-612315 &   52.932410 & -61.387599 &      1.00 &                    0.86 &                      This work \\
 DES J040822-532714 &   62.094390 & -53.453939 &      1.00 &                    0.86 &                            [2] \\
 DES J222609+004142 &  336.538760 &   0.695037 &      1.00 &                    0.86 &  [2]  [5]  [3]  [9]  [7]  [12] \\
 DES J044805-580721 &   72.022014 & -58.122584 &      1.00 &                    0.86 &                       [2]  [5] \\
 DES J011333-381312 &   18.389652 & -38.220203 &      1.00 &                    0.86 &                      This work \\
 DES J202855-523118 &  307.232456 & -52.521766 &      1.00 &                    0.86 &                            [5] \\
 DES J013542-203335 &   23.928307 & -20.559859 &      1.00 &                    0.86 &                            [5] \\
 DES J201419-575701 &  303.580760 & -57.950411 &      1.00 &                    0.86 &                       [4]  [5] \\
 DES J024911+004848 &   42.299530 &  -0.813541 &      1.00 &                    0.86 &                            [2] \\
 DES J022148-642642 &   35.453087 & -64.445142 &      1.00 &                    0.86 &                      This work \\
 DES J023016-312200 &   37.569906 & -31.366891 &      1.00 &                    0.86 &                            [5] \\
 DES J042234-280354 &   65.644648 & -28.065211 &      1.00 &                    0.86 &                           [12] \\
 DES J013002-374457 &   22.512009 & -37.749376 &      0.99 &                    0.86 &                            [5] \\
 DES J010158-491738 &   15.491818 & -49.293942 &      0.99 &                    0.86 &                       [4]  [5] \\
 DES J020505-403828 &   31.271691 & -40.641381 &      0.99 &                    0.86 &                 [2]  [5]  [11] \\
 DES J032036-162422 &   50.154139 & -16.406181 &      0.97 &                    0.86 &                            [5] \\
 DES J011646-243702 &   19.194939 & -24.617243 &      0.96 &                    0.86 &                            [5] \\
 DES J034713-453506 &   56.805347 & -45.585003 &      1.00 &                    0.71 &                       [2]  [5] \\
 DES J014433-114209 &   26.138946 & -11.702571 &      1.00 &                    0.71 &                       [5]  [8] \\
 DES J013729-103922 &   24.372653 & -10.656303 &      1.00 &                    0.71 &                      This work \\
 DES J024803-061606 &   42.013769 &  -6.268377 &      1.00 &                    0.71 &                      This work \\
 DES J045951-304324 &   74.964225 & -30.723601 &      1.00 &                    0.71 &                           [12] \\
 DES J044909-291816 &   72.289665 & -29.304562 &      0.99 &                    0.71 &                            [5] \\
 DES J040155-340520 &   60.482094 & -34.089137 &      0.98 &                    0.71 &                      This work \\
 DES J010553-053419 &   16.471353 &  -5.571992 &      0.82 &                    0.67 &                      This work \\
 DES J040205-220557 &   60.523330 & -22.099437 &      0.80 &                    0.67 &                      This work \\
 DES J052648-375125 &   81.702120 & -37.856976 &      1.00 &                    0.57 &                      This work \\
 DES J012453-144302 &   21.221090 & -14.717383 &      1.00 &                    0.57 &                       [4]  [5] \\
 DES J034021-253330 &   55.089043 & -25.558367 &      1.00 &                    0.57 &                      [5]  [12] \\
 DES J001916-413650 &    4.818033 & -41.614051 &      1.00 &                    0.57 &                            [5] \\
 DES J032711-324634 &   51.797294 & -32.776155 &      1.00 &                    0.57 &                  [4]  [5]  [6] \\
 DES J221859-451851 &  334.746758 & -45.314441 &      0.97 &                    0.57 &                      This work \\
 DES J040821-284121 &   62.090172 & -28.689265 &      0.95 &                    0.57 &                      This work \\
 DES J051047-263222 &   77.695997 & -26.539526 &      1.00 &                    0.43 &                           [12] \\
 DES J211243+000920 &  318.179744 &   0.155773 &      1.00 &                    0.43 &                            [5] \\
 DES J224221+001144 &  340.589927 &   0.195764 &      0.97 &                    0.43 &                       [3]  [9] \\
 DES J003822-255032 &    9.593161 & -25.842242 &      0.97 &                    0.43 &                       [4]  [5] \\
 DES J010826-262019 &   17.111823 & -26.338700 &      0.90 &                    0.33 &                      This work \\
 DES J045352-502234 &   73.470734 & -50.376374 &      1.00 &                    0.29 &                      This work \\
\end{longtable}
\tablefoot{
\tablefoottext{a}{Visual Inspection Score for Strong lens systems}}
\tablebib{[1]~\cite{Bettinelli2016}, [2]~\cite{Diehl2017}, [3]~\cite{Sonnenfeld2018}, [4]~\cite{Jacobs2019A}, [5]~\cite{Jacobs2019B}, [6]~\cite{Petrillo2019}, [7]~\cite{Canameras2020}, [8]~\cite{Huang2020}, [9]~\cite{Jaelani2020}, [10]~\cite{Li2020}, [11]~\cite{Nord2020}, [12]~\cite{Huang2021}}
}

\begin{figure*}[p!]
\centering
\includegraphics[width=\linewidth]{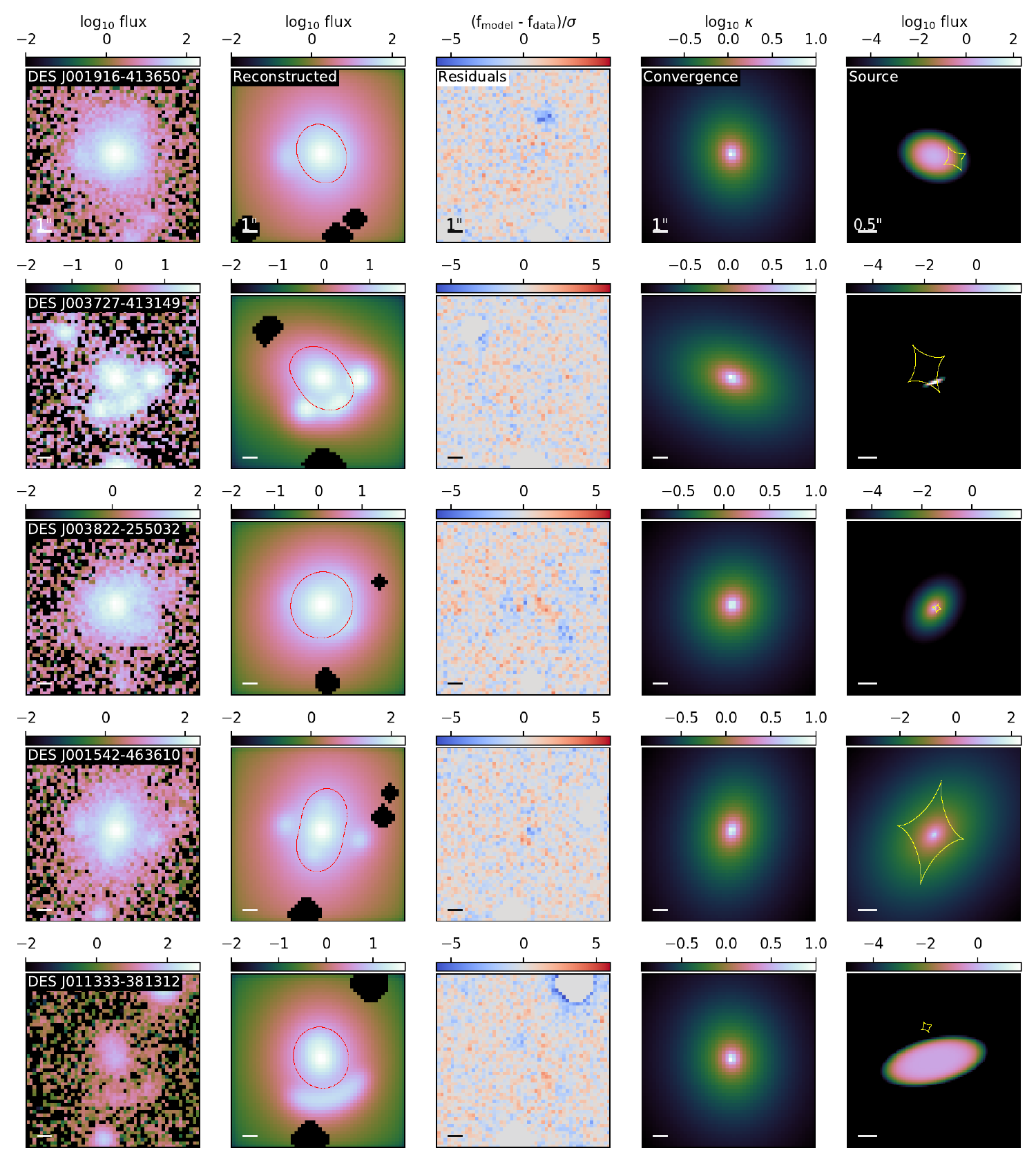}
\caption{Modeling results for the 52 lens candidates that appear to have only a single lens galaxy acting as a deflector. \textit{1st column:} Observed DES image of lens system in $r$-band. \textit{2nd column:} Reconstructed image using best-fit model parameters. The black regions are "masked" pixels that are ignored in the modeling as they contain light from contaminant objects in the image. The red curves are the critical lines of the lens model \textit{3rd column:} Normalized residual map showing the agreement between image reconstruction and the original data. \textit{4th column:} Convergence map of the lens model \textit{5th column:} Reconstructed source light profile (un-lensed). The caustic curves are shown in yellow. }
\label{Fig: modeling mosaic 1}
\end{figure*}
   
\begin{figure*}[p!]
   \centering
   \includegraphics[width=\linewidth]{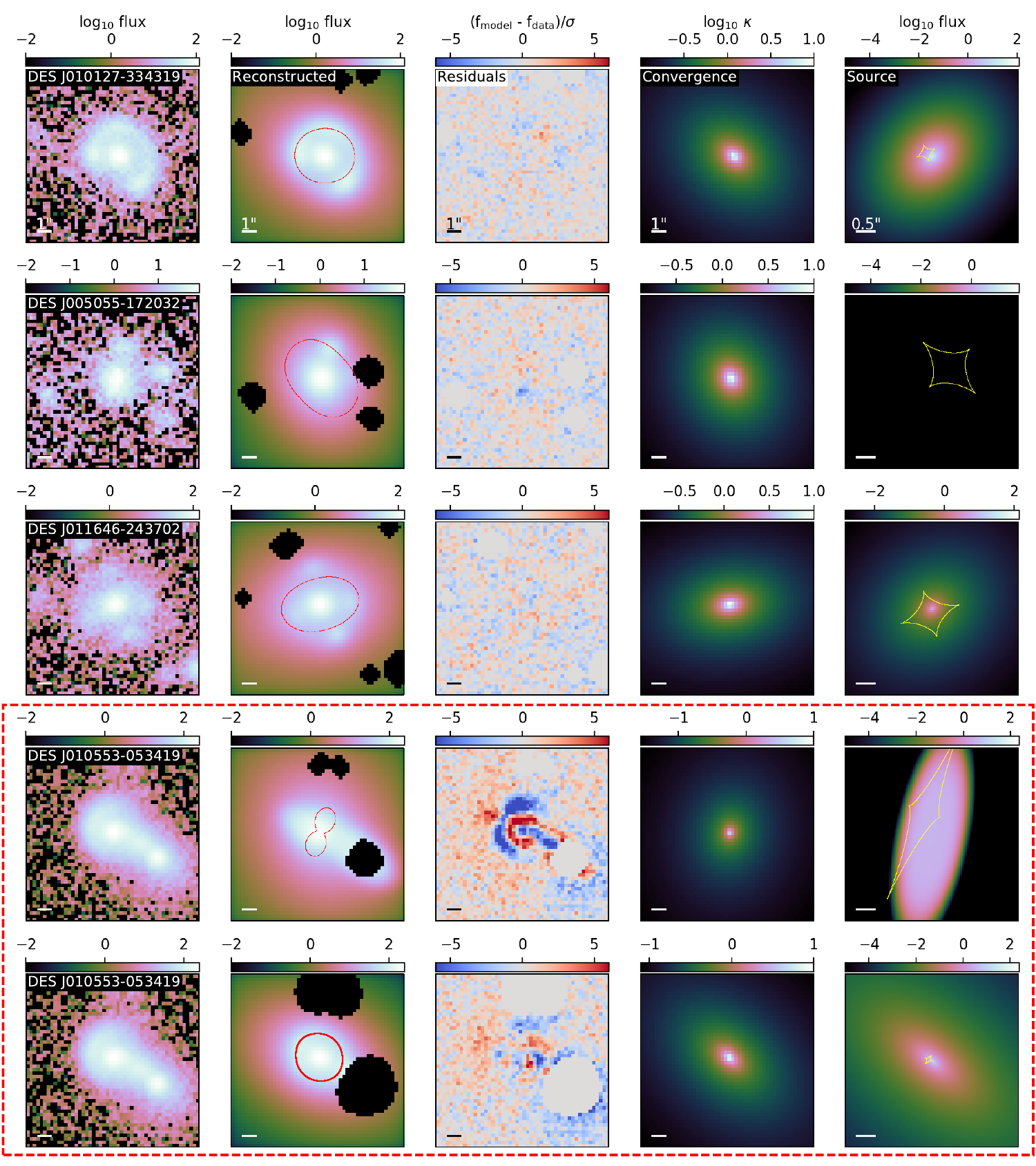}
      \caption{continued from Figure \ref{Fig: modeling mosaic 1} }
         \label{Fig: modeling mosaic 2}
   \end{figure*}
   
\begin{figure*}[p!]
   \centering
   \includegraphics[width=\linewidth]{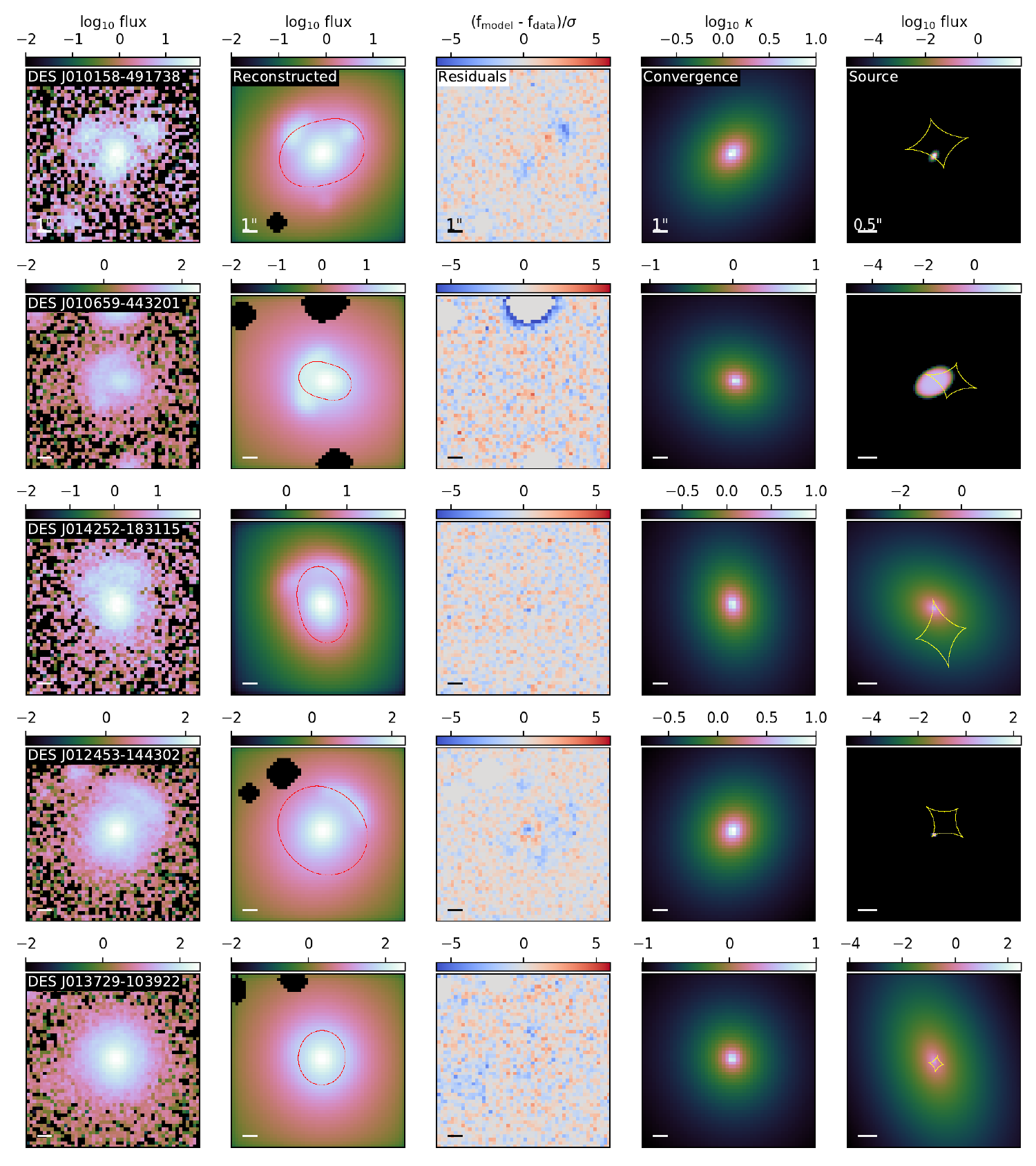}
      \caption{continued from Figure \ref{Fig: modeling mosaic 2} }
         \label{Fig: modeling mosaic 3}
   \end{figure*}
   
\begin{figure*}[p!]
   \centering
   \includegraphics[width=\linewidth]{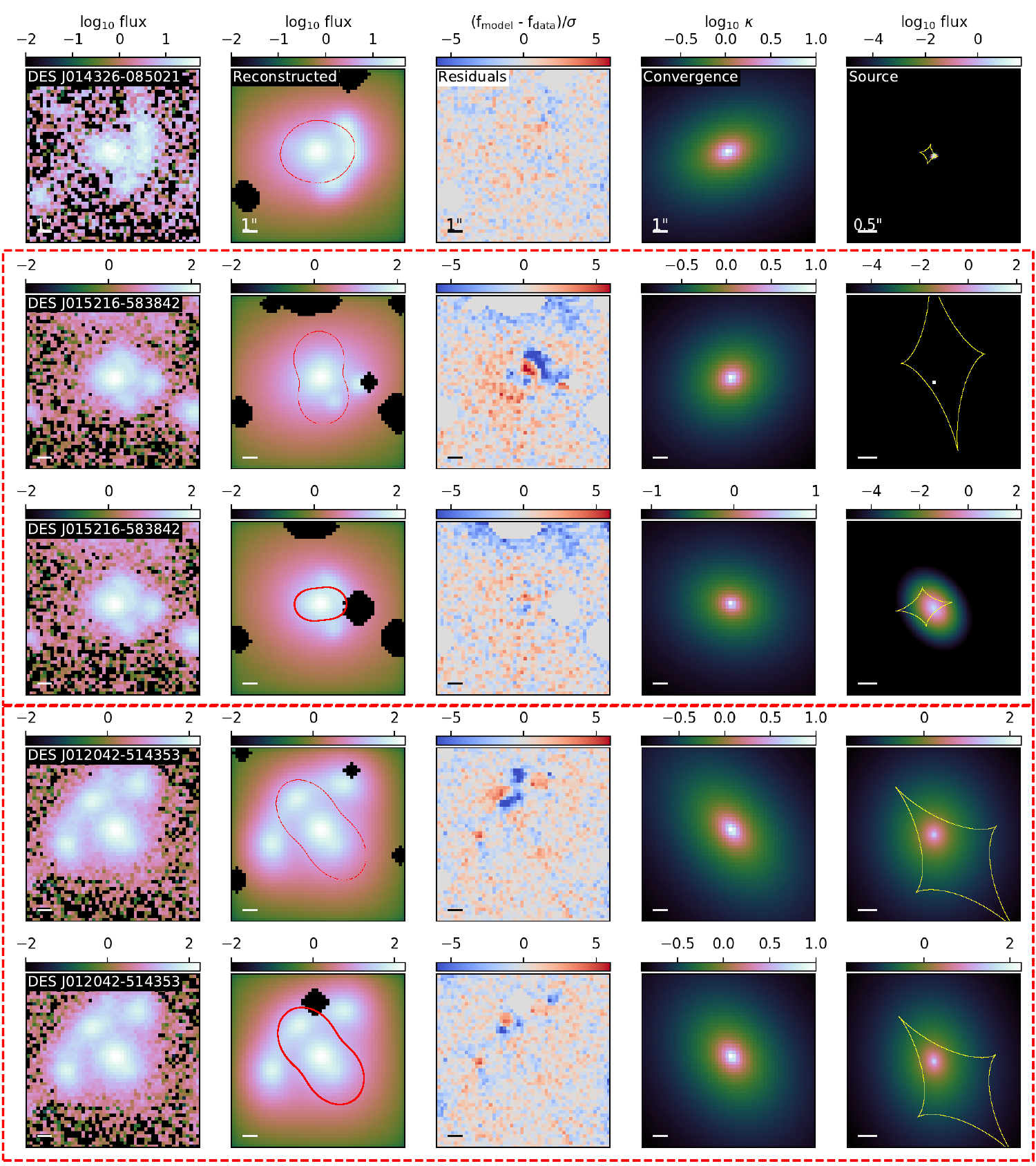}
      \caption{continued from Figure \ref{Fig: modeling mosaic 3} }
         \label{Fig: modeling mosaic 4}
   \end{figure*}
   
\begin{figure*}[p!]
   \centering
   \includegraphics[width=\linewidth]{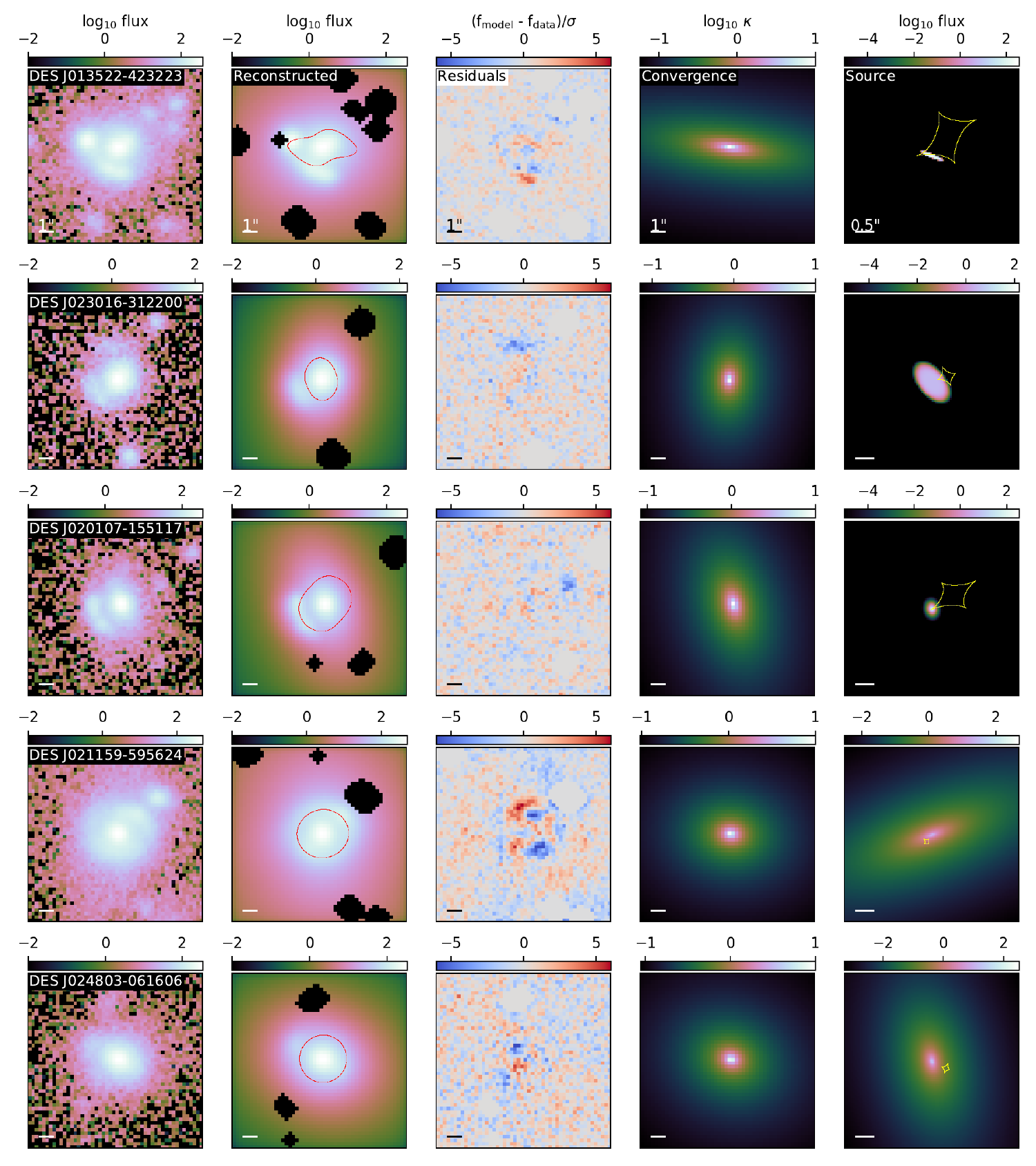}
      \caption{continued from Figure \ref{Fig: modeling mosaic 4} }
         \label{Fig: modeling mosaic 5}
   \end{figure*}
   
\begin{figure*}[p!]
   \centering
   \includegraphics[width=\linewidth]{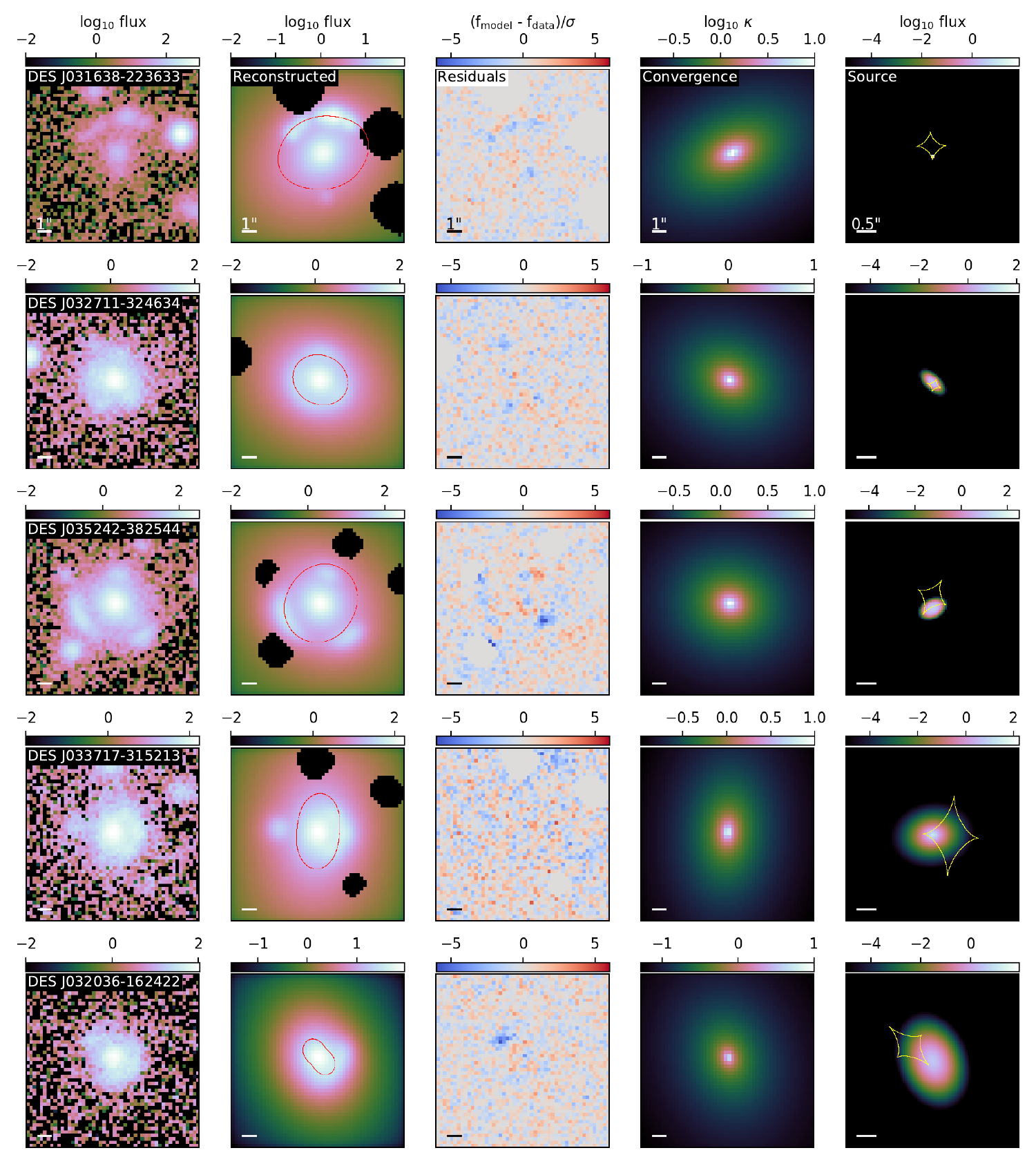}
      \caption{continued from Figure \ref{Fig: modeling mosaic 5} }
         \label{Fig: modeling mosaic 6}
   \end{figure*}

\begin{figure*}[p!]
   \centering
   \includegraphics[width=\linewidth]{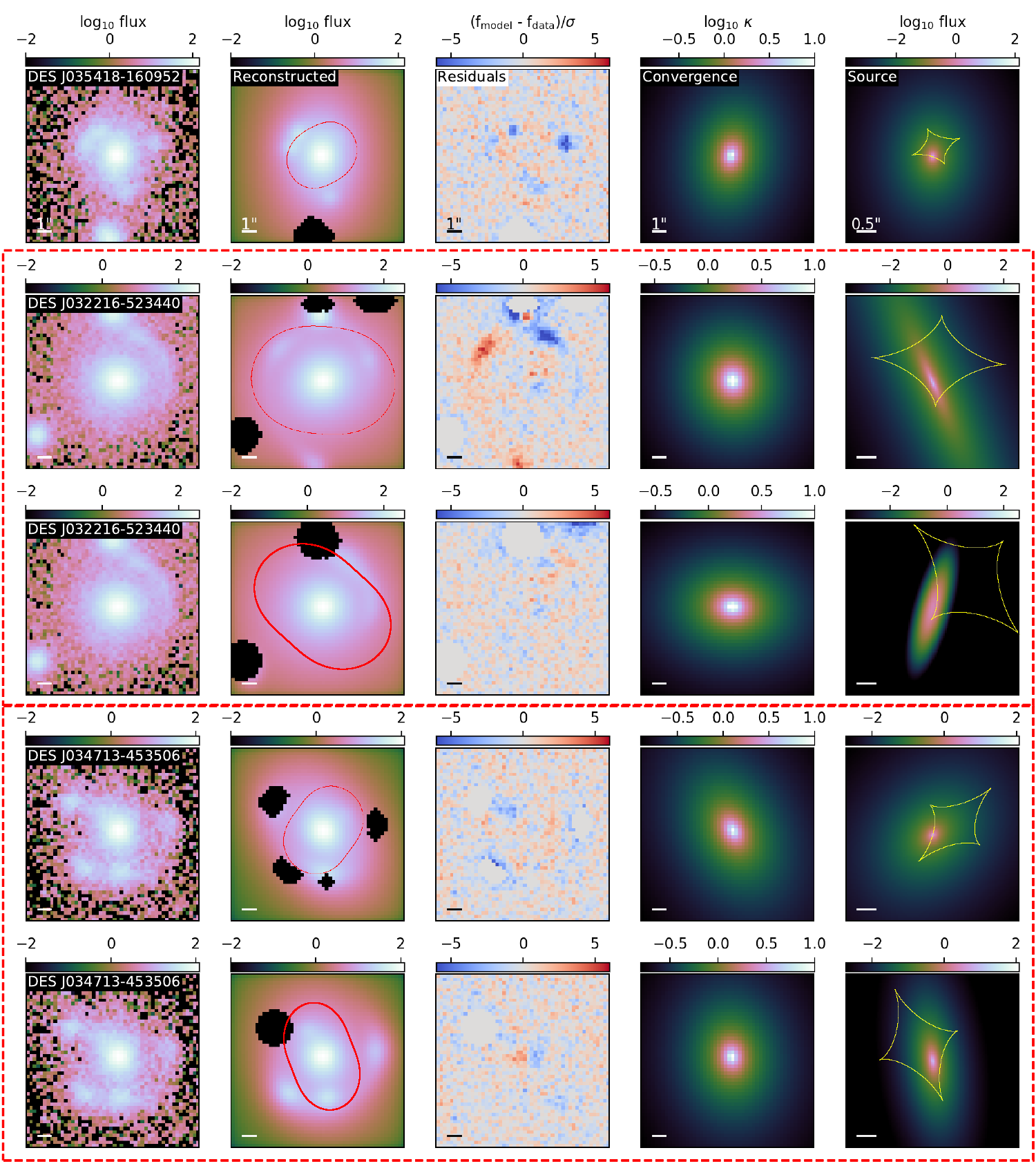}
      \caption{continued from Figure \ref{Fig: modeling mosaic 6} }
         \label{Fig: modeling mosaic 7}
   \end{figure*} 
\begin{figure*}[p!]
   \centering
   \includegraphics[width=\linewidth]{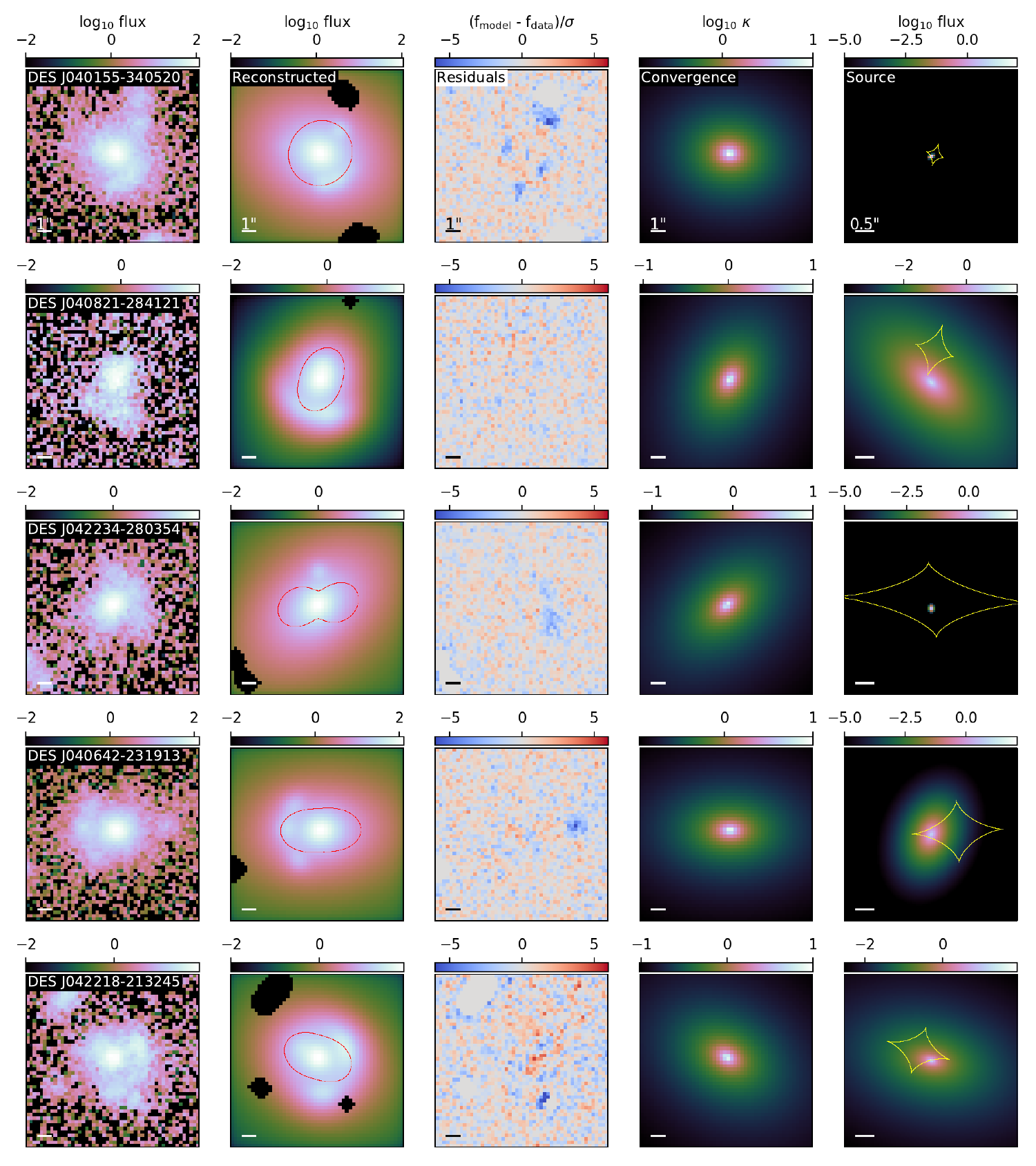}
      \caption{continued from Figure \ref{Fig: modeling mosaic 7} }
         \label{Fig: modeling mosaic 8}
   \end{figure*} 
   
\begin{figure*}[p!]
   \centering
   \includegraphics[width=\linewidth]{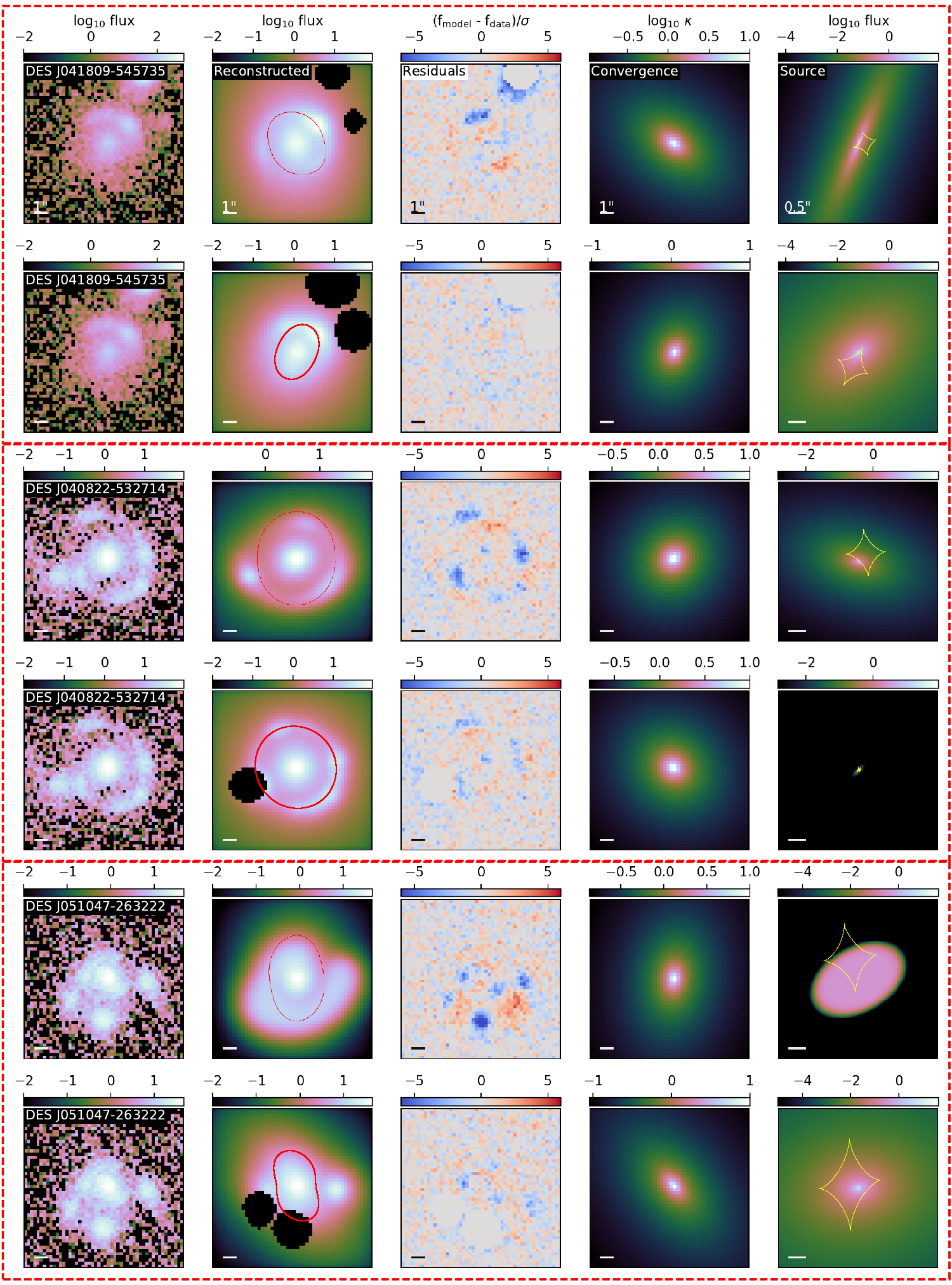}
      \caption{continued from Figure \ref{Fig: modeling mosaic 8} }
         \label{Fig: modeling mosaic 9}
   \end{figure*} 
   
\begin{figure*}[p!]
   \centering
   \includegraphics[width=\linewidth]{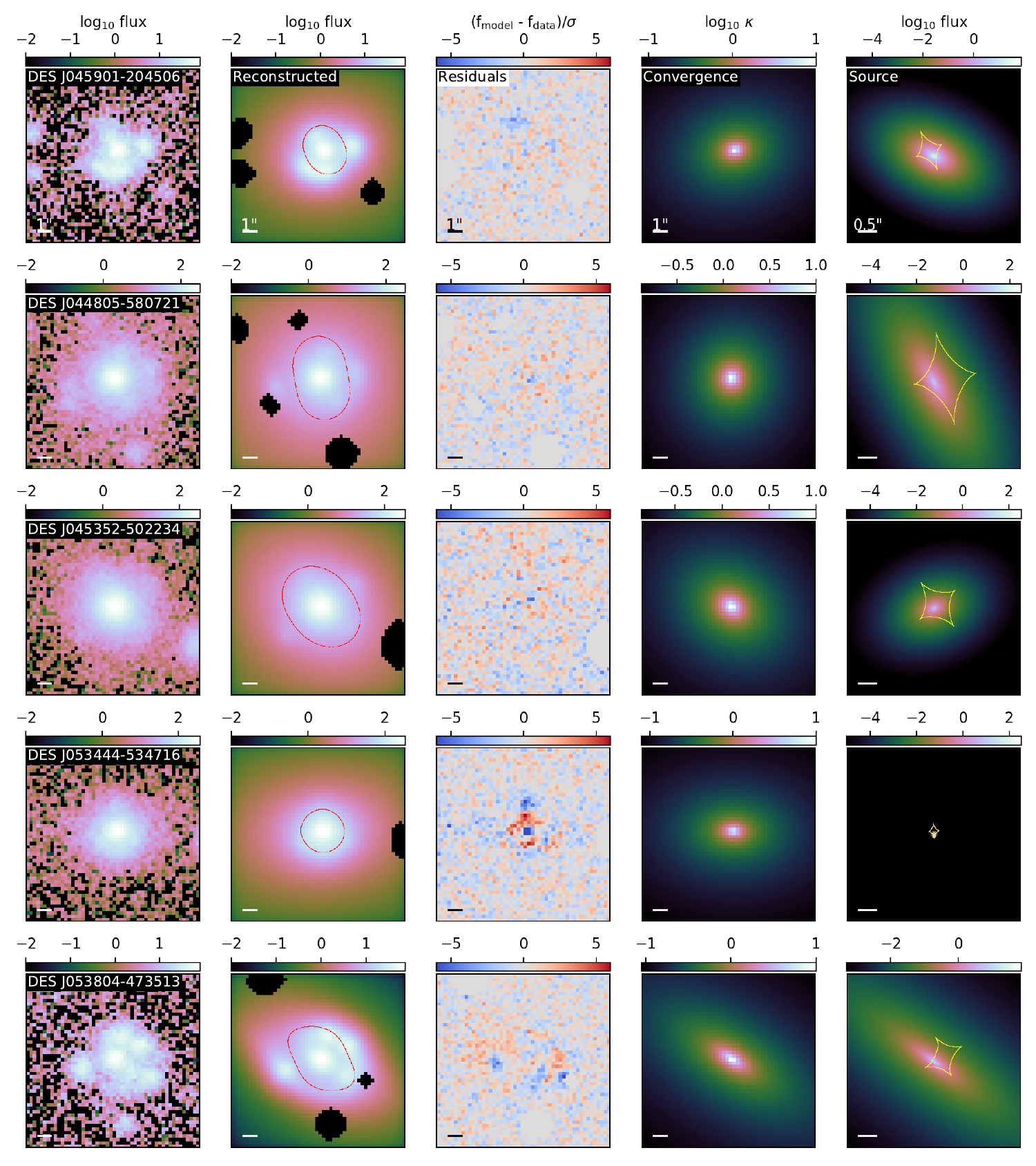}
      \caption{continued from Figure \ref{Fig: modeling mosaic 9} }
         \label{Fig: modeling mosaic 10}
   \end{figure*} 
   
\begin{figure*}[p!]
   \centering
   \includegraphics[width=\linewidth]{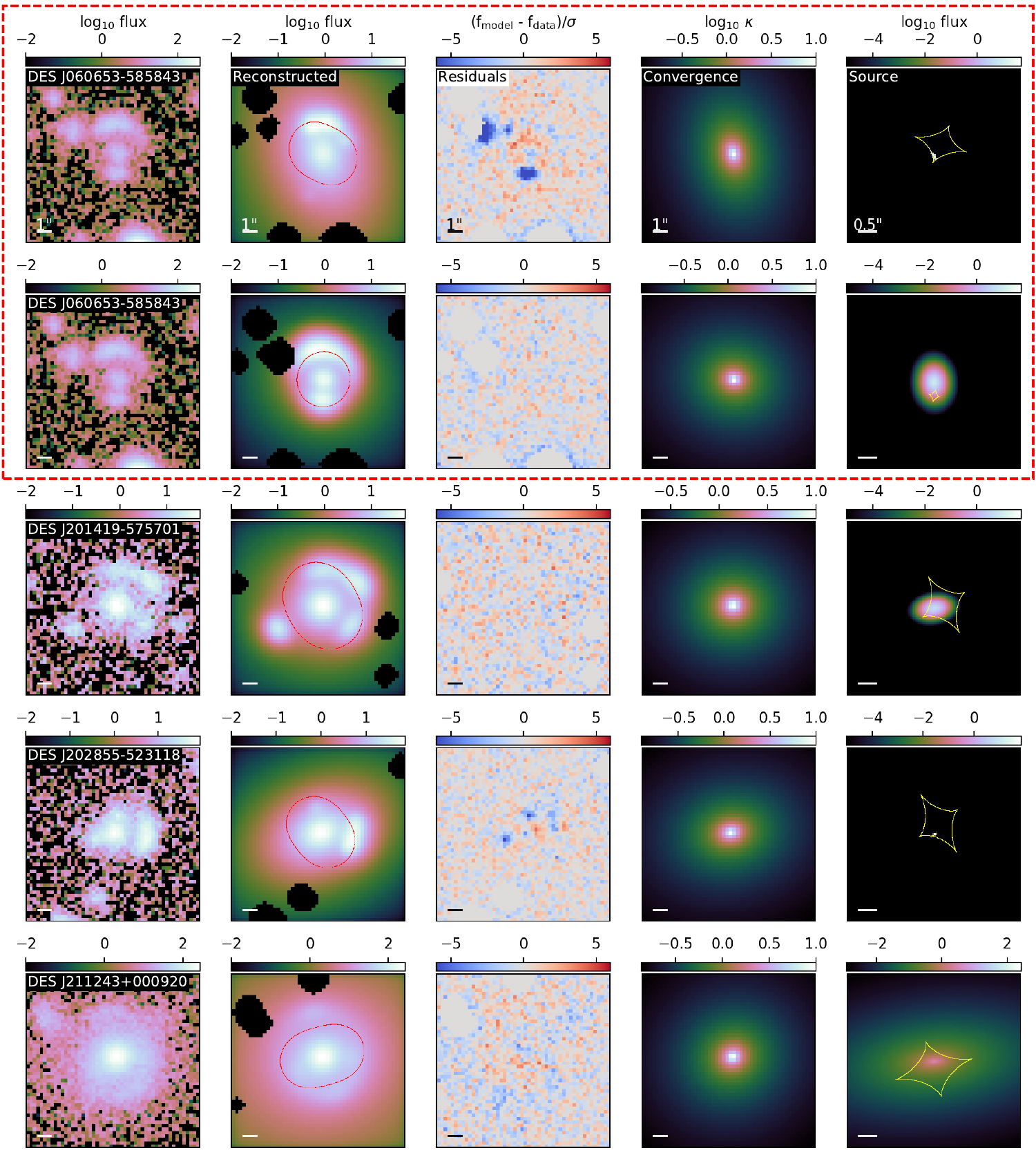}
      \caption{continued from Figure \ref{Fig: modeling mosaic 10} }
         \label{Fig: modeling mosaic 11}
   \end{figure*} 

\begin{figure*}[p!]
   \centering
   \includegraphics[width=\linewidth]{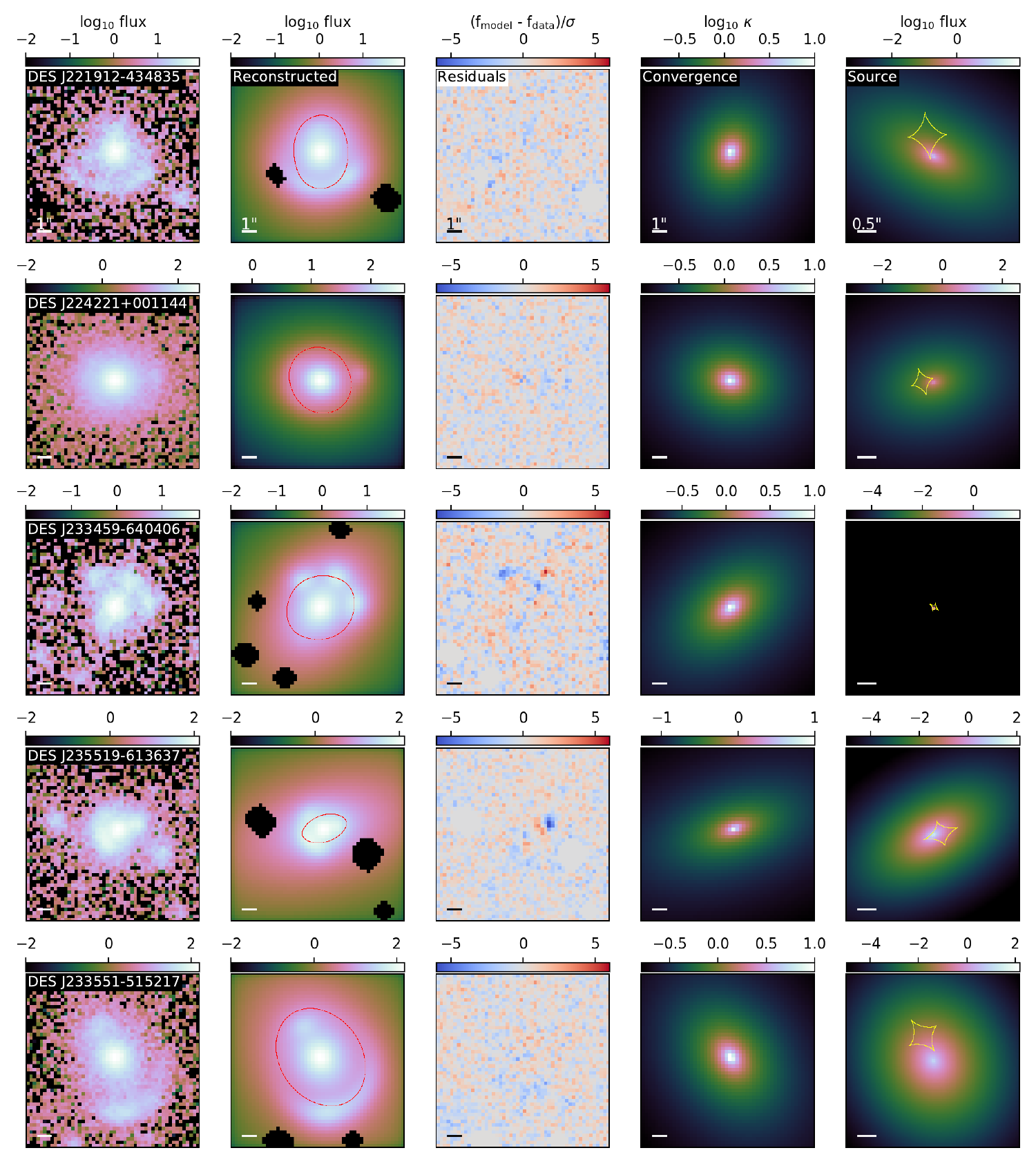}
      \caption{continued from Figure \ref{Fig: modeling mosaic 11} }
         \label{Fig: modeling mosaic 12}
   \end{figure*}

\longtab{
\renewcommand*{\arraystretch}{1.4}
\begin{longtable}{c c c c c c c c c c c}
\caption{lens mass parameters obtained from automated modeling of the single galaxy scale systems. $R_E$ is the Einstein radius of the deflector's mass profile. $q_m$ and $PA_m$ are the axis ratio and position angle of the deflector mass, respectively. The strength and angles of the external shear are given in the $\gamma_{\text{ext}}$ and $PA_{ext}$ columns, respectively.}\\
\label{tab: modeling results table}\\
\hline\hline
 Candidate & Reduced $\chi^2$ &  $R_E$  &  $q_m$ &  $PA_m$ &  $\gamma_{\text{ext}}$ & $\phi_{ext}$\\
\hline
\endfirsthead
\caption{continued.}\\
\hline\hline
Candidate & Reduced $\chi^2$ &  $R_E$  &  $q_m$ &  $PA_m$ &  $\gamma_{\text{ext}}$ & $\phi_{ext}$\\
\hline
\endhead
\hline
\endfoot
DES J001916-413650 &           $1.18$ & $2.061^{+0.020}_{-0.021}$ & $0.827^{+0.019}_{-0.019}$ &    $81^{+7}_{-7}$ & $0.053^{+0.008}_{-0.008}$ &   $-50^{+9}_{-9}$ \\
DES J003727-413149 &           $1.08$ & $2.315^{+0.004}_{-0.004}$ & $0.685^{+0.011}_{-0.013}$ &    $23^{+2}_{-3}$ & $0.159^{+0.004}_{-0.004}$ &    $-9^{+1}_{-1}$ \\
DES J003822-255032 &           $1.13$ & $2.470^{+0.008}_{-0.005}$ & $0.911^{+0.014}_{-0.022}$ &  $-74^{+9}_{-12}$ & $0.008^{+0.004}_{-0.006}$ & $-64^{+29}_{-50}$ \\
DES J001542-463610 &           $0.98$ & $2.353^{+0.007}_{-0.007}$ & $0.721^{+0.018}_{-0.020}$ &   $-79^{+4}_{-4}$ & $0.148^{+0.005}_{-0.005}$ &    $12^{+2}_{-2}$ \\
DES J011333-381312 &           $1.36$ & $2.249^{+0.029}_{-0.021}$ & $0.845^{+0.035}_{-0.025}$ &  $85^{+10}_{-11}$ & $0.038^{+0.015}_{-0.010}$ & $-76^{+16}_{-16}$ \\
DES J010127-334319 &           $1.04$ & $2.213^{+0.005}_{-0.006}$ & $0.772^{+0.014}_{-0.014}$ &    $43^{+4}_{-4}$ & $0.130^{+0.006}_{-0.006}$ &    $53^{+2}_{-2}$ \\
DES J005055-172032 &           $1.35$ & $2.714^{+0.017}_{-0.020}$ & $0.830^{+0.030}_{-0.035}$ &  $68^{+11}_{-13}$ & $0.129^{+0.009}_{-0.008}$ &   $-54^{+3}_{-3}$ \\
DES J011646-243702 &           $1.33$ & $2.465^{+0.016}_{-0.012}$ & $0.697^{+0.020}_{-0.021}$ &    $-9^{+5}_{-6}$ & $0.052^{+0.009}_{-0.010}$ &   $53^{+9}_{-11}$ \\
DES J010553-053419 &           $2.65$ & $0.547^{+0.009}_{-0.013}$ & $0.811^{+0.034}_{-0.031}$ & $-80^{+13}_{-11}$ & $0.690^{+0.004}_{-0.004}$ &    $23^{+0}_{-0}$ \\
DES J010553-053419* &           $1.69$ & $1.857^{+0.010}_{-0.005}$ & $0.671^{+0.010}_{-0.011}$ &    $43^{+2}_{-3}$ & $0.158^{+0.004}_{-0.004}$ &    $41^{+1}_{-2}$ \\
DES J010158-491738 &           $1.14$ & $2.877^{+0.041}_{-0.031}$ & $0.765^{+0.032}_{-0.032}$ &   $-48^{+7}_{-7}$ & $0.140^{+0.010}_{-0.008}$ &   $-85^{+6}_{-8}$ \\
DES J010659-443201 &           $1.74$ & $1.638^{+0.014}_{-0.015}$ & $0.872^{+0.025}_{-0.025}$ &  $23^{+12}_{-12}$ & $0.149^{+0.009}_{-0.009}$ &   $-76^{+3}_{-3}$ \\
DES J014252-183115 &           $1.19$ & $2.347^{+0.010}_{-0.009}$ & $0.729^{+0.019}_{-0.020}$ &    $79^{+3}_{-4}$ & $0.079^{+0.007}_{-0.008}$ &   $-15^{+4}_{-5}$ \\
DES J012453-144302 &           $1.21$ & $3.370^{+0.006}_{-0.006}$ & $0.868^{+0.012}_{-0.008}$ &   $-64^{+5}_{-3}$ & $0.132^{+0.004}_{-0.003}$ &   $-54^{+1}_{-1}$ \\
DES J013729-103922 &           $1.33$ & $1.947^{+0.008}_{-0.013}$ & $0.907^{+0.020}_{-0.019}$ &  $74^{+15}_{-12}$ & $0.041^{+0.009}_{-0.006}$ &   $24^{+11}_{-8}$ \\
DES J014326-085021 &           $1.12$ & $2.615^{+0.005}_{-0.003}$ & $0.695^{+0.011}_{-0.017}$ &   $-21^{+2}_{-3}$ & $0.109^{+0.003}_{-0.004}$ &   $-34^{+3}_{-4}$ \\
DES J015216-583842 &           $2.06$ & $2.294^{+0.011}_{-0.004}$ & $0.869^{+0.023}_{-0.011}$ &  $-34^{+14}_{-6}$ & $0.393^{+0.003}_{-0.002}$ &   $-10^{+0}_{-0}$ \\
DES J015216-583842* &           $1.63$ & $1.581^{+0.006}_{-0.006}$ & $0.820^{+0.019}_{-0.018}$ &    $18^{+6}_{-6}$ & $0.178^{+0.006}_{-0.005}$ &    $74^{+2}_{-2}$ \\
DES J012042-514353 &           $1.90$ & $2.950^{+0.003}_{-0.003}$ & $0.657^{+0.007}_{-0.006}$ &    $53^{+1}_{-1}$ & $0.197^{+0.002}_{-0.002}$ &   $-41^{+0}_{-0}$ \\
DES J012042-514353* &           $1.67$ & $2.870^{+0.003}_{-0.012}$ & $0.766^{+0.013}_{-0.007}$ &    $57^{+7}_{-1}$ & $0.239^{+0.002}_{-0.002}$ &   $-42^{+0}_{-0}$ \\
DES J013522-423223 &           $1.19$ & $1.715^{+0.013}_{-0.009}$ & $0.309^{+0.019}_{-0.003}$ &     $8^{+1}_{-0}$ & $0.342^{+0.002}_{-0.004}$ &    $28^{+0}_{-1}$ \\
DES J023016-312200 &           $1.22$ & $1.411^{+0.015}_{-0.017}$ & $0.666^{+0.019}_{-0.017}$ &   $-86^{+4}_{-4}$ & $0.105^{+0.009}_{-0.009}$ &   $-63^{+5}_{-5}$ \\
DES J020107-155117 &           $1.13$ & $1.920^{+0.007}_{-0.009}$ & $0.599^{+0.011}_{-0.011}$ &    $77^{+2}_{-2}$ & $0.291^{+0.005}_{-0.005}$ &    $65^{+1}_{-1}$ \\
DES J021159-595624 &           $1.63$ & $1.902^{+0.005}_{-0.006}$ & $0.796^{+0.014}_{-0.018}$ &     $6^{+4}_{-4}$ & $0.085^{+0.003}_{-0.003}$ &    $13^{+2}_{-2}$ \\
DES J024803-061606 &           $1.52$ & $1.786^{+0.009}_{-0.010}$ & $0.847^{+0.018}_{-0.018}$ &    $21^{+7}_{-7}$ & $0.088^{+0.007}_{-0.007}$ &    $21^{+5}_{-4}$ \\
DES J031638-223633 &           $1.21$ & $3.179^{+0.005}_{-0.004}$ & $0.662^{+0.007}_{-0.007}$ &   $-26^{+1}_{-1}$ & $0.104^{+0.003}_{-0.003}$ &   $-40^{+1}_{-1}$ \\
DES J032711-324634 &           $1.13$ & $2.026^{+0.009}_{-0.008}$ & $0.842^{+0.022}_{-0.020}$ &    $31^{+9}_{-8}$ & $0.023^{+0.007}_{-0.007}$ &  $55^{+20}_{-19}$ \\
DES J035242-382544 &           $1.23$ & $2.881^{+0.006}_{-0.006}$ & $0.899^{+0.013}_{-0.013}$ &    $12^{+8}_{-8}$ & $0.117^{+0.004}_{-0.004}$ &    $24^{+2}_{-2}$ \\
DES J033717-315213 &           $1.30$ & $2.184^{+0.012}_{-0.012}$ & $0.628^{+0.015}_{-0.014}$ &   $-86^{+3}_{-3}$ & $0.063^{+0.007}_{-0.008}$ &     $7^{+6}_{-6}$ \\
DES J032036-162422 &           $1.10$ & $1.101^{+0.035}_{-0.039}$ & $0.767^{+0.032}_{-0.032}$ &    $75^{+9}_{-9}$ & $0.238^{+0.018}_{-0.019}$ &   $-53^{+5}_{-5}$ \\
DES J035418-160952 &           $1.28$ & $2.517^{+0.003}_{-0.004}$ & $0.713^{+0.007}_{-0.007}$ &   $-81^{+2}_{-2}$ & $0.201^{+0.002}_{-0.002}$ &    $82^{+1}_{-1}$ \\
DES J032216-523440 &           $2.74$ & $4.738^{+0.016}_{-0.013}$ & $0.868^{+0.055}_{-0.029}$ &  $82^{+16}_{-11}$ & $0.209^{+0.004}_{-0.005}$ &   $-89^{+1}_{-1}$ \\
DES J032216-523440* &          $1.33$ & $4.628^{+0.013}_{-0.012}$ & $0.799^{+0.021}_{-0.030}$ &     $4^{+7}_{-7}$ & $0.197^{+0.004}_{-0.004}$ &   $-35^{+1}_{-1}$ \\
DES J034713-453506 &           $1.06$ & $2.951^{+0.016}_{-0.013}$ & $0.686^{+0.021}_{-0.017}$ &    $66^{+5}_{-4}$ & $0.289^{+0.008}_{-0.006}$ &    $51^{+2}_{-1}$ \\
DES J034713-453506* &          $0.96$ & $3.269^{+0.009}_{-0.010}$ & $0.762^{+0.017}_{-0.017}$ &    $85^{+3}_{-4}$ & $0.152^{+0.004}_{-0.006}$ &   $-31^{+2}_{-2}$ \\
DES J040155-340520 &           $1.04$ & $2.473^{+0.006}_{-0.007}$ & $0.824^{+0.018}_{-0.015}$ &     $6^{+5}_{-5}$ & $0.118^{+0.005}_{-0.004}$ &    $12^{+2}_{-2}$ \\
DES J040821-284121 &           $1.06$ & $2.033^{+0.034}_{-0.035}$ & $0.690^{+0.032}_{-0.031}$ &   $-63^{+8}_{-7}$ & $0.062^{+0.018}_{-0.019}$ &  $-4^{+14}_{-15}$ \\
DES J042234-280354 &           $0.91$ & $1.655^{+0.036}_{-0.036}$ & $0.650^{+0.022}_{-0.023}$ &   $-41^{+4}_{-5}$ & $0.474^{+0.013}_{-0.012}$ &   $-84^{+1}_{-1}$ \\
DES J040642-231913 &           $1.04$ & $2.296^{+0.009}_{-0.009}$ & $0.658^{+0.016}_{-0.016}$ &     $1^{+3}_{-2}$ & $0.097^{+0.007}_{-0.007}$ &    $82^{+4}_{-4}$ \\
DES J042218-213245 &           $1.43$ & $2.116^{+0.013}_{-0.012}$ & $0.732^{+0.019}_{-0.019}$ &    $29^{+5}_{-5}$ & $0.103^{+0.008}_{-0.008}$ &   $-85^{+5}_{-5}$ \\
DES J041809-545735 &           $1.54$ & $2.528^{+0.013}_{-0.010}$ & $0.675^{+0.024}_{-0.022}$ &    $39^{+6}_{-5}$ & $0.126^{+0.004}_{-0.004}$ &    $22^{+2}_{-2}$ \\
DES J041809-545735* &          $1.05$ & $2.007^{+0.017}_{-0.015}$ & $0.774^{+0.023}_{-0.021}$ &   $-67^{+7}_{-6}$ & $0.061^{+0.007}_{-0.007}$ &    $31^{+7}_{-6}$ \\
DES J040822-532714 &           $1.22$ & $3.577^{+0.008}_{-0.008}$ & $0.895^{+0.021}_{-0.019}$ & $-50^{+15}_{-13}$ & $0.099^{+0.006}_{-0.005}$ &   $-16^{+3}_{-3}$ \\
DES J040822-532714* &          $1.02$ & $3.463^{+0.002}_{-0.049}$ & $0.881^{+0.009}_{-0.099}$ &   $47^{+52}_{-4}$ & $0.038^{+0.007}_{-0.002}$ &   $43^{+3}_{-22}$ \\
DES J051047-263222 &           $1.57$ & $2.879^{+0.015}_{-0.014}$ & $0.668^{+0.018}_{-0.018}$ &   $-83^{+4}_{-4}$ & $0.086^{+0.005}_{-0.005}$ &   $-35^{+3}_{-3}$ \\
DES J051047-263222* &          $1.13$ & $2.233^{+0.013}_{-0.013}$ & $0.617^{+0.025}_{-0.028}$ &    $52^{+5}_{-5}$ & $0.242^{+0.007}_{-0.007}$ &    $14^{+2}_{-2}$ \\
DES J045901-204506 &           $1.10$ & $1.684^{+0.004}_{-0.004}$ & $0.801^{+0.014}_{-0.015}$ &   $-14^{+5}_{-5}$ & $0.213^{+0.004}_{-0.004}$ &   $-21^{+1}_{-1}$ \\
DES J044805-580721 &           $1.21$ & $2.546^{+0.014}_{-0.014}$ & $0.907^{+0.020}_{-0.020}$ & $-85^{+13}_{-12}$ & $0.184^{+0.006}_{-0.006}$ &   $-13^{+2}_{-2}$ \\
DES J045352-502234 &           $1.35$ & $2.953^{+0.005}_{-0.005}$ & $0.809^{+0.015}_{-0.014}$ &    $37^{+4}_{-4}$ & $0.067^{+0.004}_{-0.004}$ &   $-22^{+4}_{-3}$ \\
DES J053444-534716 &           $1.26$ & $1.633^{+0.003}_{-0.003}$ & $0.688^{+0.010}_{-0.014}$ &     $1^{+2}_{-2}$ & $0.176^{+0.003}_{-0.004}$ &    $-0^{+1}_{-1}$ \\
DES J053804-473513 &           $1.29$ & $2.448^{+0.013}_{-0.012}$ & $0.481^{+0.018}_{-0.017}$ &    $33^{+2}_{-3}$ & $0.212^{+0.006}_{-0.005}$ &    $18^{+2}_{-1}$ \\
DES J060653-585843 &           $1.92$ & $2.401^{+0.006}_{-0.005}$ & $0.692^{+0.011}_{-0.012}$ &    $79^{+3}_{-2}$ & $0.222^{+0.003}_{-0.002}$ &   $-84^{+1}_{-1}$ \\
DES J060653-585843* &          $1.14$ & $2.063^{+0.029}_{-0.017}$ & $0.904^{+0.024}_{-0.034}$ &  $14^{+18}_{-14}$ & $0.074^{+0.011}_{-0.004}$ &    $14^{+8}_{-3}$ \\
DES J201419-575701 &           $0.96$ & $3.085^{+0.014}_{-0.014}$ & $0.921^{+0.025}_{-0.024}$ &  $-3^{+25}_{-23}$ & $0.145^{+0.008}_{-0.008}$ &   $-25^{+3}_{-3}$ \\
DES J202855-523118 &           $1.51$ & $2.522^{+0.017}_{-0.004}$ & $0.788^{+0.011}_{-0.035}$ &  $-11^{+3}_{-10}$ & $0.218^{+0.003}_{-0.004}$ &   $-26^{+1}_{-1}$ \\
DES J211243+000920 &           $1.31$ & $2.690^{+0.015}_{-0.015}$ & $0.925^{+0.019}_{-0.019}$ & $-28^{+15}_{-16}$ & $0.153^{+0.006}_{-0.006}$ &    $77^{+2}_{-2}$ \\
DES J221912-434835 &           $1.17$ & $2.424^{+0.021}_{-0.022}$ & $0.812^{+0.022}_{-0.020}$ &   $-73^{+7}_{-7}$ & $0.089^{+0.009}_{-0.008}$ &   $-21^{+6}_{-5}$ \\
DES J224221+001144 &           $1.15$ & $2.436^{+0.012}_{-0.011}$ & $0.823^{+0.018}_{-0.018}$ &    $12^{+6}_{-6}$ & $0.119^{+0.006}_{-0.007}$ &    $-4^{+3}_{-3}$ \\
DES J233459-640406 &           $1.18$ & $2.564^{+0.002}_{-0.003}$ & $0.660^{+0.019}_{-0.011}$ &   $-38^{+3}_{-3}$ & $0.153^{+0.005}_{-0.003}$ &   $-41^{+1}_{-1}$ \\
DES J235519-613637 &           $0.99$ & $1.343^{+0.004}_{-0.004}$ & $0.527^{+0.008}_{-0.009}$ &   $-18^{+1}_{-1}$ & $0.051^{+0.004}_{-0.004}$ &   $-14^{+5}_{-4}$ \\
DES J233551-515217 &           $1.13$ & $3.597^{+0.013}_{-0.013}$ & $0.746^{+0.015}_{-0.015}$ &    $56^{+4}_{-4}$ & $0.027^{+0.007}_{-0.007}$ &  $64^{+15}_{-15}$ \\
\end{longtable}
\tablefoot{\tablefoottext{*}{Model result after manually redoing the mask.}}
\renewcommand*{\arraystretch}{1.}
}

\begin{acknowledgements}

This work is supported by the Swiss National Science Foundation (SNSF) and by the European Research Council (ERC) under the European Union’s Horizon 2020 research and innovation program (COSMICLENS: grant agreement No 787886).

RC thanks the Max Planck Society for support through the Max Planck Research Group of S.~H.~Suyu. This project has received funding from the European Research Council (ERC) under the European Unions Horizon 2020 research and innovation programme (LENSNOVA: grant agreement No 771776). 

GV has received funding from the European Union’s Horizon 2020 research and innovation programme under the Marie Sklodovska-Curie grant agreement No 897124

This research has made use of the VizieR catalogue access tool, CDS, Strasbourg, France (DOI : 10.26093/cds/vizier). The original description of the VizieR service was published in \cite{vizier}.
This research has made use of the SIMBAD database, operated at CDS, Strasbourg, France \cite{simbad}.

This material is based upon work supported by the National Science Foundation under  Cooperative  Agreement  1258333  managed  by  the  Association  of  Universities for Research in Astronomy (AURA), and the Department of Energy under Contract  No.  DE-AC02-76SF00515 with  the  SLAC  National  Accelerator  Laboratory.  Additional funding  for  {\it Rubin}  Observatory  comes  from  private  donations, grants to universities, and in-kind support from LSSTC Institutional Members.

This work rely on the following packages: {\tt numpy} \citep{VanDerWalt2011}, {\tt scipy} \citep{Virtanen2020}, {\tt matplotlib} \citep{Hunter2007}, {\tt astropy} \citep{astropy2018} and \href{https://github.com/herjy/scarlet_extensions/releases/tag/paper}{\faGithub \tt scarlet\_extensions} and {\tt csv}. A notebook detailing the procedure of deblending can be found in the following repository: \href{https://github.com/herjy/Lens_deblend/releases/tag/prototype}{\faGithub Lens-Deblend}.

\end{acknowledgements}

%
%
\bibliographystyle{aa}
\bibliography{bibliography}
\end{document}